\documentclass[a4paper,fleqn,usenatbib]{mnras}
\usepackage{graphicx}
\usepackage{amsmath}
\usepackage{amssymb}
\usepackage{txfonts}
\usepackage{subcaption}
\usepackage{upgreek}
\usepackage{rotating}
\usepackage{tabularx}
\usepackage[usenames,dvipsnames,svgnames,table]{xcolor}
\usepackage{txfonts}
\DeclareSymbolFont{cmletters}{OML}{cmm}{m}{it}
\DeclareMathSymbol{v}{\mathalpha}{cmletters}{"76}
\newcommand{\GG}[1]{}

\newcommand{\RedeclareMathOperator}[2]{\renewcommand{#1}{}\let#1\relax\DeclareMathOperator{#1}{#2}}

\newcommand{\hammer}{{\texttt{H-AMR}}}

\newcommand{\bhoss}{{\texttt{BHOSS}}}
\newcommand{\tilta}{{\texttt{T0}}}
\newcommand{\tiltb}{{\texttt{T30}}}
\newcommand{\tiltc}{{\texttt{T60}}}

\newcommand\simless\lesssim
\newcommand\simgreat\gtrsim

\title[Imaging tilted disks and jets]{Observational signatures of disk and jet misalignment in images of accreting black holes}
\author[Chatterjee, Younsi, Liska, Tchekhovskoy, Markoff, Yoon, Eijnatten, Hesp, Ingram \& van der Klis]{K. Chatterjee$^{1}$\thanks{E-mail: k.chatterjee@uva.nl}, Z. Younsi$^{2,3}$, M. Liska$^{4,1}$, A. Tchekhovskoy$^{5}$, S. B. Markoff$^{1,6}$,  \newauthor D. Yoon$^{1}$, D. van Eijnatten$^{1}$, C. Hesp$^{1}$, A. Ingram$^{7}$ \& M. B. M. van der Klis$^{1}$\\
$^{1}$Anton Pannekoek Institute for Astronomy, University of Amsterdam, Science Park 904, 1098 XH Amsterdam, The Netherlands\\
$^{2}$Mullard Space Science Laboratory, University College London, Holmbury St.~Mary, Dorking, Surrey, RH5 6NT, United Kingdom\\
$^{3}$Institut f{\"u}r Theoretische Physik, Goethe-Universit{\"a}t Frankfurt, Max-von-Laue-Stra{\ss}e 1, D-60438 Frankfurt am Main, Germany\\
$^{4}$Institute for Theory and Computation, Harvard University, 60 Garden Street, Cambridge, MA 02138, USA; John Harvard Distinguished Science and ITC\\ Fellow\\
$^{5}$Center for Interdisciplinary Exploration \& Research in Astrophysics (CIERA), Physics \& Astronomy, Northwestern University, Evanston, IL 60202, USA\\
$^{6}$Gravitation Astroparticle Physics Amsterdam (GRAPPA) Institute, University of Amsterdam, Science Park 904, 1098 XH Amsterdam, The Netherlands\\
$^{7}$Department of Physics, Astrophysics, University of Oxford, Denys Wilkinson Building, Keble Road, Oxford, OX1 3RH, UK\\
}
\date{Accepted XXX. Received YYY; in original form ZZZ}
\pubyear{2019}
\begin{document}
\label{firstpage}
\pagerange{\pageref{firstpage}--\pageref{lastpage}} 
\maketitle

\begin{abstract}
Black hole accretion is one of nature's most efficient energy extraction processes.
When gas falls in, a significant fraction of its gravitational binding energy is either converted into radiation or flows outwards in the form of black hole-driven jets and disk-driven winds.
Recently, the Event Horizon Telescope (EHT), an Earth-size sub-millimetre radio interferometer, captured the first images of M87's black hole.
These images were analysed and interpreted using general-relativistic magnetohydrodynamics (GRMHD) models of accretion disks with rotation axes aligned with the black hole spin axis.
However, since infalling gas is often insensitive to the black hole spin direction, misalignment between accretion disk and black hole spin may be a common occurrence in nature.
In this work, we use the general-relativistic radiative transfer (GRRT) code \bhoss{} to calculate the first synthetic radio images of (highly) tilted disk/jet models generated by our GPU-accelerated GRMHD code \hammer{}.
While the tilt does not have a noticeable effect on the system dynamics beyond a few tens of gravitational radii from the black hole, the warping of the disk and jet can imprint observable signatures in EHT images on smaller scales. Comparing the images from our GRMHD models to the 43 GHz and 230 GHz EHT images of M87, we find that M87 may feature a tilted disk/jet system. Further, tilted disks and jets display significant time variability in the 230 GHz flux that can be further tested by longer-duration EHT observations of M87.  
\end{abstract}


\section{Introduction}
\label{sec:introduction}

There is observational evidence for misalignment between the accretion disk and black hole (BH) spin axis in both active galactic nuclei (AGN), BH X-ray binaries (XRBs) \citep[e.g., ][]{hjellming95, greene01,maccarone2002b,caproni06,vandeneijnden17, russellT_2019} and tidal disruption events \citep[e.g.,][]{Pasham2019}, with theoretical studies of the growth of supermassive black holes (SMBHs) favouring randomly oriented accretion and affecting BH spin evolution \citep[e.g.,][]{volonteri05,king06}.
In particular, periodic variations in the jet position with respect to the line of sight have been invoked for explaining quasi-periodic oscillations (QPOs) in the emission of some sources \citep[e.g.,][]{stella98, ingram09}.
The rotational plane of BH accretion disks is therefore expected to be misaligned with respect to the BH spin axis, as infalling gas from large distances will typically not be sensitive to the direction of rotation of the central BH.
However, there is still much uncertainty over this claim; but, with recent advancements in very long baseline interferometry (VLBI) techniques, most notably the Event Horizon Telescope \citep[EHT; ][]{doeleman2008, EHT_paperI}, as well as growing interest in space VLBI \citep[e.g.,][]{Roelofs_2019, Palumbo_2019, Fish2019_space_VLBI}, imaging the near-horizon region (i.e., $r \lesssim 20~r_{\rm g}$, where $r_{\rm g}\equiv GM/c^2$ is the gravitational radius of the BH, $M$ is its mass, $G$ is the gravitational constant and $c$ is the speed of light) for SMBHs has become a reality, making it possible to directly test for misalignment in Sagittarius A* (Sgr~A*) and M87. In fact, \citet{Park_2019} briefly discussed possible misalignment in M87.

Misalignment brings about important changes in the dynamics of the system via general relativistic (GR) frame-dragging, which induces nodal Lense-Thirring precession \citep[LT; ][]{lense18} of test particles on tilted orbits around the central object, with a radially-dependent angular frequency $\Omega_{\rm LT}\propto 1/r^3$. Growing interest in the physics of accretion under the effects of LT precession led to modelling tilted accretion disks via GRMHD simulations of both thick \citep[e.g., ][]{fragile05,fragile07,mckinney_2013,polko17,liska_tilt_2018,White_2019} and thin \citep[e.g.,][]{liska_BP_2019} tilted disks, some of which carry promising indications for the origin of specific kinds of QPOs \citep[e.g.,][]{liska2019_tearing}.
Of course, the absence of QPOs due to disk precession does not rule out misalignment, since tilted geometrically-thick disks tend to have extremely long precession periods, resulting in quasi-stationary disk warps.
Due to GR warping of the disk via pressure waves \citep{papaloizou95,ivanov97,lubow00}, \citet{Dexter_2011} showed, via GRRT of GRMHD simulations, that tilting the disk brings about significant changes in the appearance of the inner $20~r_{\rm g}$ around a BH, e.g., when scaled to the mass and distance of Sagittarius A* \citep{dexter_2013}. GRMHD simulations found that tilted disks accrete onto the BH via two high density plunging streams of infalling material \citep[e.g.,][]{fragile2008,liska_tilt_2018}.
Further, it has been shown that the narrow and highly warped morphology of these plunging streams can lead to the accumulation of gas near the point of highest disk tilt, developing a pair of standing shocks \citep{fragile2008,Generozov_2013,White_2019}.
Consequently, higher inflow temperatures resulting from shock heating dominate the emission \citep[e.g., ][]{dexter_2013,White_2020_tiltedimages}. 

\begin{table}

\begin{center}
\begin{tabularx}{\columnwidth}{l c c c c c}
\hline\hline
Model & a & Resolution& $r_{\rm in}$ & $r_{\rm max}$ & $r_{\rm out}$ \\
&&($N_{\rm r}\times N_{\rm \theta}\times N_{\rm \varphi}$)&[$r_{\rm g}$]&[$r_{\rm g}$]&[$r_{\rm g}$]\\
\hline
All & 0.9375 & $448\times144\times240$& 12.5 & 25 & $10^5$  \\
\hline
\hline
Model &  $\mathcal{T}_{\rm init}$    & $Q$-factor$^{*}$ & $\phi_{\rm BH}^{*}$ & $\mathcal{T}_{\rm j}^{*}$, $\mathcal{P}_{\rm j}^{*}$& $t_{\rm sim}$\\
 & [deg]  & $(Q_{r}, Q_{\theta}, Q_{\varphi})$ & & [deg, deg]&[$10^5 t_{\rm g}]$\\
\hline
\tilta{} & 0  & (18, 14, 99) & 52.8&  0.8, --- & 1.32\\
\tiltb{} & 30  & (18, 26, 78)&45& 22.7, 28.4& 1.19\\
\tiltc{} & 60  & (13, 28, 45) &27.8&43.4, 42.2& 1.48\\
\hline\hline
\end{tabularx}
\end{center}
\caption{Top: parameters common to all models used in this work: dimensionless BH spin (a), simulation grid resolution, disk inner radius $r_{\rm in}$, disk pressure-maximum radius $r_{\rm max}$, outer grid radius $r_{\rm out}$.
Bottom: model names, initial disk misalignment, density-weighted volume-averaged MRI quality factors $Q_{r,\theta,\varphi}$ (see text for definition), dimensionless horizon magnetic flux $\phi_{\rm BH}$ [eqn.~\eqref{eqn:phibh}], jet tilt $\mathcal{T}_{\rm j}$ and precession $\mathcal{P}_{\rm j}$ angles spatially averaged over $[50,150]~r_{\rm g}$, and total simulation time $t_{\rm sim}$.
Quantities marked with $^*$ are time-averaged over [99960, 100960] $t_{\rm g}$, where $t_{\rm g} \equiv r_{\rm g}/c$.}
\label{tab:models}
\end{table}

The magnetic field strength plays a vital role in a tilted disk system because magnetic fields provide an extra torque to the GR warping, and hence help in aligning the disk.
In particular, \citet{liska_tilt_2018} found that for tilted precessing thick disks, the higher the disk magnetic field strength, the stronger the jets are and the more they tend to push the inner parts of the disk to align more closely with the BH spin axis. This can be understood because when the magnetic flux onto the BH is large enough to hinder accretion from the disk \citep{narayan03}, known as the magnetically-arrested disk (MAD) condition, the associated jet efficiently extracts the BH's rotational energy \citep{tchekhovskoy12} and becomes powerful enough to force the inner part of the disk to align with the BH spin axis \citep{mckinney_2013}, while the large-scale jet remains aligned with the disk.
Further, it was shown that in the absence of magnetic fields, the disk alignment is much weaker as compared to the MHD case \citep{sorathia2013_MHD_LT}.
Strongly magnetised jets can dictate dynamics along with jet-disk interactions, potentially developing shocks similar in nature to the pair of standing shocks present in the plunging streams.
Misalignment can therefore introduce potential degeneracies in observed images and spectra, warranting further exploration so that we are able to more clearly interpret current and future EHT horizon-scale images.
However, to date, images illustrating the warping of misaligned disk/jet systems (see also \citealt{White_2020_tiltedimages} for misaligned disk images) and corresponding observational indications of tilt in spectra have not been studied and compared to observed images \citep[e.g.,][]{EHTPaperV}. 

\begin{figure*} 
\begin{center}
    \includegraphics[width=\textwidth,trim=0cm 0cm 0cm 0cm,clip]{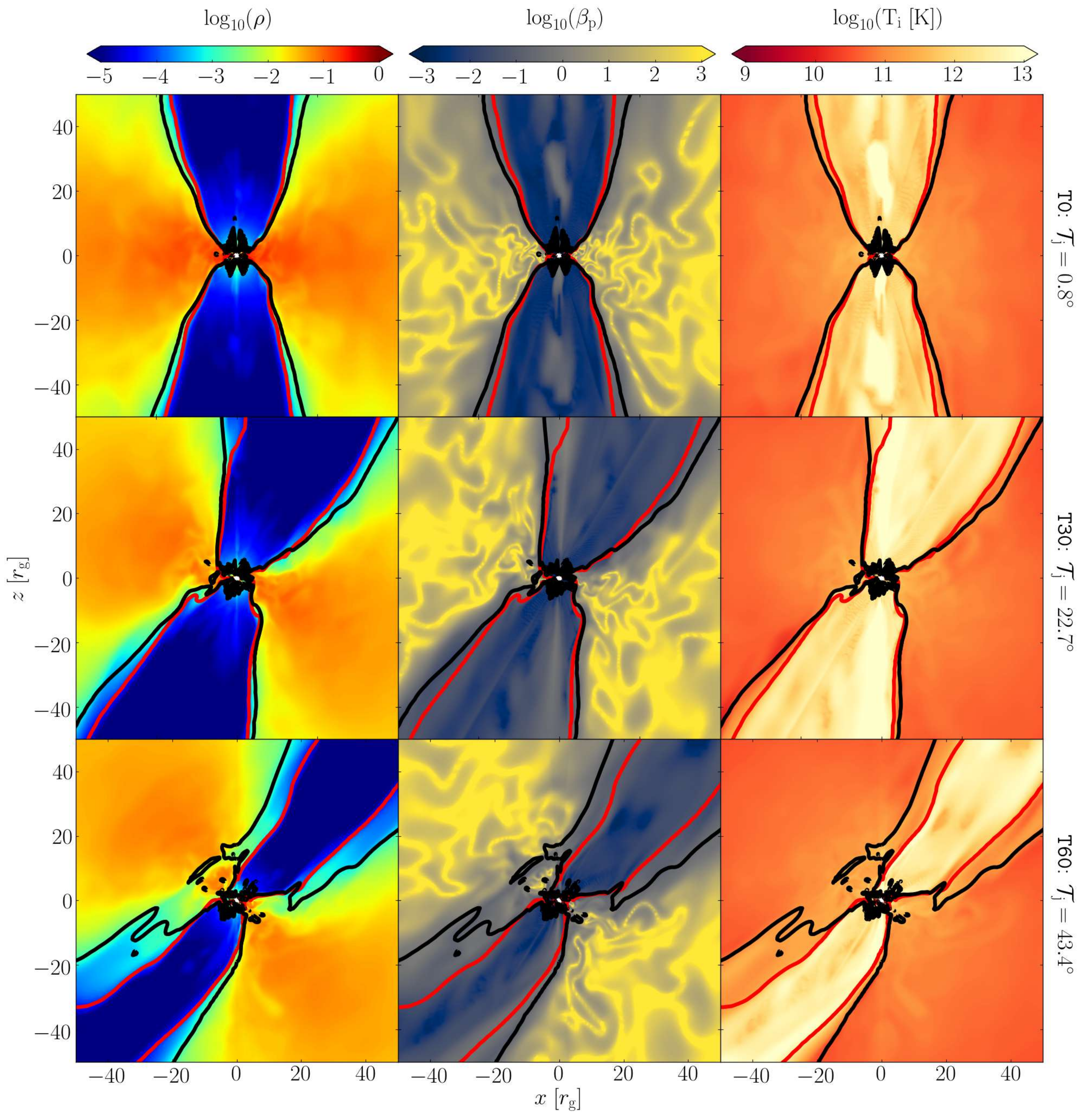}
    \caption{We show the 2D cross-sectional plots of the gas density and plasma-$\beta$ in code units, and the ion temperature $T_{\rm i}$ (in Kelvin) for each tilt model: \tilta{}, \tiltb{}, and \tiltc{}, spanning $100~r_{\rm g} \times 100~r_{\rm g}$.
    The 2D plane created by the BH spin vector and the large-scale jet angular momentum vector is displayed, rotating the x-z plane by the corresponding jet precession angle $\varphi_0=[0^{\circ},28.4^{\circ},42.2^{\circ}]$ for \tilta{}, \tiltb{}, and \tiltc{}, respectively (given in Table~\ref{tab:models}).
    The red and black lines in all plots denote the magnetisation, $b^2/\rho c^{2}=1$, and the Bernoulli parameter, $Be=1.02$, contours, respectively.
    }
    \label{fig:GRMHD_contour}
\end{center}
\end{figure*}

\begin{figure} 
    \includegraphics[width=0.96\columnwidth,trim=0cm 0cm 0cm 0cm,clip]{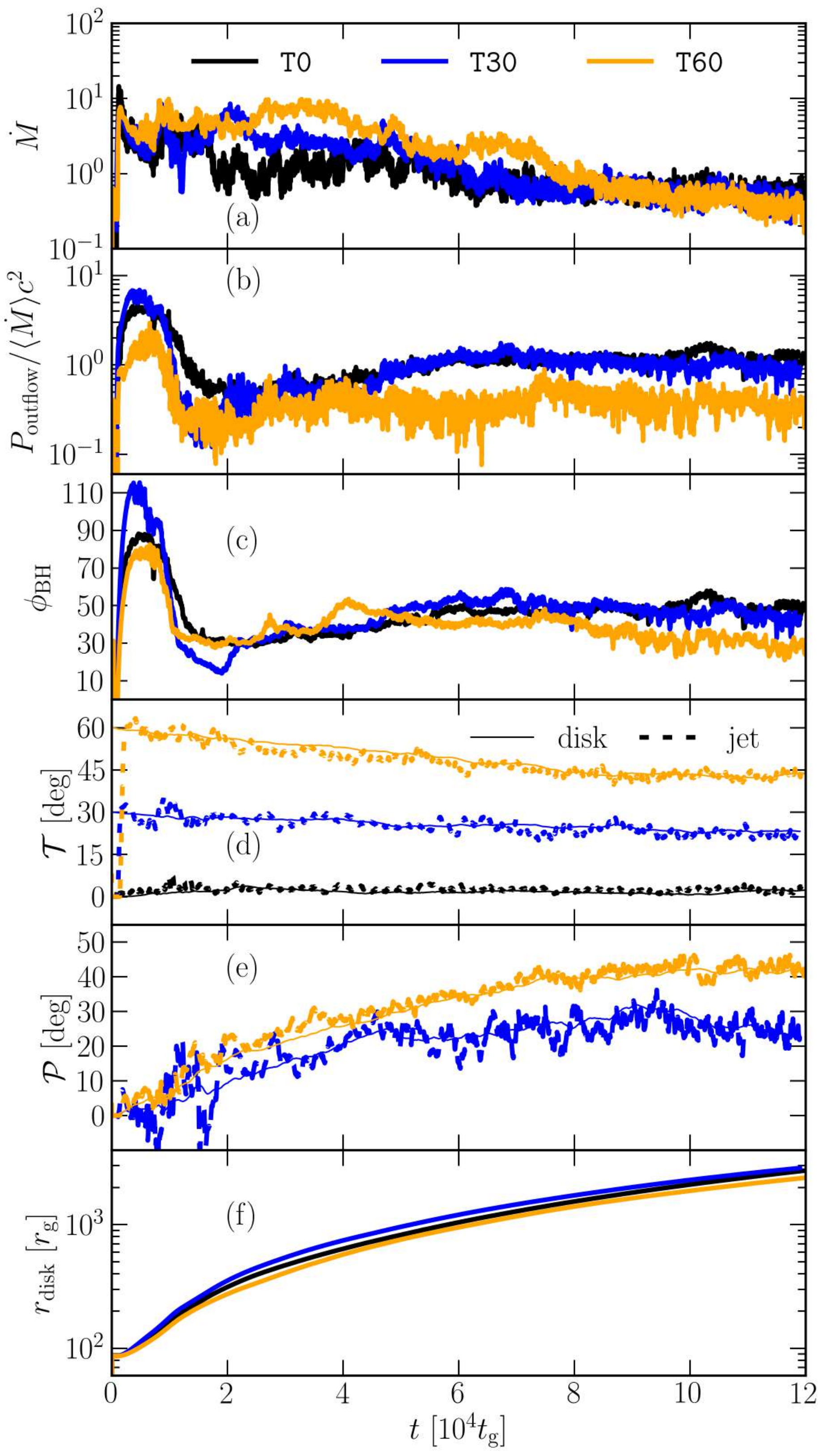}
    \caption{Time dependence of simulation quantities reveal how BH-disk systems evolve with varying initial misalignment angles, namely $\mathcal{T}_{\rm init}=0^{\circ}$ (\tilta{}), $30^{\circ}$ (\tiltb{}) and $60^{\circ}$ (\tiltc{}).
    The accretion rate, $\dot{M}$, is presented in normalised units (a), total outflow power $P_{\rm outflow}$ normalised by accretion rate (time-averaged over [99960, 100960] $t_{\rm g}$) (b), dimensionless magnetic flux $\phi_{\rm BH}$ in Gaussian units (c), with all quantities calculated at the event horizon.
    We also show the tilt angle $\mathcal{T}(t)$ and precession angle $\mathcal{P}(t)$ (d,e) of the disk, along with the jet for models \tiltb{} and \tiltc{}, as $\mathcal{P}$ for $\mathcal{T}_{\rm init}=0^{\circ}$ is not well-defined.
    Both the angles are radially averaged over [50, 150] $r_{\rm g}$. It is noteworthy that the jet orientation is more variable than that of the disk. 
    The barycentric radius of the disk $r_{\rm disk}$ [eqn.~\eqref{eqn:barycentric_radius}] (f) shows that all three models display similar viscous spreading, and therefore indicate similar rates of outward angular momentum transfer.
    }
    \label{fig:time_grmhd}
\end{figure}
In this work we explore, for the first time, a variety of initial misalignment angles for a BH disk/jet system in high resolution using our GPU-accelerated GRMHD code \hammer{} \citep{liska_hamr2020_arxiv}.
We further calculate observable images using the GRRT code \bhoss{} \citep{younsi_2016,younsi_2019_polarizedbhoss}.
In Sec.~\ref{sec:code-and-setup}, we give an overview of our methodology and simulation setup.
We present our results in Sec.~\ref{sec:results}.
In Sec.~\ref{sec:discussion}, we compare our images to the recently published EHT image of M87 \citep{EHT_paperI}.
We conclude in Sec.~\ref{sec:conclusions}.

\section{Methodology and Numerical Setup}
\label{sec:code-and-setup}

We use our state-of-the-art massively parallel, GPU-accelerated 3D GRMHD code \hammer{} \citep{liska_tilt_2018, chatterjee2019,liska_hamr2020_arxiv} to solve the GRMHD equations in a fixed Kerr spacetime.
The \hammer{} section in \citet{Porth2019} presents a description of the code attributes, such as adaptive mesh refinement (AMR), local adaptive time-stepping and a staggered mesh for the magnetic field evolution, as well as comparisons to benchmark results for a standard accretion disk problem. We adopt the geometrical unit convention, $G=c=1$, and further normalise the BH mass to $M=1$, thereby normalising the length scale to the gravitational radius $r_{\rm g}$.
We carry out the simulations in logarithmic Kerr-Schild coordinates with a numerical resolution $N_r \times N_{\theta} \times N_{\varphi}$ of $448\times 144 \times 240$ (Table~\ref{tab:models}), sufficient to resolve the magnetorotational instability \citep[MRI; ][]{bal91} in the disk. Our grid is axisymmetric and uniform in $\log (r/r_{\rm g})$, extending from $0.75~r_{\rm hor}$ to $10^{5}~r_{\rm g}$, where the event horizon radius $r_{\rm hor}\equiv r_{\rm g}\left(1+\sqrt{1-a^2}\right)$, with $a=0.9375$. A further description of the grid is given in \citet{liska_tilt_2018}. 
We use outflowing boundary conditions (BCs) at the inner and outer $r$ boundaries; transmissive polar BCs in the $\theta-$direction and periodic BCs in the $\varphi-$direction. 
To quantify the MRI resolution, we calculate the quality factors $Q_{r,\theta,\varphi}$, where $Q_{i}=\langle2\pi v_A^{i}\rangle_w/\langle\Delta^{i}\Omega\rangle_w$ measures the number of cells per MRI wavelength in direction $i=$[$r,\theta,\varphi$], volume-averaged over the disk ($r<150~r_{\rm g}$) with weight $w=\rho$, the gas density \citep[see eqns. (18-20) in][]{liska_BP_2019}.
Here $v_A^i$ is the Alfv\'en velocity, $\Delta^{i}$ the cell size, and $\Omega$ the angular velocity of the fluid.
We achieve $Q_{\theta}\gtrsim10$ (Table~\ref{tab:models}) during our chosen time period, fulfilling the numerical convergence criteria \citep[see e.g., ][]{hgk11}.

\begin{figure} 
    \includegraphics[width=3.1in,trim=0cm 0cm 0cm 0cm,clip]{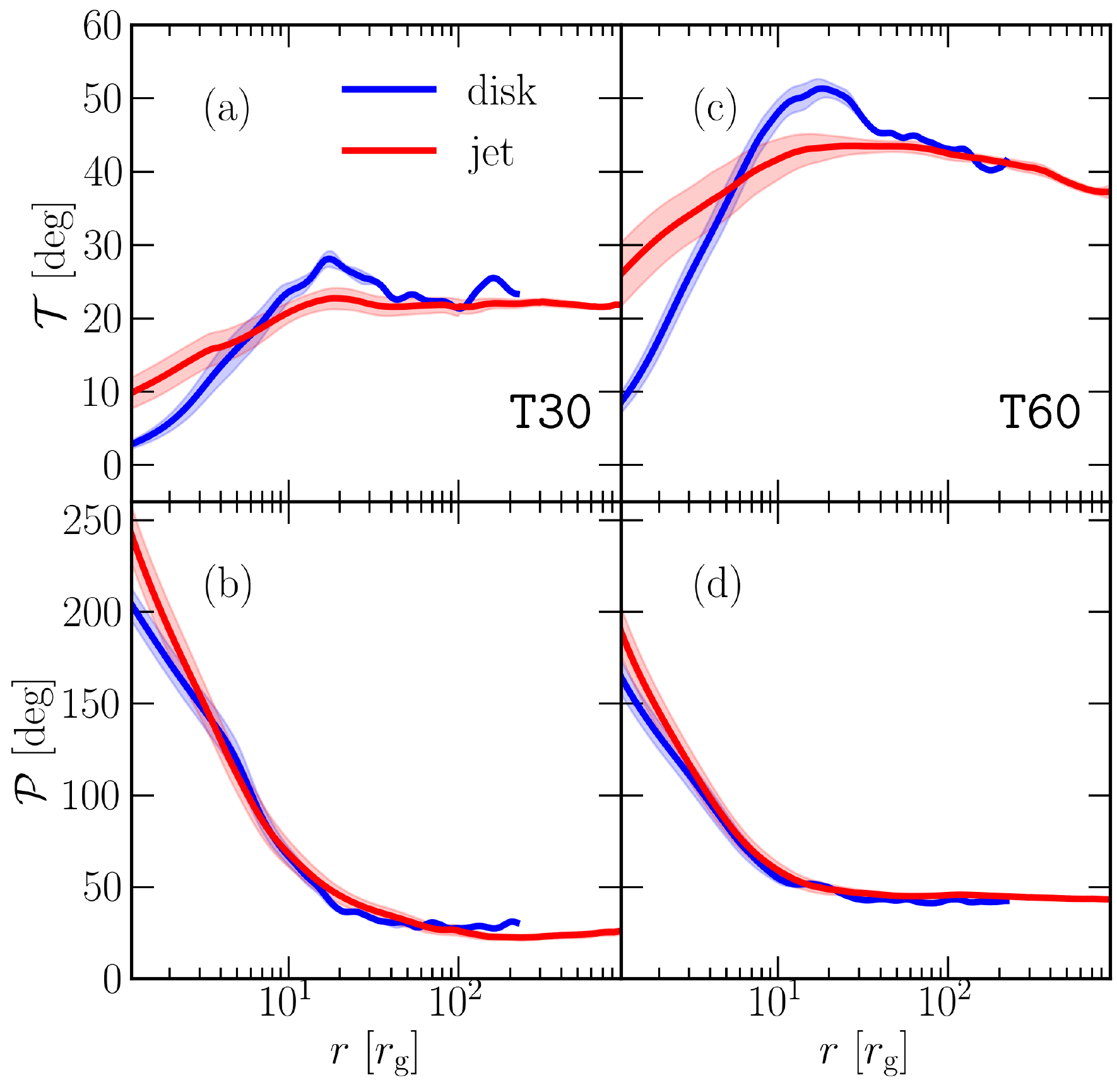}
    \caption{Misaligned disks show disk warping and jet bending.
    We show, for models \tiltb{} (a,b) and \tiltc{} (c,d), the tilt angle $\mathcal{T}$, time-averaged over [99960, 100960] $t_{\rm g}$, plotted with $1\sigma$ deviation (a,c).
    The precession angle $\mathcal{P}$, with the $1\sigma$ deviation (b,d) illustrates that the inner disk precesses much more than the outer disk and hence displays a twist in the disk. The inner parts of both the disk and the jet are quite variable as indicated by the extended shaded regions.
    We hereafter redefine the $\varphi=0^{\circ}$ plane as the plane in which the BH spin vector and the large-scale jet angular momentum vector resides for each individual simulation.
    }
    \label{fig:radius_grmhd}
\end{figure}

In all models we start with a standard \citet{fis76} torus in
hydrostatic equilibrium around the central (spinning) Kerr BH.
The torus inner edge is located at $r_{\rm in}=12.5~r_{\rm g}$ and the gas pressure ($p_{\rm g}$)-maximum is at $r_{\rm max}=25~r_{\rm g}$, with the ideal gas law adiabatic index set to $\Gamma=5/3$ (i.e., non-relativistic).
The non-zero magnetic field vector potential is given by $A_{\varphi}\propto (\rho-0.05)^2(r/r_{\rm g})^3$ and normalised to $\max(p_{\rm g})/\max(p_{\rm B})=100$, where $p_{\rm g}\equiv(\Gamma-1)u_{\rm g}$ is the gas pressure, $p_{\rm B}\equiv b^{2}/2$ is the magnetic pressure, and $u_{\rm g}$ is the fluid internal energy.
Furthermore, the magnetic field 4-vector, $b^{\mu}$, is defined in Lorentz-Heaviside units where a factor of $1/\sqrt{4\pi}$ is absorbed into its definition.
Note that our model parameters are slightly different from the tilted disk models used in \citet{fragile07}, where the authors use $r_{\rm in}=15~r_{\rm g}$ and $r_{\rm max}=25~r_{\rm g}$, an adiabatic index of $\Gamma=5/3$, and vector potential $A_{\varphi}\propto \rho$, resulting in a smaller magnetic flux content in the torus.
We also note that \citet{White_2019} use similar parameters to \citet{fragile07}, but with $\Gamma=4/3$.
We tilt our disks at 3 different angles: $\mathcal{T}_{\rm init}=0^{\circ}$
(model \tilta{}), $30^{\circ}$ \citep[model \tiltb{}; previously reported as model
``S25A93'' in ][]{liska_tilt_2018} and $60^{\circ}$ (model \tiltc{}) with
respect to the plane perpendicular to the BH spin axis (see
Table~\ref{tab:models}).
GRMHD simulations suffer from numerical errors in the jet funnel as pockets of low density gas are created when matter either falls inwards due to the BH's gravity or is expelled outwards via magnetic forces, leading to truncation errors in the solution.
In our simulations, we replenish the near-vacuum regions by ad-hoc mass-loading the jet funnel in the drift frame of the magnetic field by following \citet{ressler_2017} and further employ a floor model with $\rho c^2 \geq \max\left[p_{\rm B}/50,\,2\times10^{-6}c^2(r/r_{\rm g})^{-2}\right]$ and $u_{\rm g} \geq \max\left[p_{\rm B}/150,\,10^{-7}c^2(r/r_{\rm g})^{-2\Gamma}\right]$. 

In order to capture the radiation output from our GRMHD simulations and calculate synthetic radio-frequency images which are comparable with observational data, we perform GRRT calculations in post-processing using \bhoss{}: by ray-tracing the GRMHD data and integrating the GRRT equations, the spectral properties of the emergent light produced in the strong gravitational field around the BH are calculated.
The geodesic equations themselves are integrated backwards (from observer-to-source) using a 5\textsuperscript{th} order adaptive timestep Runge-Kutta-Fehlberg scheme (to maintain computational speed and accuracy), and the GRRT equations are integrated in tandem, using the reverse-integration approach introduced in \citet{younsi_2012}, including all emission, absorption and optical depth effects of the plasma.

We set the camera latitude $\theta_0$ (otherwise known as the observer inclination angle) and longitude $\varphi_0$ at a distance of $1000~r_{\rm g}$ from the BH.
The radiative transfer requires the BH mass and the mass accretion rate to set the physical scales of the simulation.
We choose our target central BH to be that of the galaxy M87, dubbed ``M87*'', with a mass $M=6.5\times 10^9 M_{\odot}$ at a distance $D=16.8$ Mpc \citep{EHTPaperVI}. For the rest of the text, we only use ``M87'' when referring to the BH and jet properties. In this work, we consider a relativistic thermal Maxwell-J\"{u}ttner electron distribution function for the synchrotron absorption and emission \citep[given by][]{Leung_2011}, with the ion-electron temperature ratio set according to the \citet{Moscibrodzka_2016} prescription, namely:
\begin{equation}
    T_{\rm i}/T_{\rm e}:=\frac{R_{\ell}+\beta_{\rm p}^2 R_{\rm h}}{1+\beta_{\rm p}^2}\,,
    \label{eqn:temp_ratio}
\end{equation}
\noindent where $\beta_{\rm p}\equiv p_{\rm g}/p_{\rm B}$ is the plasma-$\beta$ parameter, and $T_{\rm i}$ and $T_{\rm e}$ are the ion and electron temperatures, in Kelvin, respectively. This assumption for the electron temperature model is the same as used in \citet[][]{EHTPaperV}. The dimensionless electron temperature is defined as:
\begin{equation}
    \Theta_{\rm e}:=\frac{p_{\rm g}}{\rho} \frac{m_{\rm p}/m_{\rm e} }{T_{\rm i}/T_{\rm e}}\,,
    \label{eqn:e_temp}
\end{equation}

\noindent where $m_{\rm p}$ and $m_{\rm e}$ are the proton and electron masses, respectively.
The electron temperature in c.g.s. units is therefore simply $T_{\rm e}= m_{\rm e} c^2 \Theta_{\rm e}/k_{\rm B}$, where $k_{\rm B}$ is the Boltzmann constant.
In practice, eqn.~\eqref{eqn:temp_ratio} sets the temperature ratio in the jet ($\beta_{\rm p}\ll 1$) to be $T_{\rm i}/T_{\rm e} \sim R_{\ell}$ and in the disk ($\beta_{\rm p}\gg 1$), $T_{\rm i}/T_{\rm e} \sim R_{\rm h}$. 
For our models, we fix $R_{\ell}=1$, while we vary $R_{\rm h}=(1,10,100)$.
Additionally, we exclude all emission from the jet spine (defined where $b^2/\rho c^2 \geq 1$, where $b$ is the fluid-frame magnetic field strength) as the gas density and temperature in this region may be affected by the simulation floors and are therefore unreliable.

\section{Results}
\label{sec:results}

\begin{figure*}
\begin{center}
\centering
\includegraphics[height=3.42in,trim=0cm 0cm 0cm 0cm,clip]{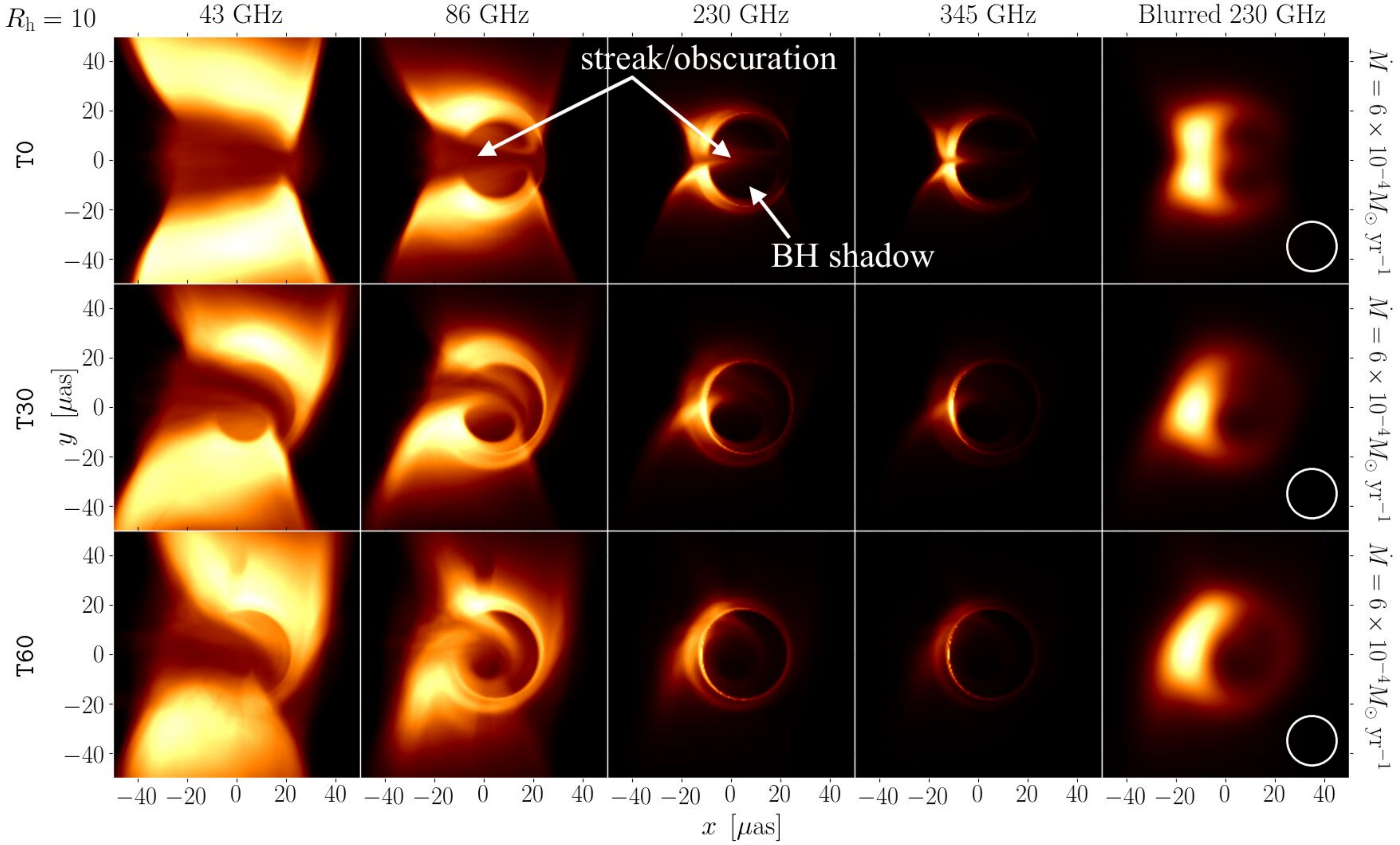}\hfill
\includegraphics[height=3.42in,trim=0cm 0cm 0cm 0cm,clip]{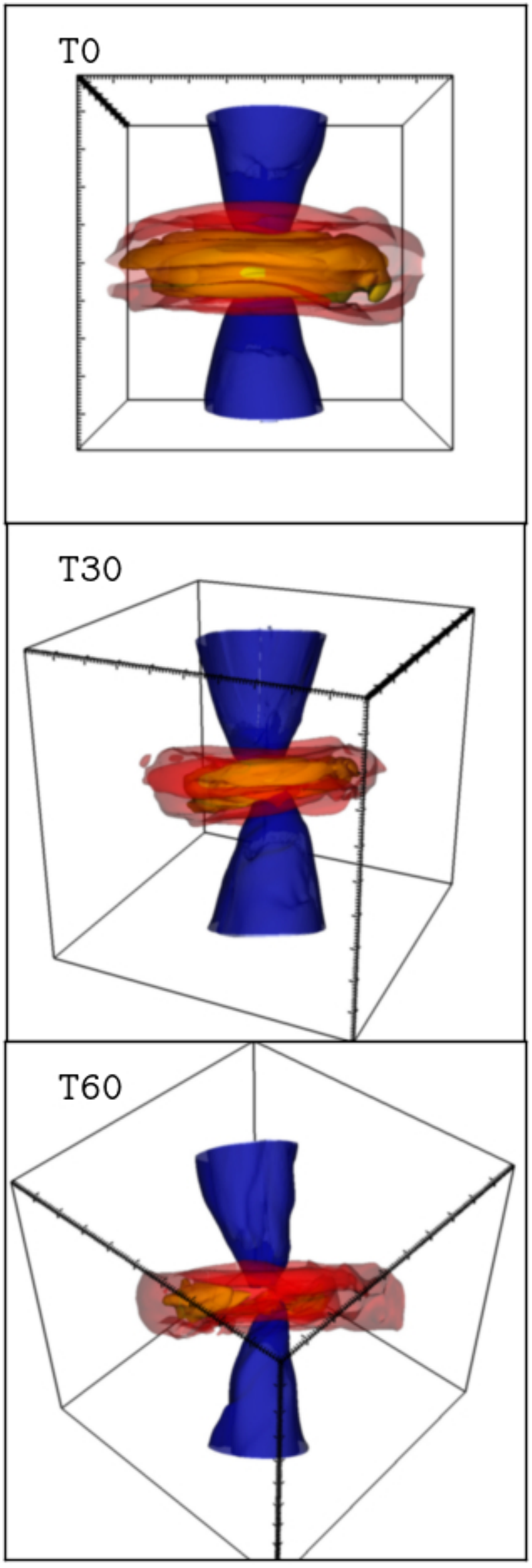}

    \caption{Left figure: time-averaged, edge-on synthetic GRRT images of misaligned BH disk/jet systems using $R_{\rm h}$=10 for the electron temperature prescription (see eqn.~\eqref{eqn:temp_ratio}), scaled to M87 at a constant accretion rate $\dot{M}=6\times 10^{-4}~M_{\odot} \, {\rm yr}^{-1}$, calculated over [99960, 100960]~$t_{\rm g}$ and zoomed-in to the inner $100\times 100~\upmu {\rm as}^2$ region.
    The observer is positioned edge-on relative to the large-scale disk using the disk tilt and precession angle values from Fig.~\ref{fig:radius_grmhd}, i.e., ($\theta_0, \varphi_0$)=($90^{\circ},0^{\circ}$). The colour scale is linear and normalised to unity for each image.
    From left to right: images at frequencies of 43 GHz, 86 GHz, 230 GHz and 345 GHz, along with the 230 GHz image convolved with a gaussian beam (full-width half-maximum of $20 \, \upmu {\rm as}$).
    From top to bottom: synthetic observable GRRT images with initial tilt angles of 0$^{\circ}$ (\tilta{}), 30$^{\circ}$ (\tiltb{}) and 60$^{\circ}$ (\tiltc{}).
    Rightmost figure: 3D density contour plot visualisations for (top to bottom): \tilta{}, \tiltb{} and \tiltc{}, respectively, as viewed at the same observer position as for the GRRT images.
    Dimensions are indicated by a bounding cube of size $50~r_{\rm g}$.
    The corresponding cases for $R_{\rm h}=(1,100)$ are shown in Appendix~\ref{sec:other_Rhigh_images}.
    }
    \label{fig:images_avg_rh10}
    \end{center}
\end{figure*}

Figure~\ref{fig:GRMHD_contour} shows the gas density $\rho$, the plasma-$\beta$ ($\beta_{\rm p}$) and the ion temperature $T_{\rm i}$ (in Kelvin) of the simulations at $t=10^5~t_{\rm g}$, clearly illustrating the misalignment of both the high-density disk and the low-density jet with respect to the BH spin direction, which points upwards in this figure.
Each BH system is shown in the plane of the BH spin and the jet angular momentum vector, or, in other words, the $\varphi=0^{\circ}$ plane rotated by the corresponding jet precession angle $\mathcal{P}_{\rm j}$ (see Table~\ref{tab:models}).
Figure~\ref{fig:GRMHD_contour} also delineates the jet funnel (dark blue in the figure), given by $b^2/\rho c^{2}=1$ (red line) and the unbound material given by the Bernoulli parameter $Be:= -h u_t = 1.02$ (black line), with the latter being taken as a proxy for the disk-wind region (light blue-green region). In the definition of $Be$, $h$ and $u_t$ are the enthalpy and the temporal component of the co-variant velocity (interpreted as the conserved particle energy) respectively. Over time, in all three models the accretion disk develops turbulence via the MRI, leading to gas accretion onto the BH [Fig.~\ref{fig:time_grmhd}(a)] in the form of plunging streams \citep{fragile07,liska_tilt_2018}.
Figure~\ref{fig:time_grmhd} (b and c) shows that our chosen initial field configuration evolves to create a near-MAD disk in each case, with the dimensionless magnetic flux through the horizon $\phi_{\rm BH}=\Phi_{\rm BH}/(\langle\dot{M}\rangle r_{\rm g} c^2)^{1/2}\lesssim \phi_{\rm max}\approx 50$ \citep{tch11}, producing highly efficient jets with $P_{\rm outflow}/(\langle \dot{M}\rangle c^2)\sim 1$, powered by BH rotational energy extracted via the Blandford-Znajek mechanism \citep{blandford77}.
Here we make use of the following definitions.
The BH accretion rate (positive for inflow of gas towards the BH):
\begin{equation}
    \dot{M}:=-\iint \rho u^r \, \! dA_{\theta\varphi}\,,
    \label{eqn:mdot}
\end{equation}
the magnetic flux at the event horizon: 
\begin{equation}
    \Phi_{\rm BH}:=\frac{1}{2}\iint |B^{r}| \, dA_{\theta\varphi}\,,
    \label{eqn:phibh}
\end{equation}
and the outflow power:
\begin{equation}
    P_{\rm outflow}:=\dot{M}c^2-\dot{E}\,,
    \label{eqn:power}
\end{equation}
\noindent where the energy accretion rate is defined as $\dot{E}=\iint T^r_t \, dA_{\theta\varphi}$ (taken to be positive for inflow of energy towards the BH), $u^r$ and $B^r$ are the radial velocity and magnetic field respectively, $T^r_t$ is the total radial energy flux, $dA_{\theta\varphi} =\sqrt{-g}\, d\theta \, d\varphi$ is the surface area element and $g\equiv|g_{\mu\nu}|$ is the metric determinant.

Figure~\ref{fig:time_grmhd} (d,e) shows the disk/jet tilt and precession angles, spatially-averaged over [50, 150] $r_{\rm g}$, and demonstrates that the large-scale jet is perpendicular to the large-scale disk on average \citep[in agreement with][]{liska_tilt_2018}, with some oscillatory behaviour of the jet tilt and precession angles illustrating the jet's dynamic nature.
Over time, the inner part of the misaligned disks tends to align with the plane perpendicular to the BH spin axis, with the tilt angle decreasing by roughly $25\%$ compared to the initial tilt.
Further, even though we started with a compact disk, subsequent disk evolution causes the disk to puff up due to viscous spreading, as illustrated by the disk barycentric radius \citep[Fig.~\ref{fig:time_grmhd} (f); see also][]{Porth2019}, given by:
\begin{equation}
    r_{\rm disk}=\dfrac{\iint r \, \rho \,  dA_{\theta\varphi}}{\iint \rho \, dA_{\theta\varphi}}\,.
    \label{eqn:barycentric_radius}
\end{equation}
\noindent As the disk becomes substantially larger, it stops precessing and the disk and jet tilt angle become roughly constant at $t\gtrsim9\times 10^4~t_{\rm g}$.
Note that even though the disk and the jet no longer precess, they remain misaligned with respect to the BH spin vector, clearly illustrating that the absence of QPOs or any indication of precession does not rule out the presence of misalignment.
Effectively, we end up with three disk/jet models with time- and spatially-averaged jet tilt and precession angles $\left\{\mathcal{T}_{\rm j},\mathcal{P}_{\rm j}\right\}$ as follows: \tilta{}: $\left\{0.8^{\circ},0^{\circ}\right\}$, \tiltb{}: $\left\{22.7^{\circ},28.4^{\circ}\right\}$, and \tiltc{}: $\left\{43.4^{\circ},42.2^{\circ}\right\}$ (see Table~\ref{tab:models}). 

Figure~\ref{fig:radius_grmhd} shows the radial profiles of the tilt $\mathcal{T}$ (top row) and precession $\mathcal{P}$ (bottom row) for the models \tiltb{} (left column) and \tiltc{} (right column), time-averaged over [99960, 100960] $t_{\rm g}$, along with their $1\sigma$ standard deviation.
This interval at such late times is chosen to allow the simulations to reach a quasi-steady state at least within the inner $100~r_{\rm g}$.
We average the radial profiles over $1000~t_{\rm g}$ in order to average over the short time-scale variations in the flow that otherwise cause the angles to fluctuate at small radii. LT torques affect the near-BH region, with radial tilt oscillations in
both the disk and the jet peaking at $r\sim 10-20~r_{\rm g}$
[Fig.~\ref{fig:radius_grmhd} (a,c)].
This behaviour was shown previously for model \tiltb{}\ by \citet{liska_tilt_2018} and is consistent with simulations of small misalignment \citep{fragile07,White_2019}.
The peak $\mathcal{T}$ values are close to $30^{\circ}$, instead of $\sim 40^{\circ}$ at earlier times \citep[$5\times 10^4~t_{\rm g}$;][]{liska_tilt_2018}, suggesting gradual alignment of the disk over time, until the viscous spreading of the disk saturates.
Figure~\ref{fig:radius_grmhd} [a,c] clearly shows that
the jet follows the disk orientation at larger radii, and therefore, the large scale jet is always misaligned with respect to the BH spin direction, while the inner jet undergoes a small degree of alignment.
This result is similar to the conclusions of \citet{mckinney_2013}, who considered misalignment under MAD conditions.
However, this contradicts the assumption by \citet{White_2020_tiltedimages} that
the jets follow the BH spin direction at large radii. This assumption can significantly affect the
orientation of the BH images as the inferred observer inclination depends on the jet orientation in the sky (see Sec.~\ref{sec:bestfit}). In this work, we have considered relatively powerful jets with $\phi_{\rm BH}\gtrsim 30$ (Table~\ref{tab:models}) as compared to \citet[][]{White_2019}, where $\phi_{\rm BH}\lesssim 8$, assuming that all the magnetic flux goes into powering the jets \citep[$P_{\rm outflow}\propto \phi_{\rm BH}^2$;][]{blandford77, tchekhovskoy10}.

Henceforth, we will refer to the plane of the BH spin and the jet angular momentum vector, i.e., the 2D cross-section shown in Fig.~\ref{fig:GRMHD_contour}, as the reference $\varphi=0^{\circ}$ plane, which would be useful as a notation for the BH images in the next section.

\begin{figure} 
    \includegraphics[width=\columnwidth]{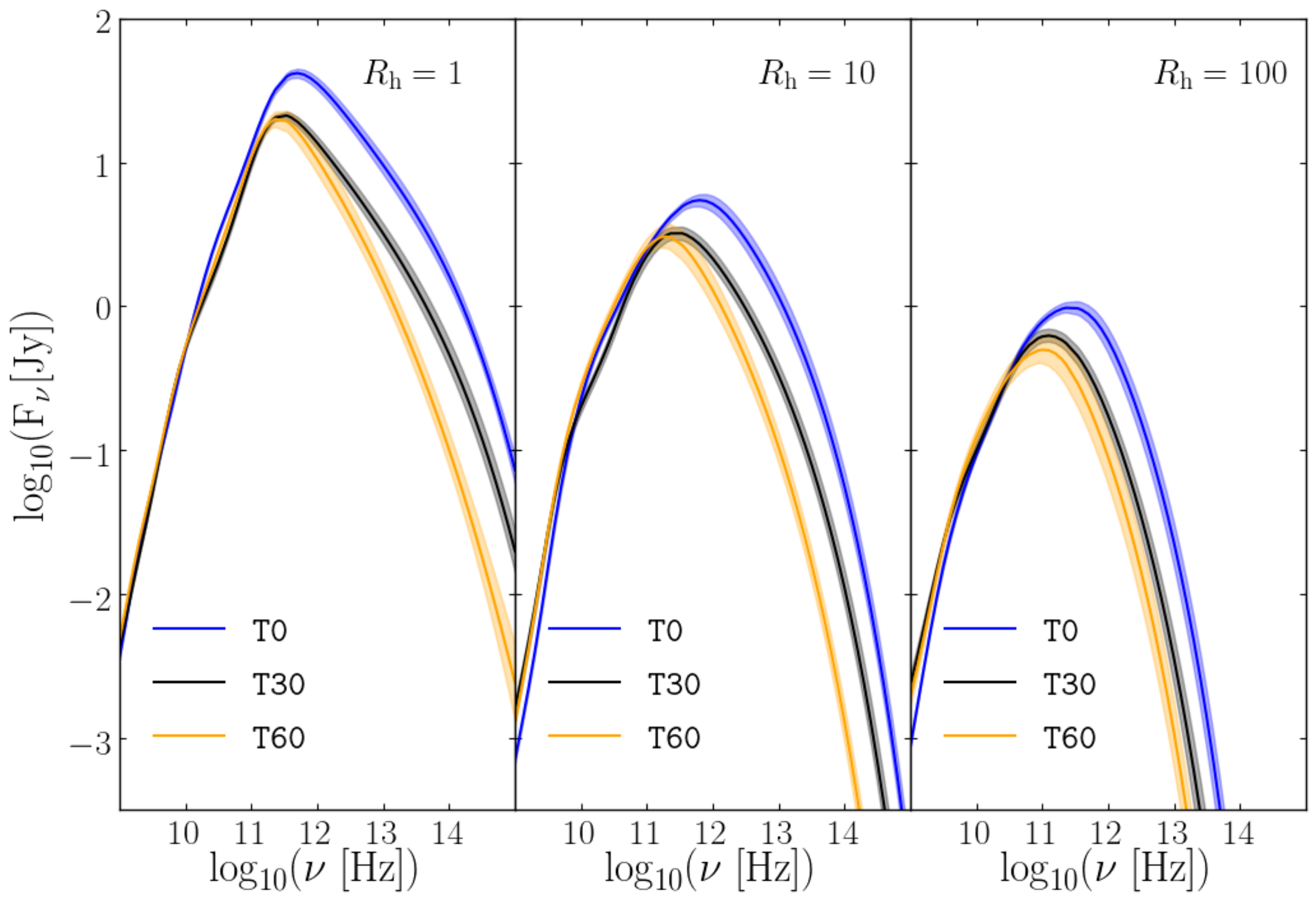}
    \caption{The resultant spectral energy distributions (SEDs) for misaligned BH disk/jet systems viewed edge-on to the outer disk (Fig.~\ref{fig:images_avg_rh10}) show that with increasing $R_{\rm h}$, the peak flux decreases, but the low-frequency emission does not vary with increasing misalignment even though misalignment introduces a Doppler-boosted inner jet (Fig.~\ref{fig:images_avg_rh10}).
    The GRRT images are scaled to M87 at an accretion rate of $6\times 10^{-4} M_{\odot} \, {\rm yr}^{-1}$.
    From left to right: SEDs for increasing $R_{\rm h}$ values for each of models \tilta{}, \tiltb{} and \tiltc{}.
    }
    \label{fig:tilt_spectrum}
\end{figure}

\subsection{GRRT imaging: synchrotron maps and SEDs}
\label{sec:images_edgeon}

In jet observations, the inclination ($\theta_0$) is usually defined as the angle between the observer's line of sight and the large-scale jet.
We therefore align our camera position according to the jet tilt $\mathcal{T}$ and precession $\mathcal{P}$ angles, spatially and temporally averaged between [50, 150] $r_{\rm g}$ and [99960, 100960] $t_{\rm g}$.
For our observer camera, we redefine the spherical grid: the $\varphi_0=0^{\circ}$ plane is equivalent to the GRMHD $y=0$ plane rotated by the corresponding precession $\mathcal{P}_{\rm j}$ angle (Table~\ref{tab:models}), and the polar axis, i.e., the $x=y=0$ line, is rotated by the corresponding tilt angle $\mathcal{T}_{\rm j}$.
Therefore, we now have a camera grid which is aligned with the the large-scale jet. 
Note that the precession angle is undefined for non-tilted jets \citep[e.g.,][]{White_2019}, hence we took $\mathcal{P}_{\rm j}=0^{\circ}$ for \tilta{}.
The camera field-of-view (FOV) is set to be 75 $r_{\rm g}$ $\times$ 75 $r_{\rm g}$, which is sufficient since warping occurs within $r\lesssim20~r_{\rm g}$, while the image resolution is $1024\times 1024$ pixels.
For a BH with the same mass and at the same distance as M87, $100~r_{\rm g}$ corresponds to 382 micro-arcseconds ($\upmu {\rm as}$), which corresponds to a FOV of $286.5 ~\upmu {\rm as}$. The simplest example to showcase the alignment of the camera grid with the large scale jet is when we observe our disk-jet system at an inclination of $\theta_0=90^{\circ}$ (i.e, edge-on to the outer disk).

\begin{figure*}
\begin{center}

\centering
\includegraphics[height=3.42in,trim=0cm 0cm 0cm 0cm,clip]{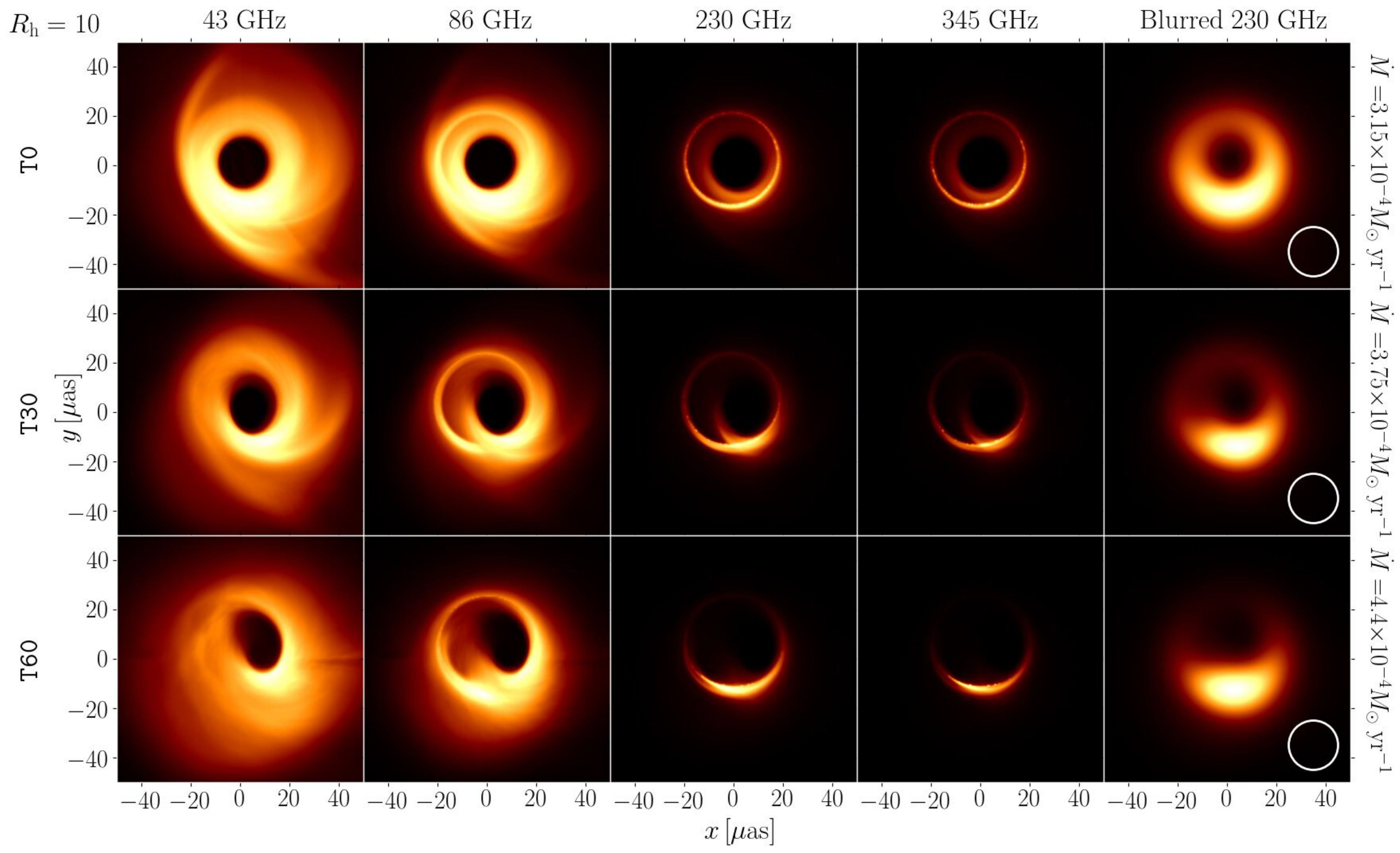}\hfill
\includegraphics[height=3.42in,trim=0cm 0cm 0cm 0cm,clip]{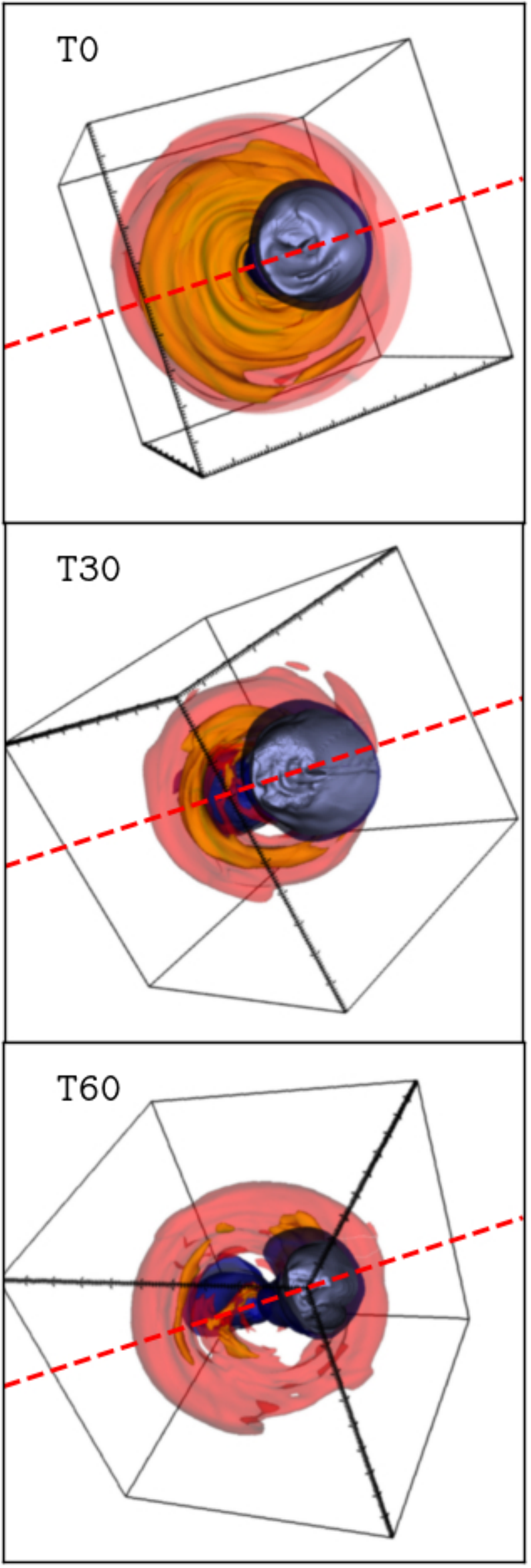}
    \caption{Left figure: time-averaged, synthetic radio images of misaligned BH disk/jet systems using $R_{\rm h}$=10.
    The observer position is chosen such that the viewing angle is $17^{\circ}$ offset from the direction of the large-scale bottom jet ($r\gtrsim50~r_{\rm g}$), i.e., $(\theta_0, \varphi_0)=(163^{\circ}, 180^{\circ})$, calculated over [99960, 100960]~$t_{\rm g}$.
    Images are rotated to fit the position angle PA$=288^{\circ}$ of the large-scale jet in M87.
    From left to right: images at frequencies of 43 GHz, 86 GHz, 230 GHz and 345 GHz, along with the blurred 230 GHz image (see Fig.~\ref{fig:images_avg_rh10}).
    From top to bottom: underlying GRMHD simulations with initial tilt angles of $0^{\circ}$, $30^{\circ}$ and $60^{\circ}$.
    Rightmost figure: density contour plot visualisations for (top to bottom) \tilta{}, \tiltb{} and \tiltc{} viewed at the same observer position as for the synthetic GRRT images and rotated to fit the large-scale jet position angle PA$=288^{\circ}$ (red dashed line).
    }
    \label{fig:images_avg_m87_rh10}
    \end{center}
\end{figure*}

\begin{figure} 
    \includegraphics[width=\columnwidth]{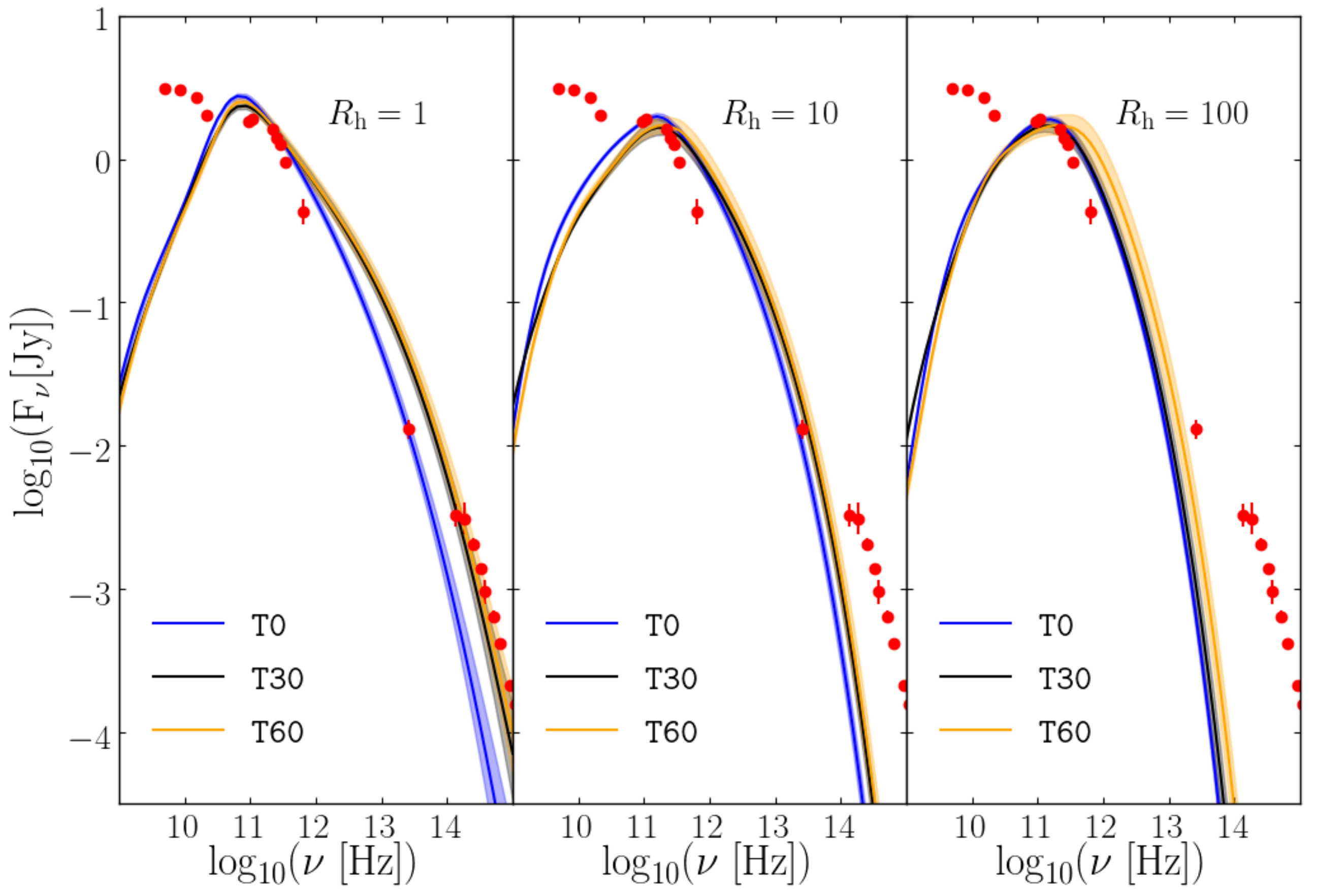}
    \caption{SEDs for misaligned BH disk/jet systems applied to M87 (Fig.~\ref{fig:images_avg_rh10}) show little change with increasing $R_{\rm h}$.
    The GRRT images are scaled to M87 with an accretion rate chosen to fit for the 221 GHz data point in the SED. 
    From left to right: SEDs for increasing $R_{\rm h}$ values for each of models \tilta{}, \tiltb{} and \tiltc{}.
    The radio, sub-mm, infra-red and optical data points are taken from \citet{Prieto2016_M87}.
    }
    \label{fig:tilt_spectrum_M87}
\end{figure}

Figure~\ref{fig:images_avg_rh10} shows synthetic mm-radio images for $R_{\rm h}=10$ applied to a M87-like disk/jet system at an accretion rate of $6\times 10^{-4} M_{\odot} {\rm yr}^{-1}$, with the observer positioned perpendicular to the jet axis (or edge-on to the large-scale disk) at four frequencies: 43 GHz, 86 GHz, 230 GHz and 345 GHz.
We have chosen $R_{\rm h}=10$, which yields contributions from both the disk and the jet to the total emission.
We also show the 3D visualisations of the disk/jet system in the rightmost column of Fig.~\ref{fig:images_avg_rh10}, as observed from the same direction as the GRRT camera.
The rotating accretion flow surrounding the BH is clearly visible in the 43 GHz and 86 GHz images (Fig.~\ref{fig:images_avg_rh10}, left: first and second columns).
At higher tilts, the inner part of the bottom jet bends towards the observer and is Doppler-boosted.
This bending is visible as observing frequency is increased to 230 GHz and 345 GHz, where the inner regions of the accretion flow closest to the BH are probed (Fig.~\ref{fig:images_avg_rh10}, third and fourth columns). 
The dark (or flux depressed) circular region in the image centre is due to gravitational lensing by the BH and is a characteristic of images of accreting BHs \citep{Narayan_2019}.
Hereafter, we refer to this flux depressed region as the BH ``shadow" (see Fig.~\ref{fig:images_avg_rh10}). 
If we focus at the centre of the \tilta{} 230 GHz image (third from left in the top row of Fig.~\ref{fig:images_avg_rh10}), we see that the BH shadow is obscured by a thin ``streak"-like shaped feature. This feature, hereafter referred to as the ``streak", is the flux originating from the plunging streams, which begin at the innermost stable circular orbit (or ISCO) for an aligned disk (\tilta{}) and further out for the case of the tilted runs \citep{fragile07}. This streak shines more brightly for \tilta{} as the emission is boosted towards the observer, whereas in the case of \tiltb{} and \tiltc{}, the inner disk is bent with respect to the outer disk, with the flow no longer being directed towards the observer.
Such a curved streak is therefore an indication of disk tilt, given that we know the jet direction, and could potentially be important for interpreting future EHT images of Sgr~A*, for which the inclination is a matter of much debate, with semi-analytical modelling (assuming a jetted model) favouring higher inclinations \citep[e.g.,][]{Markoff_2007, Connors_2017}, GRMHD simulations opting for both high \citep[e.g.,][]{Moscibrodzka_2009, Shcherbakov_2012, Drappeau_2013} as well as intermediate ($\sim 30^{\circ}-60^{\circ}$) inclinations \citep[e.g.,][]{Dexter_2010, moscibrodzka_2013, Davelaar2018} and, more recently, \citet{Gravity:18_hotspot} inferred low inclinations $\lesssim 30^{\circ}$. 

We see that the streak becomes progressively more optically thick at lower frequencies.
Imaging at 86 GHz, where this feature is most distinguishable among the selected frequencies, may be essential for capturing possible disk/jet warps due to misalignment.
At 86 GHz, \citet{Issaoun_2019} has shown that combining the imaging power of the Global Millimeter VLBI Array (GMVA) and the Atacama Large Millimeter/submillimeter Array (ALMA) results in constraining the size and structure of the central emitting region in Sgr~A* to an impressive limit.
Further constraints from the anticipated EHT Sgr~A* 230 GHz images will be vital in capturing possible disk misalignment.
We also show GRRT images of our models at different inclination angles later on in Sec.~\ref{sec:flyby} and Appendix~\ref{sec:flyby_T0_T60}.
In Appendix~\ref{sec:other_Rhigh_images_edgeon}, we show sets of images for disk-dominated emission ($R_{\rm h}=1$; Fig.~\ref{fig:images_avg_rh1}) and jet-dominated emission ($R_{\rm h}=100$; Fig.~\ref{fig:images_avg_rh100}).
With higher $R_{\rm h}$ values, the disk electron temperature drops and the jet becomes more visible with respect to the disk.
Due to the decrease in disk electron temperature, the streak feature also diminishes in brightness at 230 GHz and 345 GHz. 

Figure~\ref{fig:tilt_spectrum} shows the corresponding spectral energy distribution (SED) generated for each model for the three different values of $R_{\rm h}$.
Since the mass and distance scales, as well as the accretion rate, are fixed for this set of GRRT images, higher $R_{\rm h}$ values result in lower peak synchrotron fluxes.
Viewing from edge-on with respect to the disk, for higher tilts the aforementioned warping causes one of the inner jets ($r\lesssim 20~r_{\rm g}$) to point towards the observer and the other jet to point away, as seen from Figs.~\ref{fig:images_avg_rh10} and \ref{fig:images_avg_rh100}.
Hence, the net radio emission remains roughly similar for the three tilted models at the same $R_{\rm h}$.
For larger misalignment angles, the models have lower synchrotron peaks with a shift of the peak towards the radio as well as lower near-infrared (NIR) and optical emission. Hence for an edge-on BH disk/jet system, the disk/jet warp creates a substantial change in the image as well as the SED.
Note that, instead of using a constant accretion rate, if we were to fit to the same flux at the EHT frequency (i.e., 230 GHz), the accretion rate would have to be higher for the tilted models and hence the radio flux would also be higher, while the NIR/optical emission will be quite similar between the three models. Additionally, for the case of Sgr~A*, a higher accretion rate would change the predicted X-ray emission due to synchrotron self-Compton (SSC) emission.
Consequently, such tilted models may be ruled out by comparing to quiescent Sgr~A* X-ray spectra.

\subsection{The case of M87}
\label{sec:images_M87}

In this section, we apply our tilted models to M87 \citep{EHT_paperI}.
To match the crescent position in the M87 image as well as the direction of the jet, \citet{EHTPaperV} found that for a positive BH spin ($a>0$), the observer inclination angle $\theta_0>90^{\circ}$. From the jet/counter-jet flux intensity ratio at 43 GHz, \citet{mertens2016} estimated that the jet is $\sim 17^{\circ}$ offset to our line-of-sight (i.e., the viewing angle). Using these two results, we chose an inclination angle of $\theta_0=163^{\circ}$, i.e., $17^{\circ}$ offset from the jet moving in the opposite direction to the black hole spin vector, such that we view the disk material to rotate clockwise. 
Hence, for M87, we shift our camera position to $\theta_0=163^\circ$ for each of the different tilt cases, with the jet direction given by the time-averaged tilt $\mathcal{T}_{\rm j}$ and precession $\mathcal{P}_{\rm j}$ angles from Table~\ref{tab:models}. 
As a first exploration of the simulations, in this section we restrict ourselves to the observer's line of sight being in the plane shared by the large scale jets and BH spin axes (later we will relax this condition). This still leaves two possible configurations: one where the BH spin vector and the bottom jet reside in the same half of the image plane ($\varphi_0=0^\circ$) and the other where the BH spin vector and the bottom jet are in opposite halves ($\varphi_0=180^\circ$). We find that $\varphi_0=180^\circ$ is more favourable since the forward jet must appear on the right hand side of the synthetic image plane while keeping the crescent shape in the bottom half of the image.
This choice does not make a difference for the \tilta{} case, since the jet is roughly axisymmetric.
Furthermore, we rotate the image to match the M87 outer-jet position angle of PA $=288^{\circ}$ \citep{walker_2018_M87}.

Figure~\ref{fig:images_avg_m87_rh10} shows the GRRT M87 images for $R_{\rm h}=10$, time-averaged over [99960, 100960]~$t_{\rm g}$, with the accretion rate for each model set to match the 221 GHz flux shown in the M87 SED (Fig.~\ref{fig:tilt_spectrum_M87}).
The synchrotron maps and SEDs have the same FOV and resolution as quoted in the previous section.
The shadow size is about $40~\upmu {\rm as}$ as expected.
However, due to the disk warp, some flux is present in front of the shadow for both misaligned models, which may itself be a crucial diagnostic since the brightness ratio from the ring to the depression (i.e., the shadow region) is expected to be increasingly better measured with future EHT observations.
The extent of the southern bright crescent in the blurred 230 GHz image becomes smaller with increasing tilt angle.
It is particularly noteworthy that the asymmetry in the upper and lower halves of the photon ring increases with increase in tilt.
However, both of these features depend on the underlying electron temperature distribution, and therefore investigation of more sophisticated treatments, e.g., two-temperature plasma physics \citep[e.g.,][]{Chael2018}, is required.

The $R_{\rm h}=1$ and $R_{\rm h}=100$ cases are shown in Appendix~\ref{sec:other_Rhigh_images_M87}, namely, Figs.~\ref{fig:images_avg_m87_rh1} and \ref{fig:images_avg_m87_rh100}.
The plunging region becomes more visible with increasing tilt angle as the streak in front of the shadow grows more distinct, similar to tilted models seen in Sec.~\ref{sec:images_edgeon} with larger value of $R_{\rm h}$.
Since the disk and jet at low radii display a smaller tilt and larger precession angle as compared to the large scale jet (see Fig.~\ref{fig:radius_grmhd}), a part of the disk/jet is slightly warped towards the observer and lensing effects no longer completely dominate the 230 GHz synthetic image. A notable difference from previous work is the absence of the bright ``double" crescent feature, one on each side of the shadow, seen in face-on images of tilted disks \citep[][Fig.~5]{dexter_2013}. This double crescent structure was attributed to standing shocks in the plunging region for tilted disk models in \citet{fragile07}.
Considering that \citet{White_2020_tiltedimages}, who used similar parameters to \citet{fragile07}, also found the double crescent feature to be absent, this leads us to speculate that such strong standing shock features as seen by \citet[][]{dexter_2013} might stem from a difference in grid resolution and/or the numerical methods, e.g., the shock-capturing scheme, employed in the different GRMHD codes, or could be a result of our assumed electron temperature model \citep[used in the EHT M87 papers;][]{EHTPaperV}, as suggested by \citet{White_2020_tiltedimages}.
In our models, the flow in the plunging streams is not subject to strong shock-heating, and therefore, the ion temperature does not steeply rise in this region and there is no enhanced synchrotron emission.
Hence, the plunging streams do not form the double crescent feature, but instead we see a single crescent feature consistent with Doppler boosting, without significant additional heating. Similar to the results of \citet{White_2020_tiltedimages}, the change in the shape of the crescent feature with tilt translates to a change in the BH shadow shape as is evident from the blurred 230~GHz images in Fig.~\ref{fig:images_avg_m87_rh10}. Further, we note that the key element in calculating the above M87 images is the orientation of the jet, which undergoes some alignment towards the BH spin direction, but at larger scales ($r\gtrsim50~r_{\rm g}$), clearly follows the disk orientation (see Fig.~\ref{fig:radius_grmhd}).
Not accounting for the misalignment of the jet can have definite consequences for the 43 GHz jet position angle PA, which \citet{walker_2018_M87} determined to be PA$=288^{\circ}$.
As will be seen in Sec.~\ref{sec:flyby}, one can obtain excellent matches to the EHT M87 image for a variety of camera positions if the 43 GHz image is not used as a constraint. 

The accretion rate used for each $R_{\rm h}$ tilt model is set by fitting the corresponding time-averaged SED to the 221 GHz data point (Fig.~\ref{fig:tilt_spectrum_M87}).
The SED shows that the \tilta{} case displays higher emission at low frequencies as the jet at low radii is pointing towards the observer, whereas for the tilted models, the jet undergoes bending towards the observer over $r\lesssim ~20 \, r_{\rm g}$.

\section{Discussion}
\label{sec:discussion}
\subsection{Time series analysis of the M87 GRRT images}
\label{sec:timing}
\begin{figure} 
    \includegraphics[width=\columnwidth]{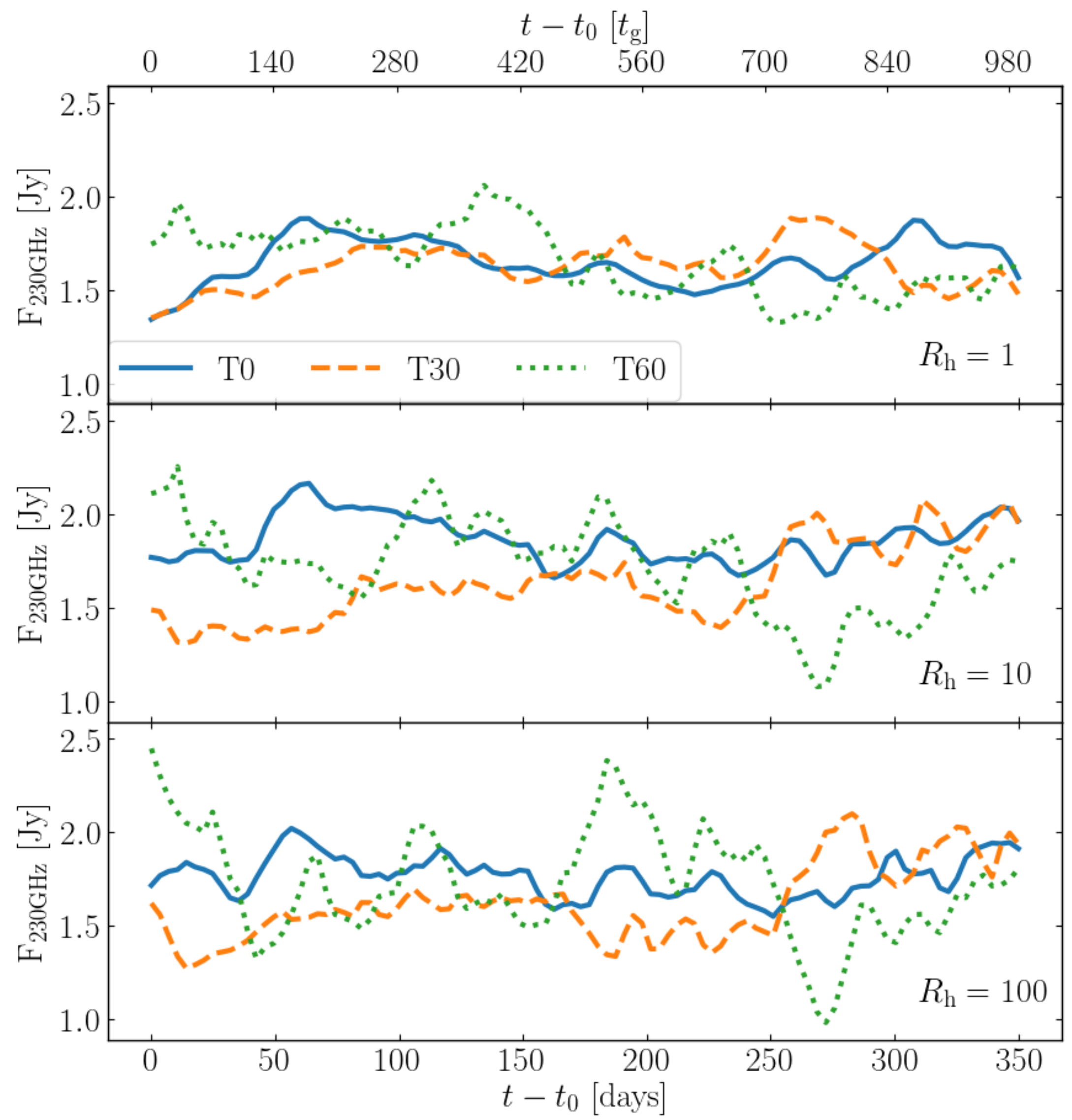}
    \caption{Higher tilts show more variability in the near-horizon flux when applied to the M87 BH.
    The 230 GHz lightcurve is shown over 1000~$t_{\rm g}$, equivalent to 353.55 days when scaled to M87, or roughly one year, for models: \tilta{}, \tiltb{} and \tiltc{} using (from top to bottom) $R_{\rm h}$ values of 1, 10 and 100, respectively.
    The start time $t_0=99970.58~t_{\rm g}\equiv 35354.92~$days. 
    \tilta{} shows the least amount of variation as most of the 230 GHz flux is dominated by gravitational lensing effects rather than disk turbulence, while \tiltc{} shows a flux change by a factor of more than 2 over the entire time period.
    }
    \label{fig:tilt_lightcurves_M87}
\end{figure}

Figure~\ref{fig:tilt_lightcurves_M87} shows the 230 GHz lightcurve derived from the core-unresolved spectrum (Fig.~\ref{fig:tilt_spectrum_M87}) for each tilt model at $R_{\rm h}= 1$, $10$ and $100$, applied to the case of M87.
The \tilta{} model shows the least amount of variation at 230 GHz due to strong gravitational lensing effects, irrespective of the temperature prescription, and is consistent with the results of \citet{EHTPaperV}.
For the \tiltb{} case, the variability is also small, except for a roughly monotonic flux increment for $t>245$ days due to a flaring event.
It is interesting to see that this flux increase in \tiltb{} is more prominent for $R_{\rm h}=10$ and $100$, since the jet plays a greater role in the net flux at higher $R_{\rm h}$ values.
Overall, \tiltc{} consistently displays the largest 1$\sigma$-deviation ($>10\%$) over the chosen time period for a given $R_{\rm h}$ model, with a maximum factor of two change in the 230 GHz flux.
This is a result of the turbulent inflow (outflow) in the disk (jet) being partially directed towards the observer.
Hence the variability amplitude of the 230 GHz light curve may provide a diagnostic of disk misalignment in M87 and, perhaps, other AGN.
Future timing analysis of M87 EHT data may shed further light on whether there is indeed a tilted disk present. 

In Appendix~\ref{sec:timing_extended}, we perform further analysis on the flux variability timescales for each lightcurve and construct the power spectrum as well as the structure function. All of the models show substantial variability over timescales ranging from days to months, with the fractional root-mean-square (rms) amplitude $\approx 7-16\%$. We then model the power spectra and the structure function to find that the characteristic timescale, i.e., the timescale at which the variability transitions from red noise on short timescales to white noise on long timescales. The best-fit characteristic timescales from the power spectra and the structure functions are roughly of the order of a hundred days and a few tens of days respectively, as compared to the timescale of $\sim 45$ days obtained by \citet{Bower_2015_submm_M87} where the 230 GHz lightcurve of M87 spanned over 10 years. Due to the short duration of our lightcurves, we only capture the short timescale variability, with the power spectra well modelled by the characteristic slope $=-2$ of red noise (see Appendix~\ref{sec:timing_extended} for more details).

\begin{figure} 
    \includegraphics[width=\columnwidth]{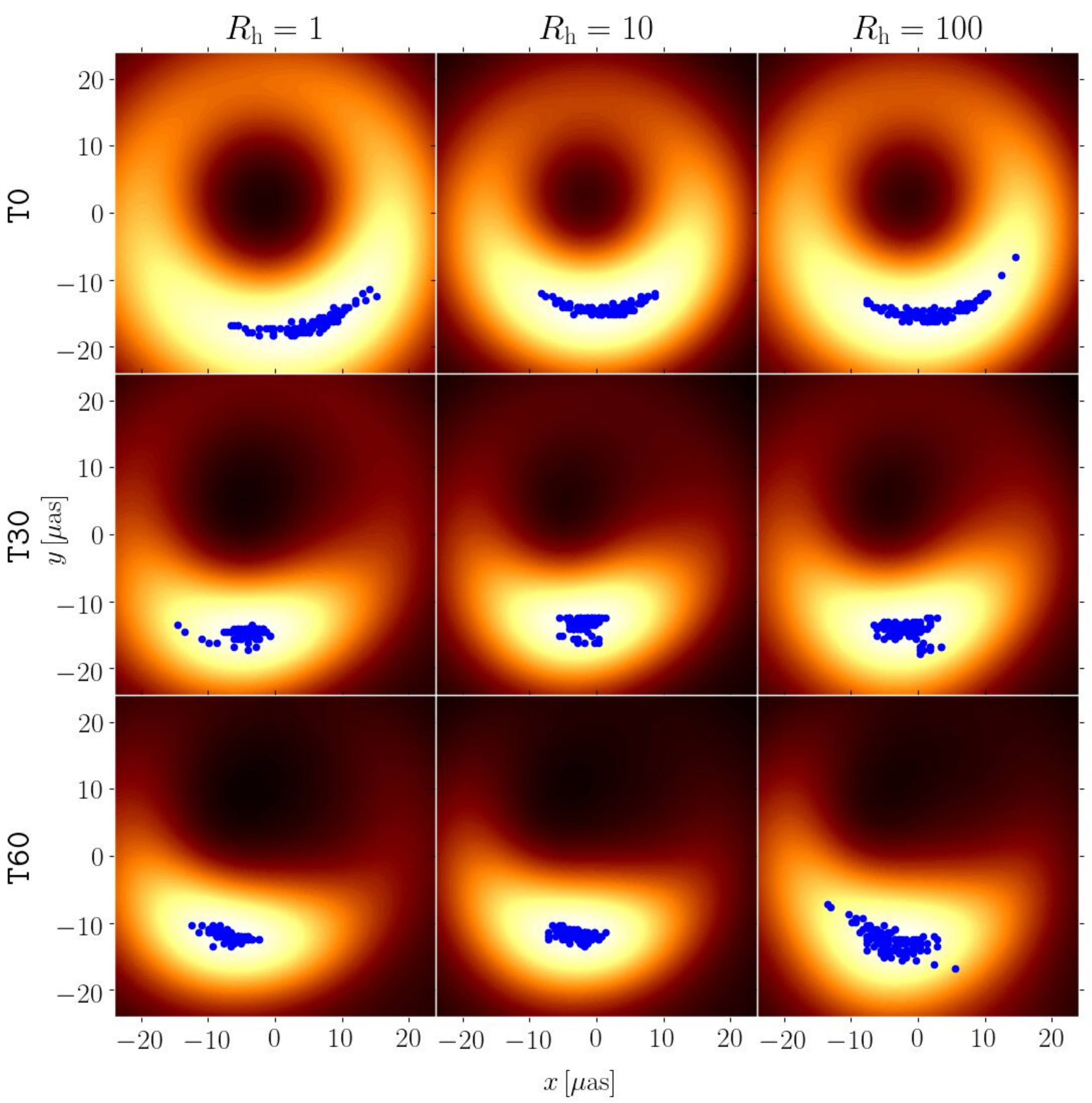}
    \caption{The location of the brightest pixel in the synthetic 230 GHz M87 images varies the most for \tilta{}.
    We show the brightest pixel locations (blue circles) over time against a background of the average blurred 230 GHz image for each of the nine models considered in this work. Note that each blue dot represents a different time snapshot.
    Top to bottom: tilt models \tilta{}, \tiltb{} and \tiltc{}.
    Left to right: $R_{\rm h}=$1, 10 and 100 models for the electron temperature prescription.
    For the \tilta{} case, gravitational lensing dominates the emission and hence turbulence in the flow changes the brightest spot location considerably.
    Doppler boosting of the plunging stream dominates the tilted images, which creates a preferential zone that points towards the observer where the boost is the maximum, and hence restricts the brightest spot location.}
    \label{fig:bright_pixel_M87}
\end{figure}

Figure~\ref{fig:bright_pixel_M87} shows the spread of the brightest pixel of the blurred 230 GHz image for each combination of initial tilt and the $R_{\rm h}$ parameter.
For this figure, each time snapshot is convolved with a gaussian filter of ${\rm FWHM}=20~\upmu {\rm as}$ and the position of the brightest pixel in each resulting image is calculated.
From this figure, \tilta{} displays the maximum positional variability, spread across the time-averaged southern crescent, which is a result of the rotation of the disk as well as the gravitationally-lensed emission from the counter-jet.
Misalignment introduces some constraint on the positional spread as seen in the second and third rows in Fig.~\ref{fig:bright_pixel_M87}, displaying models \tiltb{} and \tiltc{}. $R_{\rm h}=10$ results in the smallest spread in both radial and azimuthal directions on the image plane, since the inner regions of both the disk and jet dominates the emission at 230 GHz and hence the location of the jet-disk boundary along the plunging streams dictates the position of the brightest spot. 

\begin{table}
\begin{center}
\renewcommand{\arraystretch}{1.6}
\begin{tabularx}{\columnwidth}{l | c c c}
\hline\hline
\vspace*{0mm}
& & $\langle F_{\rm 230 GHz} \rangle^{\max[F_{\rm 230 GHz}]-\langle F_{\rm 230 GHz} \rangle}_{\min[F_{\rm 230 GHz}]-\langle F_{\rm 230 GHz} \rangle}$ & \\
Model & $R_{\rm h}$=1 & $R_{\rm h}$=10 & $R_{\rm h}$=100\\
\hline
\tilta{} & $1.649^{+0.232}_{-0.307}$ & $1.864^{+0.302}_{-0.203}$ & $1.765^{+0.257}_{-0.214}$\\
\tiltb{} & $1.624^{+0.26}_{-0.274}$ & $1.633^{+0.443}_{-0.32}$ & $1.622^{+0.477}_{-0.349}$\\
\tiltc{} & $1.656^{+0.402}_{-0.329}$ & $1.723^{+0.532}_{-0.644}$ & $1.727^{+0.721}_{-0.744}$\\
\hline\hline
\end{tabularx}
\end{center}
\caption{Higher disk-BH misalignment results in a more variable 230 GHz lightcurve.
We show the time-averaged fluxes from the 230 GHz lightcurve for each tilt and $R_{\rm h}$ model (see Fig.~\ref{fig:tilt_lightcurves_M87}) as well as the 1-$\sigma$ standard deviation normalised by the averaged flux.
\tilta{} shows the lowest deviation, while \tiltc{}, the highest, indicating that deviations $\gtrsim10\%$ might be a hint of possible misalignment.}
\label{tab:lightcurve_dev}
\end{table}

\begin{figure} 
    \includegraphics[width=\columnwidth]{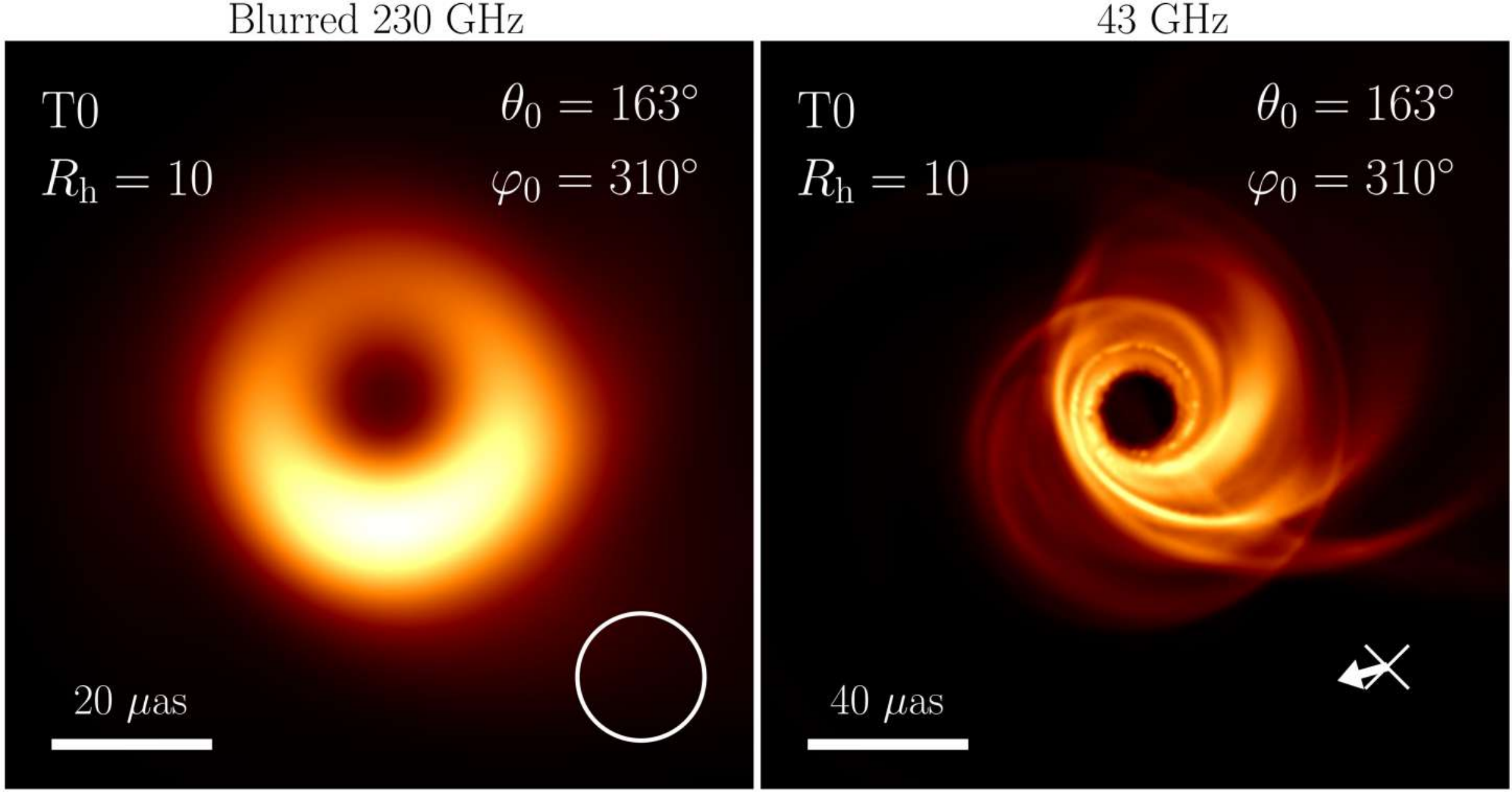}
    \includegraphics[width=\columnwidth]{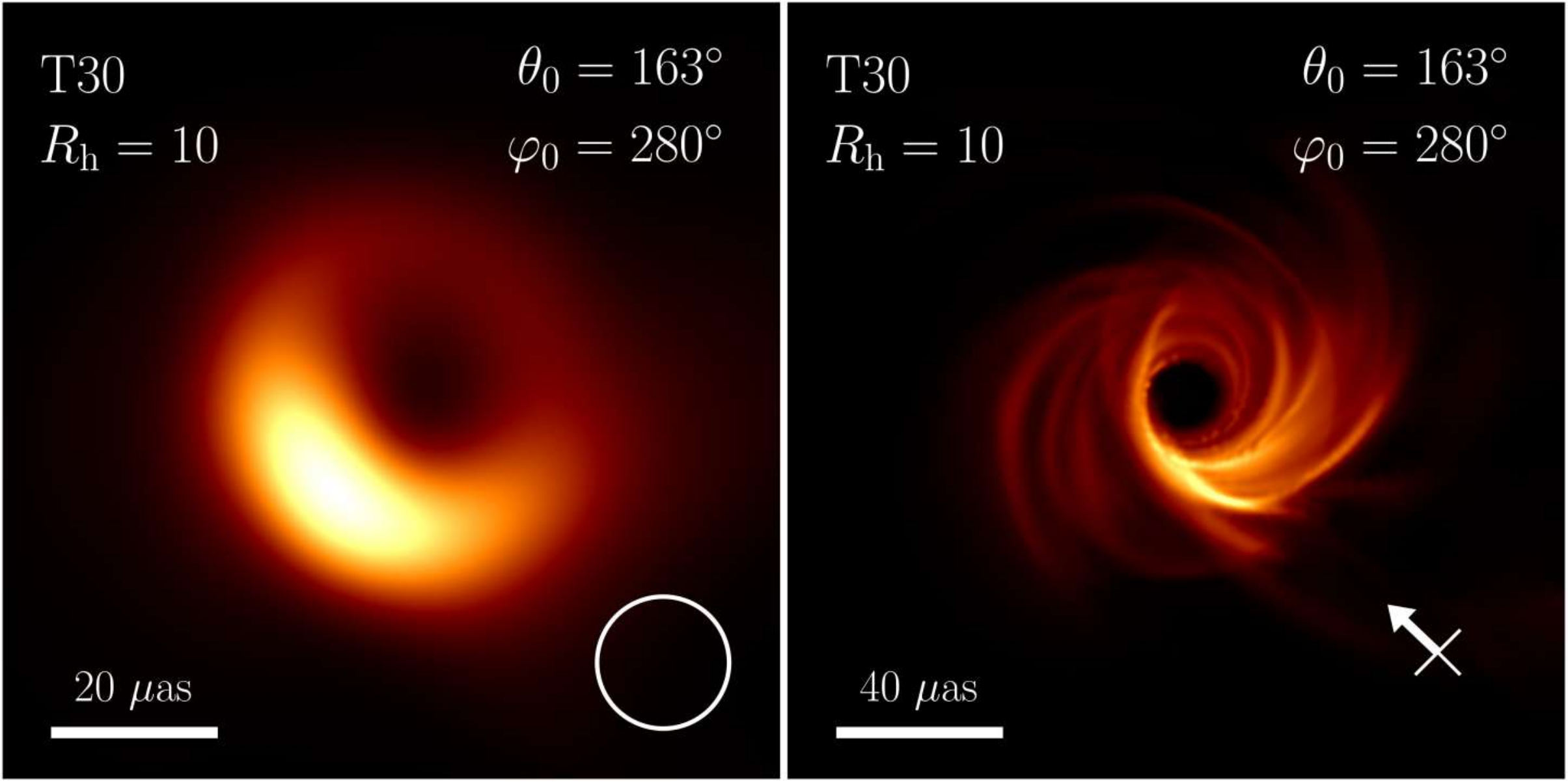}
    \includegraphics[width=\columnwidth]{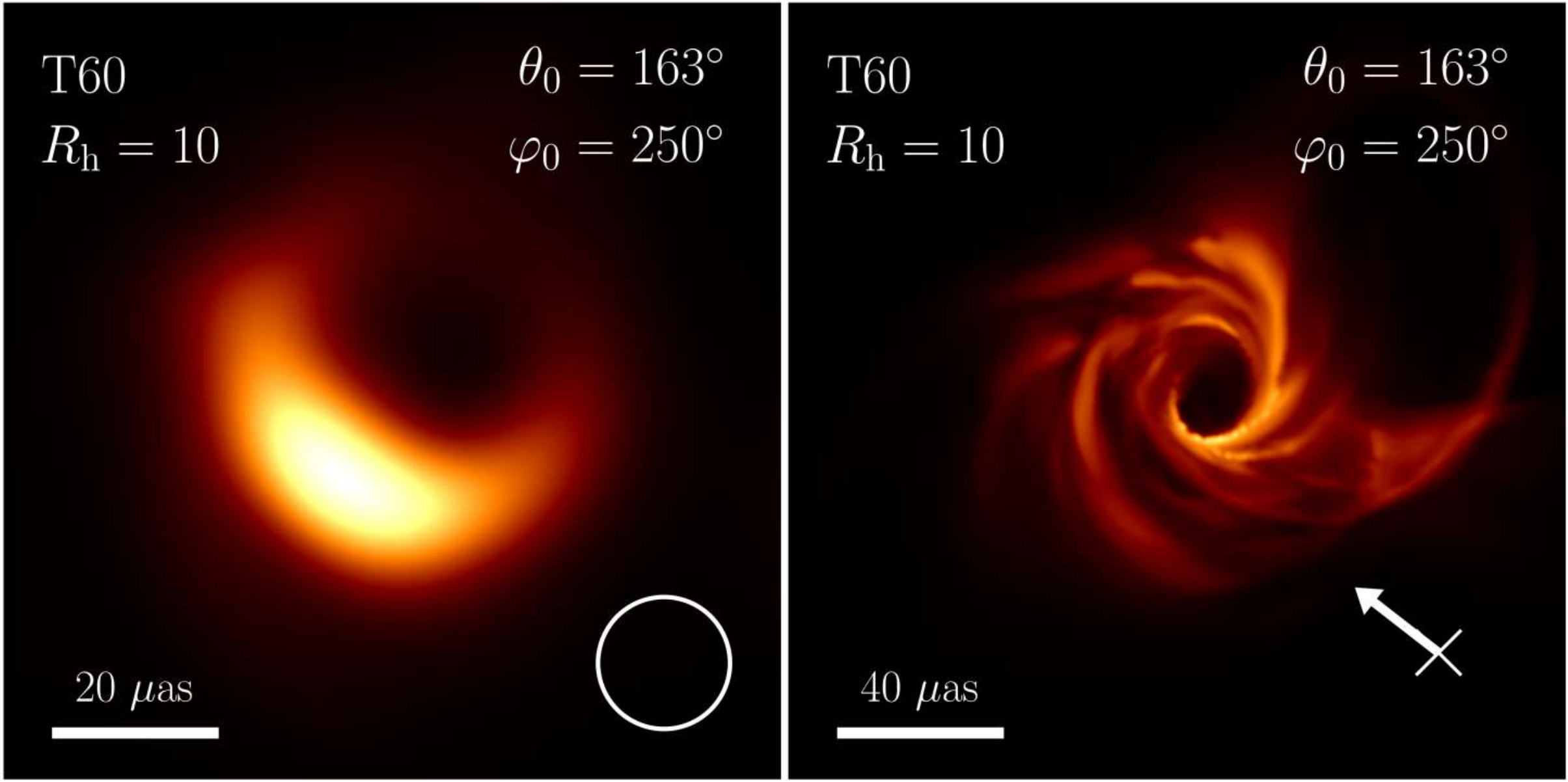}
    \caption{Comparing our models with the EHT M87 images indicates that M87 favourably hosts a misaligned disk and jet.
    We show the best fit-by-eye GRRT snapshot images from all three models when compared to the 230 GHz M87 BH image, with the 43 GHz jet pointed at PA $\sim$ $288^{\circ}$ with $R_{\rm h}=10$. The BH spin direction and projection onto the image plane is also shown, with X meaning the BH spin is pointing into the plane. Best-fit 230 GHz GRRT images of aligned BH models are consistent with the EHT M87 image to within $1.5\sigma$. If this deviation persists with future EHT observations of M87, a misaligned disk/jet could become a more viable model for M87.
    }
    \label{fig:M87_fit_by_eye}
\end{figure}

\subsection{Best-fit images for M87}
\label{sec:bestfit}
In Sec.~\ref{sec:images_M87}, we considered GRRT maps of misaligned jets observed at an inclination of $163^{\circ}$ offset from the large-scale jet direction, fixing the observer position with respect to the BH spin and the jet.
In this subsection, we relax this assumption and move the observer's position azimuthally around the bottom jet at an inclination of $163^{\circ}$, looking for the best fit-by-eye convolved images for each model relative to the 230 GHz M87 observations \citep{EHT_paperI}, while ensuring that the 43 GHz jet image has a PA $\sim 288^{\circ}$ \citep[][]{walker_2018_M87}.
The accretion rate is taken to be the same as in Fig.~\ref{fig:images_avg_m87_rh10}.
Figure~\ref{fig:M87_fit_by_eye} shows the best fits for a single representative time snapshot (at $t=100460~t_{\rm g}$) in the case of each tilt model. Firstly, the tilted model images show significant dependence on the camera longitude $\varphi_0$ (see Sec.~\ref{sec:flyby}), since the underlying warped disk/jet is not axisymmetric.
On the other hand, the aligned disk/jet model is roughly axisymmetric and so, the images are independent of the camera-$\varphi_0$, as expected. 

Secondly, the best-fit \tilta{} 230 GHz image (Fig.~\ref{fig:M87_fit_by_eye}, first row, left column) has a PA offset with respect to the best-fit EHT GRRT image \citep[][Fig.~1, right panel]{EHTPaperV}.
\citet{EHTPaperV} found that models with the spin vector of the BH oriented away from the observer ($\theta_0 \gtrsim 90^{\circ}$) are favoured by comparing to both 230 and 43 GHz images of M87, which we also find.
The statistically best-fit 230 GHz images were found to have a mean PA in the range of $203^{\circ}-209^{\circ}$ \citep[see Fig.~9 in][]{EHTPaperV} with a standard deviation of $54^{\circ}$, which means that aligned GRMHD simulations are consistent with the PA $\sim 288^{\circ}$ found for the M87 jet at 43 GHz \citep{walker_2018_M87} within $\sim 1.5\sigma$.
If we rotate our \tilta{} 230 GHz image so that we match the orientation of the 43 GHz M87 jet to the 43 GHz GRRT image, we find that the southern crescent is shifted to the bottom right quarter of the image, and hence does not match well with the EHT observed M87 image.
Interestingly, disk misalignment provides more flexibility to fit the jet PA better as the crescent position is dependent on the choice of the camera longitude $\varphi_0$, as we mentioned above.
Figure~\ref{fig:M87_fit_by_eye} (second row) shows that for the \tiltb{} model, which possesses an average tilt angle of $22.7^{\circ}$, the image has a bright crescent shape located in the bottom left of the image at $\varphi_0=280^{\circ}$ and fits remarkably well with the M87 jet PA, possibly hinting at the presence of a tilted disk in M87. Indeed, if the 230 GHz best-fit aligned disk/jet model always resides at a $1.5\sigma$ deviation in PA away from the 43 GHz M87 jet PA over future EHT M87 observations, disk/jet misalignment in M87 could become a significant possibility.

Thirdly, the brightness asymmetry in the upper and lower halves of the photon ring increases with increasing misalignment (as mentioned above in Sec.~\ref{sec:images_M87}).
Using the crescent position and the ring brightness asymmetry in the 230 GHz image along with the jet PA orientation might help us to pinpoint the tilt angle of the jet, which can lead to better BH spin estimates.
As an example, the \tiltc{} case appears more asymmetric by eye than the reconstructed images of M87, which leads us to favour smaller tilts.
In this work, we propose that M87 possesses a disk with a misalignment $\lesssim 60^{\circ}$, given the assumption that the BH spin is $a=0.9375$ and the inclination angle is $17^{\circ}$ offset to the large scale bottom jet (i.e., the viewing angle).
Images at frequencies lower than 230 GHz might provide more definitive proof of misalignment by directly capturing the disk warping region, located within roughly $20~r_{\rm g}$ of the BH (see Fig.~\ref{fig:radius_grmhd}).
However, the typical beam sizes for radio interferometric images at 43 GHz and 86 GHz are presently too large to adequately resolve the inner $20~r_{\rm g}$.
Imaging the jet base region with sufficient resolution to capture the warping of the disk and the jet as well as checking for time variability in the EHT observations (see Sec.~\ref{sec:timing}) will go a long way towards testing our prediction of a tilted disk in M87.

\begin{figure*}
    \includegraphics[width=\textwidth]{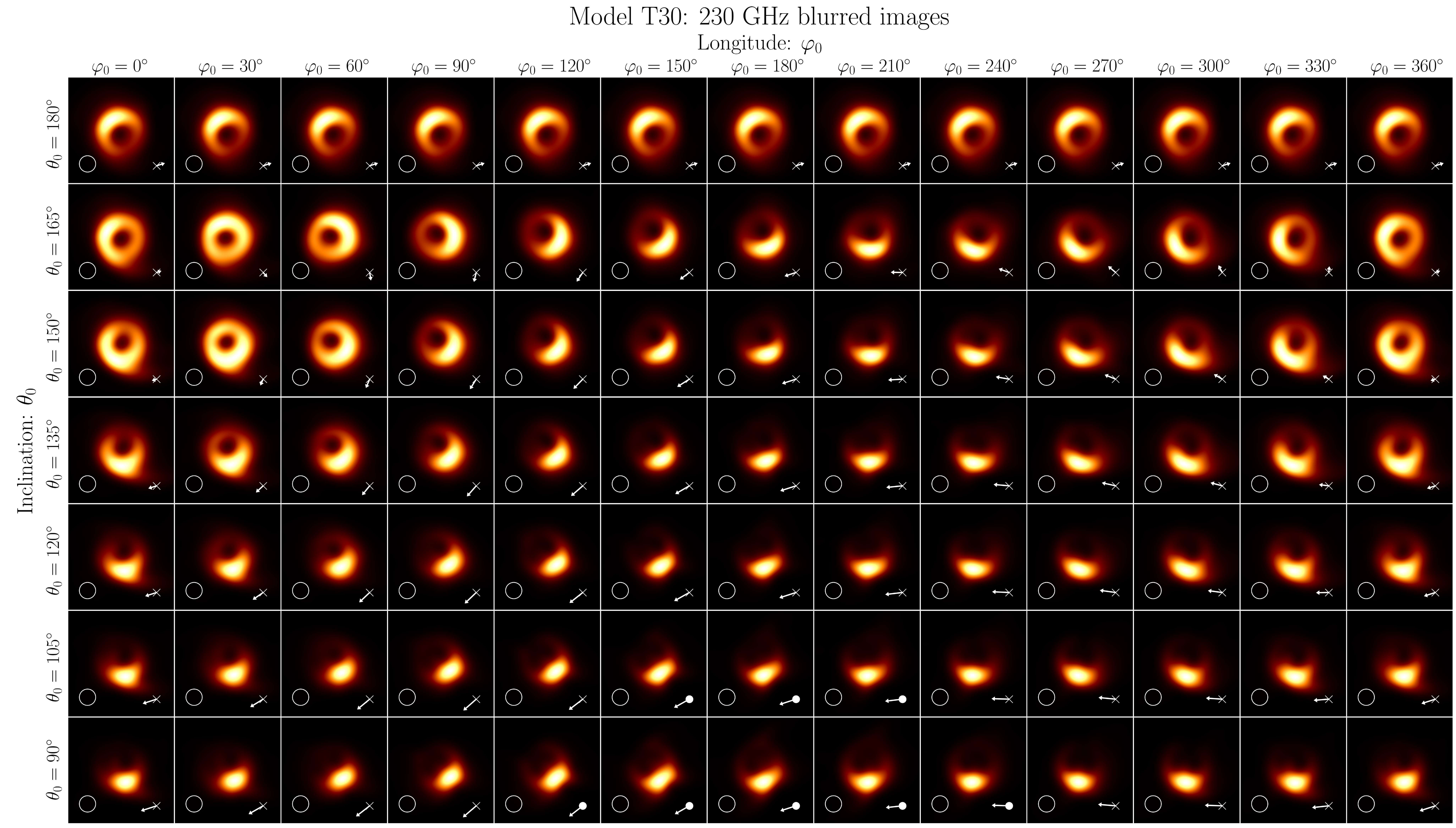}\vspace*{5mm}
    \includegraphics[width=\textwidth,trim=3cm 0cm 0cm 0cm,clip]{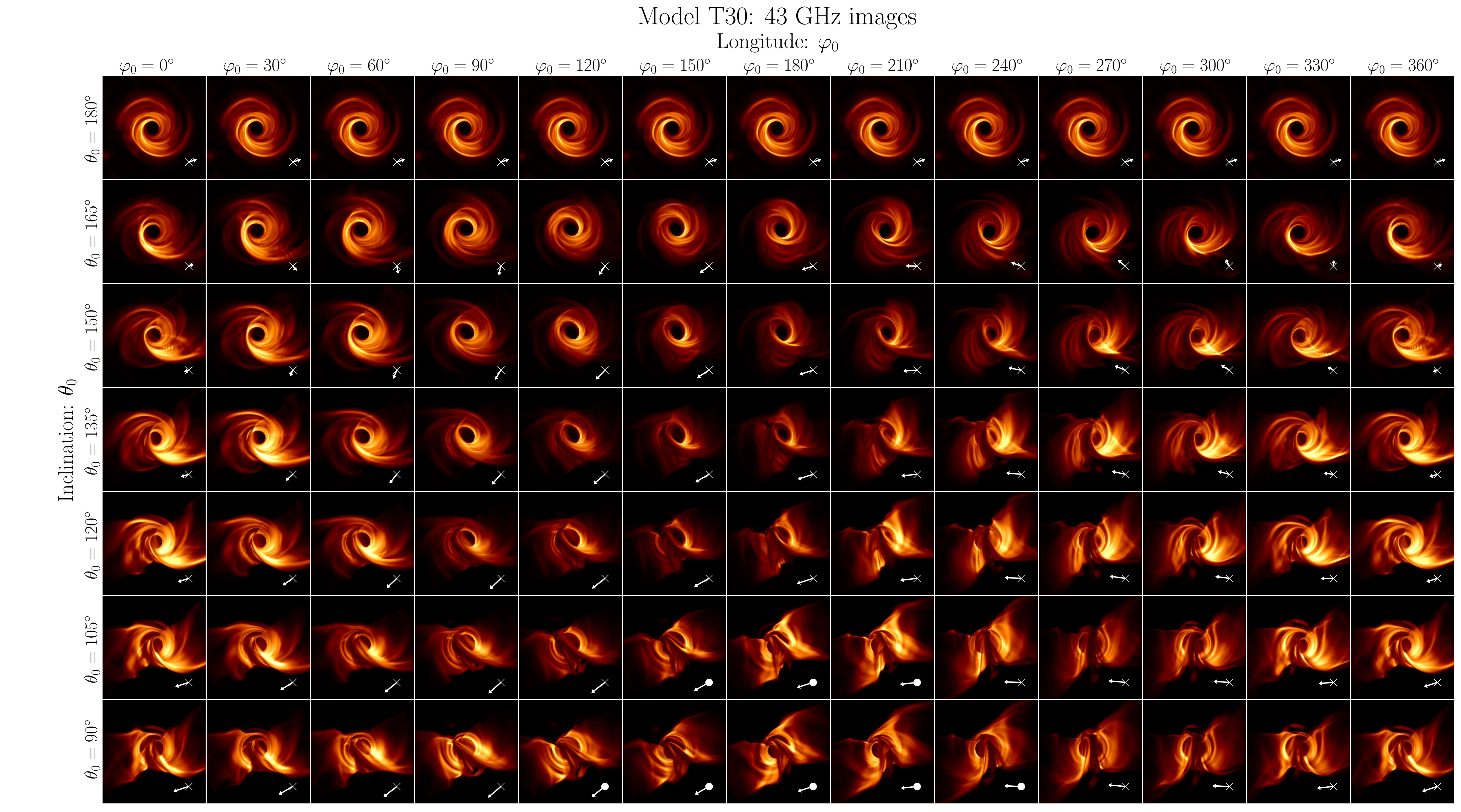}
    \caption{Images at 230 GHz and 43 GHz for different values of the observer inclination $\theta_0$ and longitude $\varphi_0$, for a single snapshot of model \tiltb{} with $R_{\rm h}=10$. The box sizes are $110\times 110 ~\upmu {\rm as}^2$ and $150\times 150 ~\upmu {\rm as}^2$ for the 230 and the 43 GHz images respectively.
    A strong dependence of the images on $\varphi_0$ due to the non-axisymmetric nature of the disk/jet warping is clearly seen. The BH spin vector direction is also shown with a white arrow and the X (O) indicating whether the BH spin vector is pointing into (out of) the image plane. The arrow length is indicative of the BH spin projection onto the image plane.
    The images have been rotated to fit the 43 GHz jet PA from \citet{walker_2018_M87}.
    }
    \label{fig:T30_flyby}
\end{figure*}

\subsection{Varying the observer inclination and longitude}
\label{sec:flyby}
Figure~\ref{fig:T30_flyby} shows the blurred 230 GHz and 43 GHz images for model \tiltb{} observed over a range of inclination angles, $\theta_0=180^{\circ}-90^{\circ}$, as well as also azimuthally rotating the observer around the large scale bottom jet with $\varphi_0=0^{\circ}-360^{\circ}$, assuming $R_{\rm h}=10$ and scaled to M87 at an accretion rate of $3.75\times 10^{-4} M_{\odot} {\rm yr}^{-1}$ (see middle row of Fig.~\ref{fig:images_avg_m87_rh10}). 
Using this catalogue of images, we can pick out the best-fit images as compared to the EHT M87 image.
Firstly, we see that ring brightness asymmetry constraint favours smaller inclinations, along with a few acceptable images at $\theta_0\sim 45^{\circ}$ and $\varphi_0\lesssim30^{\circ}$ or $\gtrsim330^{\circ}$.
If matching the jet PA to $288^{\circ}$ and requiring the bright crescent to be in the bottom-left of the image, higher $\varphi_0$ values are clearly favoured.
This result shows that accounting for the large-scale jet orientation is crucial in constraining the acceptable values of $\theta_0$ and $\varphi_0$.
Secondly, at higher inclinations, the spectral shape for each tilt model is sufficiently distinguishable (see Fig.~\ref{fig:tilt_spectrum}) and perhaps the M87 SED might be able to rule out some observer inclinations.
Not only can we use these images to fit for M87, but given that Sgr~A* extends over a similar angular size in the sky (due to comparatively close values for the ratio of the BH mass and the distance), we can also fit these images to upcoming EHT observations of Sgr~A* in the same manner.

\section{Conclusions}
\label{sec:conclusions}

In this work, using our GPU-accelerated GRMHD code \hammer{}, we have simulated three BH accretion disks, with the disk mid-plane initially misaligned by $0^{\circ}$, $30^{\circ}$ and $60^{\circ}$ with respect to the $z=0$ plane.
Using these three models, we calculated the first synthetic radio images of misaligned accretion disk/jet models scaled to the M87 black-hole mass, performing radiation transfer using the GRRT code \bhoss{}.
Our analysis of the 230 GHz synthetic images suggests that M87$^{*}$ could host an accretion disk with a reasonably large misalignment angle ($\lesssim 60^{\circ}$). In light of these results, the misalignment angle, neglected in the current EHT theoretical models \citet{EHTPaperV}, may therefore be an important additional parameter within the observational fitting procedure of GRMHD simulation models worthy of further exploration. Tilt brings about a change in the black hole shadow appearance, which could be verified with future EHT observations of M87$^{*}$. We note that, recently, \citet[][]{White_2020_tiltedimages} considered the case of weakly magnetised misaligned disks and found similar results to our work.

The present study has assumed that the emission originates solely from thermal synchrotron radiation,  further assuming an electron temperature prescription which calculates the electron temperature from the GRMHD-based ion temperature.
These assumptions play an important role in the appearance of the M87 230 GHz synthetic images and warrant further study in order to fully address the question of possible misalignment in M87.
From semi-analytic modelling of the M87 spectra \citep[e.g.,][]{lucchini19}, the 230 GHz emission is expected to have a significant non-thermal synchrotron contribution, describable by a hybrid thermal-nonthermal electron distribution function \citep[e.g.,][]{Davelaar2019}. As \citet{lucchini19} mentions, the X-ray spectrum is more likely to be dominated by the non-thermal emission rather than by a synchrotron self-Compton inner jet.
Therefore, self-consistently accounting for both the EHT image characteristics and the X-ray spectra will be the most constraining avenue for modelling the mixed thermal/non-thermal synchrotron emission in M87.
Our future work will include two-temperature physics, which would allow independent evolution of the electron and ion temperatures, and perhaps shall add an additional layer of complexity in the system \citep[e.g.,][]{ressler_2015,Ryan2018,Chael2018}.
Finally, with promising new telescope additions to the EHT array in the near future and the possible advent of space-VLBI telescopes over the next few decades, resolving smaller-scale structures close to the base of the jet will gradually become a reality, playing a key role in the search for a misaligned disk in M87 and other SMBH systems.

\section*{Acknowledgements}
\label{sec:acks}
We thank Jason Dexter for interesting discussions and useful comments. We thank the anonymous referee for providing helpful suggestions on the text. 
This research was made possible by NSF PRAC awards no. 1615281 and OAC-1811605 at the Blue Waters sustained-petascale computing project and supported in part under grant no. NSF PHY-1125915.
KC, SM and DY are supported by the Netherlands Organization for Scientific Research (NWO) VICI grant (no. 639.043.513), ZY is supported by a Leverhulme Trust Early Career Research Fellowship, ML and MK were supported by the NWO Spinoza Prize, AT by Northwestern University and by National Science Foundation grants AST-1815304, AST-1911080, CH by the Amsterdam Science Talent Scholarship, and AI by a Royal Society University Research Fellowship. This research has made use of NASA’s Astrophysics Data System.

\section*{Data Availability}

Data used to plot the images in this work is uploaded at:
\href{http://doi.org/10.5281/zenodo.3757013}{http://doi.org/10.5281/zenodo.3757013}.

\bibliographystyle{mnras}
\bibliography{mybib}

\begin{thebibliography}{}
\makeatletter
\relax
\def\mn@urlcharsother{\let\do\@makeother \do\$\do\&\do\#\do\^\do\_\do\%\do\~}
\def\mn@doi{\begingroup\mn@urlcharsother \@ifnextchar [ {\mn@doi@}
  {\mn@doi@[]}}
\def\mn@doi@[#1]#2{\def\@tempa{#1}\ifx\@tempa\@empty \href
  {http://dx.doi.org/#2} {doi:#2}\else \href {http://dx.doi.org/#2} {#1}\fi
  \endgroup}
\def\mn@eprint#1#2{\mn@eprint@#1:#2::\@nil}
\def\mn@eprint@arXiv#1{\href {http://arxiv.org/abs/#1} {{\tt arXiv:#1}}}
\def\mn@eprint@dblp#1{\href {http://dblp.uni-trier.de/rec/bibtex/#1.xml}
  {dblp:#1}}
\def\mn@eprint@#1:#2:#3:#4\@nil{\def\@tempa {#1}\def\@tempb {#2}\def\@tempc
  {#3}\ifx \@tempc \@empty \let \@tempc \@tempb \let \@tempb \@tempa \fi \ifx
  \@tempb \@empty \def\@tempb {arXiv}\fi \@ifundefined
  {mn@eprint@\@tempb}{\@tempb:\@tempc}{\expandafter \expandafter \csname
  mn@eprint@\@tempb\endcsname \expandafter{\@tempc}}}

\bibitem[\protect\citeauthoryear{{Balbus} \& {Hawley}}{{Balbus} \&
  {Hawley}}{1991}]{bal91}
{Balbus} S.~A.,  {Hawley} J.~F.,  1991, \mn@doi [\apj] {10.1086/170270}, \href
  {http://adsabs.harvard.edu/cgi-bin/nph-bib_query?bibcode=1991ApJ...376..214B&db_key=AST}
  {376, 214}

\bibitem[\protect\citeauthoryear{{Blandford} \& {Znajek}}{{Blandford} \&
  {Znajek}}{1977}]{blandford77}
{Blandford} R.~D.,  {Znajek} R.~L.,  1977, \mn@doi [\mnras]
  {10.1093/mnras/179.3.433}, \href
  {http://adsabs.harvard.edu/abs/1977MNRAS.179..433B} {179, 433}

\bibitem[\protect\citeauthoryear{{Bower}, {Dexter}, {Markoff}, {Gurwell}, {Rao}
   \& {McHardy}}{{Bower} et~al.}{2015}]{Bower_2015_submm_M87}
{Bower} G.~C.,  {Dexter} J.,  {Markoff} S.,  {Gurwell} M.~A.,  {Rao} R.,
  {McHardy} I.,  2015, \mn@doi [\apjl] {10.1088/2041-8205/811/1/L6}, \href
  {https://ui.adsabs.harvard.edu/abs/2015ApJ...811L...6B} {811, L6}

\bibitem[\protect\citeauthoryear{{Caproni}, {Livio}, {Abraham}  \& {Mosquera
  Cuesta}}{{Caproni} et~al.}{2006}]{caproni06}
{Caproni} A.,  {Livio} M.,  {Abraham} Z.,   {Mosquera Cuesta} H.~J.,  2006,
  \mn@doi [\apj] {10.1086/508508}, \href
  {http://adsabs.harvard.edu/abs/2006ApJ...653..112C} {653, 112}

\bibitem[\protect\citeauthoryear{{Chael}, {Rowan}, {Narayan}, {Johnson}  \&
  {Sironi}}{{Chael} et~al.}{2018}]{Chael2018}
{Chael} A.,  {Rowan} M.,  {Narayan} R.,  {Johnson} M.,   {Sironi} L.,  2018,
  \mn@doi [\mnras] {10.1093/mnras/sty1261}, \href
  {https://ui.adsabs.harvard.edu/abs/2018MNRAS.478.5209C} {478, 5209}

\bibitem[\protect\citeauthoryear{{Chael}, {Narayan}  \& {Johnson}}{{Chael}
  et~al.}{2019}]{Chael_2019}
{Chael} A.,  {Narayan} R.,   {Johnson} M.~D.,  2019, \mn@doi [\mnras]
  {10.1093/mnras/stz988}, \href
  {https://ui.adsabs.harvard.edu/abs/2019MNRAS.486.2873C} {486, 2873}

\bibitem[\protect\citeauthoryear{{Chatterjee}, {Liska}, {Tchekhovskoy}  \&
  {Markoff}}{{Chatterjee} et~al.}{2019}]{chatterjee2019}
{Chatterjee} K.,  {Liska} M.,  {Tchekhovskoy} A.,   {Markoff} S.~B.,  2019,
  \mn@doi [\mnras] {10.1093/mnras/stz2626}, \href
  {https://ui.adsabs.harvard.edu/abs/2019MNRAS.490.2200C} {490, 2200}

\bibitem[\protect\citeauthoryear{{Connors} et~al.,}{{Connors}
  et~al.}{2017}]{Connors_2017}
{Connors} R.~M.~T.,  et~al., 2017, \mn@doi [\mnras] {10.1093/mnras/stw3150},
  \href {https://ui.adsabs.harvard.edu/abs/2017MNRAS.466.4121C} {466, 4121}

\bibitem[\protect\citeauthoryear{{Davelaar}, {Mo{\'s}cibrodzka}, {Bronzwaer}
  \& {Falcke}}{{Davelaar} et~al.}{2018}]{Davelaar2018}
{Davelaar} J.,  {Mo{\'s}cibrodzka} M.,  {Bronzwaer} T.,   {Falcke} H.,  2018,
  \mn@doi [\aap] {10.1051/0004-6361/201732025}, \href
  {https://ui.adsabs.harvard.edu/abs/2018A&A...612A..34D} {612, A34}

\bibitem[\protect\citeauthoryear{{Davelaar} et~al.,}{{Davelaar}
  et~al.}{2019}]{Davelaar2019}
{Davelaar} J.,  et~al., 2019, \mn@doi [\aap] {10.1051/0004-6361/201936150},
  \href {https://ui.adsabs.harvard.edu/abs/2019A&A...632A...2D} {632, A2}

\bibitem[\protect\citeauthoryear{Dexter \& Fragile}{Dexter \&
  Fragile}{2011}]{Dexter_2011}
Dexter J.,  Fragile P.~C.,  2011, \mn@doi [\apj] {10.1088/0004-637x/730/1/36},
  730, 36

\bibitem[\protect\citeauthoryear{Dexter \& Fragile}{Dexter \&
  Fragile}{2013}]{dexter_2013}
Dexter J.,  Fragile P.~C.,  2013, \mn@doi [\mnras] {10.1093/mnras/stt583}, 432,
  2252

\bibitem[\protect\citeauthoryear{{Dexter}, {Agol}, {Fragile}  \&
  {McKinney}}{{Dexter} et~al.}{2010}]{Dexter_2010}
{Dexter} J.,  {Agol} E.,  {Fragile} P.~C.,   {McKinney} J.~C.,  2010, \mn@doi
  [\apj] {10.1088/0004-637X/717/2/1092}, \href
  {https://ui.adsabs.harvard.edu/abs/2010ApJ...717.1092D} {717, 1092}

\bibitem[\protect\citeauthoryear{{Dexter}, {Kelly}, {Bower}, {Marrone}, {Stone}
   \& {Plambeck}}{{Dexter} et~al.}{2014}]{Dexter_2014}
{Dexter} J.,  {Kelly} B.,  {Bower} G.~C.,  {Marrone} D.~P.,  {Stone} J.,
  {Plambeck} R.,  2014, \mn@doi [\mnras] {10.1093/mnras/stu1039}, \href
  {https://ui.adsabs.harvard.edu/abs/2014MNRAS.442.2797D} {442, 2797}

\bibitem[\protect\citeauthoryear{{Doeleman} et~al.,}{{Doeleman}
  et~al.}{2008}]{doeleman2008}
{Doeleman} S.~S.,  et~al., 2008, \mn@doi [\nat] {10.1038/nature07245}, \href
  {https://ui.adsabs.harvard.edu/abs/2008Natur.455...78D} {455, 78}

\bibitem[\protect\citeauthoryear{{Drappeau}, {Dibi}, {Dexter}, {Markoff}  \&
  {Fragile}}{{Drappeau} et~al.}{2013}]{Drappeau_2013}
{Drappeau} S.,  {Dibi} S.,  {Dexter} J.,  {Markoff} S.,   {Fragile} P.~C.,
  2013, \mn@doi [\mnras] {10.1093/mnras/stt388}, \href
  {https://ui.adsabs.harvard.edu/abs/2013MNRAS.431.2872D} {431, 2872}

\bibitem[\protect\citeauthoryear{{EHTC} et~al.,}{{EHTC}
  et~al.}{2019a}]{EHT_paperI}
{EHTC} et~al., 2019a, \mn@doi [\apjl] {10.3847/2041-8213/ab0ec7}, \href
  {http://adsabs.harvard.edu/abs/2019ApJ...875L...1E} {875, L1}

\bibitem[\protect\citeauthoryear{{EHTC} et~al.,}{{EHTC}
  et~al.}{2019b}]{EHTPaperV}
{EHTC} et~al., 2019b, \mn@doi [\apjl] {10.3847/2041-8213/ab0e85}, \href
  {https://ui.adsabs.harvard.edu/abs/2019ApJ...875L...4E} {875, L5}

\bibitem[\protect\citeauthoryear{{EHTC} et~al.,}{{EHTC}
  et~al.}{2019c}]{EHTPaperVI}
{EHTC} et~al., 2019c, \mn@doi [\apjl] {10.3847/2041-8213/ab0e85}, \href
  {https://ui.adsabs.harvard.edu/abs/2019ApJ...875L...4E} {875, L6}

\bibitem[\protect\citeauthoryear{{Emmanoulopoulos}, {McHardy}  \&
  {Uttley}}{{Emmanoulopoulos} et~al.}{2010}]{Emmanoulopoulos_2010}
{Emmanoulopoulos} D.,  {McHardy} I.~M.,   {Uttley} P.,  2010, \mn@doi [\mnras]
  {10.1111/j.1365-2966.2010.16328.x}, \href
  {https://ui.adsabs.harvard.edu/abs/2010MNRAS.404..931E} {404, 931}

\bibitem[\protect\citeauthoryear{{Fish}, {Shea}  \& {Akiyama}}{{Fish}
  et~al.}{2020}]{Fish2019_space_VLBI}
{Fish} V.~L.,  {Shea} M.,   {Akiyama} K.,  2020, \mn@doi [Advances in Space
  Research] {10.1016/j.asr.2019.03.029}, \href
  {https://ui.adsabs.harvard.edu/abs/2020AdSpR..65..821F} {65, 821}

\bibitem[\protect\citeauthoryear{{Fishbone} \& {Moncrief}}{{Fishbone} \&
  {Moncrief}}{1976}]{fis76}
{Fishbone} L.~G.,  {Moncrief} V.,  1976, \apj, \href
  {http://adsabs.harvard.edu/cgi-bin/nph-bib_query?bibcode=1976ApJ...207..962F&db_key=AST}
  {207, 962}

\bibitem[\protect\citeauthoryear{{Fragile} \& {Anninos}}{{Fragile} \&
  {Anninos}}{2005}]{fragile05}
{Fragile} P.~C.,  {Anninos} P.,  2005, \mn@doi [\apj] {10.1086/428433}, \href
  {http://adsabs.harvard.edu/abs/2005ApJ...623..347F} {623, 347}

\bibitem[\protect\citeauthoryear{{Fragile} \& {Blaes}}{{Fragile} \&
  {Blaes}}{2008}]{fragile2008}
{Fragile} P.~C.,  {Blaes} O.~M.,  2008, \mn@doi [\apj] {10.1086/591936}, \href
  {https://ui.adsabs.harvard.edu/abs/2008ApJ...687..757F} {687, 757}

\bibitem[\protect\citeauthoryear{{Fragile}, {Blaes}, {Anninos}  \&
  {Salmonson}}{{Fragile} et~al.}{2007}]{fragile07}
{Fragile} P.~C.,  {Blaes} O.~M.,  {Anninos} P.,   {Salmonson} J.~D.,  2007,
  \mn@doi [\apj] {10.1086/521092}, \href
  {http://adsabs.harvard.edu/abs/2007ApJ...668..417F} {668, 417}

\bibitem[\protect\citeauthoryear{Generozov, Blaes, Fragile  \&
  Henisey}{Generozov et~al.}{2013}]{Generozov_2013}
Generozov A.,  Blaes O.,  Fragile P.~C.,   Henisey K.~B.,  2013, \mn@doi [\apj]
  {10.1088/0004-637x/780/1/81}, 780, 81

\bibitem[\protect\citeauthoryear{{Gravity Collaboration} et~al.,}{{Gravity
  Collaboration} et~al.}{2018}]{Gravity:18_hotspot}
{Gravity Collaboration} et~al., 2018, \mn@doi [\aap]
  {10.1051/0004-6361/201834294}, \href
  {https://ui.adsabs.harvard.edu/abs/2018A&A...618L..10G} {618, L10}

\bibitem[\protect\citeauthoryear{{Greene}, {Bailyn}  \& {Orosz}}{{Greene}
  et~al.}{2001}]{greene01}
{Greene} J.,  {Bailyn} C.~D.,   {Orosz} J.~A.,  2001, \mn@doi [\apj]
  {10.1086/321411}, \href {http://adsabs.harvard.edu/abs/2001ApJ...554.1290G}
  {554, 1290}

\bibitem[\protect\citeauthoryear{{Hawley}, {Guan}  \& {Krolik}}{{Hawley}
  et~al.}{2011}]{hgk11}
{Hawley} J.~F.,  {Guan} X.,   {Krolik} J.~H.,  2011, \mn@doi [\apj]
  {10.1088/0004-637X/738/1/84}, \href
  {https://ui.adsabs.harvard.edu/abs/2011ApJ...738...84H} {738, 84}

\bibitem[\protect\citeauthoryear{{Hjellming} \& {Rupen}}{{Hjellming} \&
  {Rupen}}{1995}]{hjellming95}
{Hjellming} R.~M.,  {Rupen} M.~P.,  1995, \mn@doi [\nat] {10.1038/375464a0},
  \href {http://adsabs.harvard.edu/abs/1995Natur.375..464H} {375, 464}

\bibitem[\protect\citeauthoryear{{Ingram}, {Done}  \& {Fragile}}{{Ingram}
  et~al.}{2009}]{ingram09}
{Ingram} A.,  {Done} C.,   {Fragile} P.~C.,  2009, \mn@doi [\mnras]
  {10.1111/j.1745-3933.2009.00693.x}, \href
  {http://adsabs.harvard.edu/abs/2009MNRAS.397L.101I} {397, L101}

\bibitem[\protect\citeauthoryear{{Issaoun}, {Johnson}, {Blackburn}
  et~al.}{{Issaoun} et~al.}{2019}]{Issaoun_2019}
{Issaoun} S.,  {Johnson} M.~D.,  {Blackburn} L.,   et~al., 2019, \mn@doi [\apj]
  {10.3847/1538-4357/aaf732}, \href
  {https://ui.adsabs.harvard.edu/abs/2019ApJ...871...30I} {871, 30}

\bibitem[\protect\citeauthoryear{{Ivanov} \& {Illarionov}}{{Ivanov} \&
  {Illarionov}}{1997}]{ivanov97}
{Ivanov} P.~B.,  {Illarionov} A.~F.,  1997, \mn@doi [\mnras]
  {10.1093/mnras/285.2.394}, \href
  {http://adsabs.harvard.edu/abs/1997MNRAS.285..394I} {285, 394}

\bibitem[\protect\citeauthoryear{{King} \& {Pringle}}{{King} \&
  {Pringle}}{2006}]{king06}
{King} A.~R.,  {Pringle} J.~E.,  2006, \mn@doi [\mnras]
  {10.1111/j.1745-3933.2006.00249.x}, \href
  {http://adsabs.harvard.edu/abs/2006MNRAS.373L..90K} {373, L90}

\bibitem[\protect\citeauthoryear{{Lense} \& {Thirring}}{{Lense} \&
  {Thirring}}{1918}]{lense18}
{Lense} J.,  {Thirring} H.,  1918, Physikalische Zeitschrift, \href
  {http://adsabs.harvard.edu/abs/1918PhyZ...19..156L} {19}

\bibitem[\protect\citeauthoryear{Leung, Gammie  \& Noble}{Leung
  et~al.}{2011}]{Leung_2011}
Leung P.~K.,  Gammie C.~F.,   Noble S.~C.,  2011, \mn@doi [\apj]
  {10.1088/0004-637x/737/1/21}, 737, 21

\bibitem[\protect\citeauthoryear{{Liska}, {Hesp}, {Tchekhovskoy}, {Ingram},
  {van der Klis}  \& {Markoff}}{{Liska} et~al.}{2018}]{liska_tilt_2018}
{Liska} M.,  {Hesp} C.,  {Tchekhovskoy} A.,  {Ingram} A.,  {van der Klis} M.,
  {Markoff} S.,  2018, \mn@doi [\mnras] {10.1093/mnrasl/slx174}, 474, L81

\bibitem[\protect\citeauthoryear{{Liska} et~al.,}{{Liska}
  et~al.}{2019a}]{liska_hamr2020_arxiv}
{Liska} M.,  et~al., 2019a, arXiv e-prints, \href
  {https://ui.adsabs.harvard.edu/abs/2019arXiv191210192L} {p. arXiv:1912.10192}

\bibitem[\protect\citeauthoryear{Liska, Tchekhovskoy, Ingram  \&
  van der Klis}{Liska et~al.}{2019b}]{liska_BP_2019}
Liska M.,  Tchekhovskoy A.,  Ingram A.,   van der Klis M.,  2019b, \mn@doi
  [\mnras] {10.1093/mnras/stz834}, 487, 550

\bibitem[\protect\citeauthoryear{{Liska}, {Hesp}, {Tchekhovskoy}, {Ingram},
  {van der Klis}, {Markoff}  \& {Van Moer}}{{Liska}
  et~al.}{2020}]{liska2019_tearing}
{Liska} M.,  {Hesp} C.,  {Tchekhovskoy} A.,  {Ingram} A.,  {van der Klis} M.,
  {Markoff} S.~B.,   {Van Moer} M.,  2020, \mn@doi [\mnras]
  {10.1093/mnras/staa099}, \href
  {https://ui.adsabs.harvard.edu/abs/2020MNRAS.tmp..707L} {}

\bibitem[\protect\citeauthoryear{{Lubow} \& {Ogilvie}}{{Lubow} \&
  {Ogilvie}}{2000}]{lubow00}
{Lubow} S.~H.,  {Ogilvie} G.~I.,  2000, \mn@doi [\apj] {10.1086/309101}, \href
  {http://adsabs.harvard.edu/abs/2000ApJ...538..326L} {538, 326}

\bibitem[\protect\citeauthoryear{{Lucchini}, {Krau{\ss}}  \&
  {Markoff}}{{Lucchini} et~al.}{2019}]{lucchini19}
{Lucchini} M.,  {Krau{\ss}} F.,   {Markoff} S.,  2019, \mn@doi [\mnras]
  {10.1093/mnras/stz2125}, \href
  {https://ui.adsabs.harvard.edu/abs/2019MNRAS.489.1633L} {489, 1633}

\bibitem[\protect\citeauthoryear{{Maccarone}}{{Maccarone}}{2002}]{maccarone2002b}
{Maccarone} T.~J.,  2002, \mn@doi [\mnras] {10.1046/j.1365-8711.2002.05876.x},
  \href {http://adsabs.harvard.edu/abs/2002MNRAS.336.1371M} {336, 1371}

\bibitem[\protect\citeauthoryear{{Markoff}, {Bower}  \& {Falcke}}{{Markoff}
  et~al.}{2007}]{Markoff_2007}
{Markoff} S.,  {Bower} G.~C.,   {Falcke} H.,  2007, \mn@doi [\mnras]
  {10.1111/j.1365-2966.2007.12071.x}, \href
  {https://ui.adsabs.harvard.edu/abs/2007MNRAS.379.1519M} {379, 1519}

\bibitem[\protect\citeauthoryear{{McKinney}, {Tchekhovskoy}  \&
  {Blandford}}{{McKinney} et~al.}{2013}]{mckinney_2013}
{McKinney} J.~C.,  {Tchekhovskoy} A.,   {Blandford} R.~D.,  2013, \mn@doi
  [Science] {10.1126/science.1230811}, \href
  {http://adsabs.harvard.edu/abs/2013Sci...339...49M} {339, 49}

\bibitem[\protect\citeauthoryear{{Mertens}, {Lobanov}, {Walker}  \&
  {Hardee}}{{Mertens} et~al.}{2016}]{mertens2016}
{Mertens} F.,  {Lobanov} A.~P.,  {Walker} R.~C.,   {Hardee} P.~E.,  2016,
  \mn@doi [\aap] {10.1051/0004-6361/201628829}, 595, A54

\bibitem[\protect\citeauthoryear{{Mo{\'s}cibrodzka} \&
  {Falcke}}{{Mo{\'s}cibrodzka} \& {Falcke}}{2013}]{moscibrodzka_2013}
{Mo{\'s}cibrodzka} M.,  {Falcke} H.,  2013, \mn@doi [\aap]
  {10.1051/0004-6361/201322692}, \href
  {https://ui.adsabs.harvard.edu/abs/2013A&A...559L...3M} {559, L3}

\bibitem[\protect\citeauthoryear{{Mo{\'{s}}cibrodzka}, {Gammie}, {Dolence},
  {Shiokawa}  \& {Leung}}{{Mo{\'{s}}cibrodzka}
  et~al.}{2009}]{Moscibrodzka_2009}
{Mo{\'{s}}cibrodzka} M.,  {Gammie} C.~F.,  {Dolence} J.~C.,  {Shiokawa} H.,
  {Leung} P.~K.,  2009, \mn@doi [\apj] {10.1088/0004-637x/706/1/497}, 706, 497

\bibitem[\protect\citeauthoryear{{Mo{\'s}cibrodzka}, {Falcke}  \&
  {Shiokawa}}{{Mo{\'s}cibrodzka} et~al.}{2016}]{Moscibrodzka_2016}
{Mo{\'s}cibrodzka} M.,  {Falcke} H.,   {Shiokawa} H.,  2016, \mn@doi [A\&A]
  {10.1051/0004-6361/201526630}, 586, A38

\bibitem[\protect\citeauthoryear{{Narayan}, {Igumenshchev}  \&
  {Abramowicz}}{{Narayan} et~al.}{2003}]{narayan03}
{Narayan} R.,  {Igumenshchev} I.~V.,   {Abramowicz} M.~A.,  2003, \mn@doi
  [\pasj] {10.1093/pasj/55.6.L69}, \href
  {http://adsabs.harvard.edu/abs/2003PASJ...55L..69N} {55, L69}

\bibitem[\protect\citeauthoryear{{Narayan}, {Johnson}  \& {Gammie}}{{Narayan}
  et~al.}{2019}]{Narayan_2019}
{Narayan} R.,  {Johnson} M.~D.,   {Gammie} C.~F.,  2019, \mn@doi [\apjl]
  {10.3847/2041-8213/ab518c}, \href
  {https://ui.adsabs.harvard.edu/abs/2019ApJ...885L..33N} {885, L33}

\bibitem[\protect\citeauthoryear{Palumbo, Doeleman, Johnson, Bouman  \&
  Chael}{Palumbo et~al.}{2019}]{Palumbo_2019}
Palumbo D. C.~M.,  Doeleman S.~S.,  Johnson M.~D.,  Bouman K.~L.,   Chael
  A.~A.,  2019, \mn@doi [\apj] {10.3847/1538-4357/ab2bed}, 881, 62

\bibitem[\protect\citeauthoryear{{Papaloizou} \& {Lin}}{{Papaloizou} \&
  {Lin}}{1995}]{papaloizou95}
{Papaloizou} J.~C.~B.,  {Lin} D.~N.~C.,  1995, \mn@doi [\araa]
  {10.1146/annurev.aa.33.090195.002445}, \href
  {http://adsabs.harvard.edu/abs/1995ARA%26A..33..505P} {33, 505}

\bibitem[\protect\citeauthoryear{Park, Hada, Kino, Nakamura, Ro  \&
  Trippe}{Park et~al.}{2019}]{Park_2019}
Park J.,  Hada K.,  Kino M.,  Nakamura M.,  Ro H.,   Trippe S.,  2019, \mn@doi
  [\apj] {10.3847/1538-4357/aaf9a9}, 871, 257

\bibitem[\protect\citeauthoryear{{Pasham} et~al.,}{{Pasham}
  et~al.}{2019}]{Pasham2019}
{Pasham} D.~R.,  et~al., 2019, \mn@doi [Science] {10.1126/science.aar7480},
  \href {https://ui.adsabs.harvard.edu/abs/2019Sci...363..531P} {363, 531}

\bibitem[\protect\citeauthoryear{{Polko} \& {McKinney}}{{Polko} \&
  {McKinney}}{2017}]{polko17}
{Polko} P.,  {McKinney} J.~C.,  2017, \mn@doi [\mnras] {10.1093/mnras/stw1875},
  \href {http://adsabs.harvard.edu/abs/2017MNRAS.464.2660P} {464, 2660}

\bibitem[\protect\citeauthoryear{Porth, Chatterjee, Narayan  et~al.}{Porth
  et~al.}{2019}]{Porth2019}
Porth O.,  Chatterjee K.,  Narayan R.,   et~al., 2019, \mn@doi [\apjs]
  {10.3847/1538-4365/ab29fd}, \href {https://arxiv.org/abs/1904.04923} {243(2),
  40pp}

\bibitem[\protect\citeauthoryear{{Prieto}, {Fern{\'a}ndez-Ontiveros},
  {Markoff}, {Espada}  \& {Gonz{\'a}lez-Mart{\'\i}n}}{{Prieto}
  et~al.}{2016}]{Prieto2016_M87}
{Prieto} M.~A.,  {Fern{\'a}ndez-Ontiveros} J.~A.,  {Markoff} S.,  {Espada} D.,
   {Gonz{\'a}lez-Mart{\'\i}n} O.,  2016, \mn@doi [\mnras]
  {10.1093/mnras/stw166}, \href
  {https://ui.adsabs.harvard.edu/abs/2016MNRAS.457.3801P} {457, 3801}

\bibitem[\protect\citeauthoryear{{Ressler}, {Tchekhovskoy}, {Quataert},
  {Chandra}  \& {Gammie}}{{Ressler} et~al.}{2015}]{ressler_2015}
{Ressler} S.~M.,  {Tchekhovskoy} A.,  {Quataert} E.,  {Chandra} M.,   {Gammie}
  C.~F.,  2015, \mn@doi [\mnras] {10.1093/mnras/stv2084}, \href
  {https://ui.adsabs.harvard.edu/abs/2015MNRAS.454.1848R} {454, 1848}

\bibitem[\protect\citeauthoryear{{Ressler}, {Tchekhovskoy}, {Quataert}  \&
  {Gammie}}{{Ressler} et~al.}{2017}]{ressler_2017}
{Ressler} S.~M.,  {Tchekhovskoy} A.,  {Quataert} E.,   {Gammie} C.~F.,  2017,
  \mn@doi [\mnras] {10.1093/mnras/stx364}, \href
  {http://adsabs.harvard.edu/abs/2017MNRAS.467.3604R} {467, 3604}

\bibitem[\protect\citeauthoryear{{Roelofs} et~al.,}{{Roelofs}
  et~al.}{2019}]{Roelofs_2019}
{Roelofs} F.,  et~al., 2019, \mn@doi [A\&A] {"10.1051/0004-6361/201732423"},
  625, A124

\bibitem[\protect\citeauthoryear{{Russell} et~al.,}{{Russell}
  et~al.}{2019}]{russellT_2019}
{Russell} T.~D.,  et~al., 2019, \mn@doi [\apj] {10.3847/1538-4357/ab3d36},
  \href {https://ui.adsabs.harvard.edu/abs/2019ApJ...883..198R} {883, 198}

\bibitem[\protect\citeauthoryear{{Ryan}, {Ressler}, {Dolence}, {Gammie}  \&
  {Quataert}}{{Ryan} et~al.}{2018}]{Ryan2018}
{Ryan} B.~R.,  {Ressler} S.~M.,  {Dolence} J.~C.,  {Gammie} C.,   {Quataert}
  E.,  2018, \mn@doi [\apj] {10.3847/1538-4357/aad73a}, \href
  {https://ui.adsabs.harvard.edu/abs/2018ApJ...864..126R} {864, 126}

\bibitem[\protect\citeauthoryear{{Shcherbakov}, {Penna}  \&
  {McKinney}}{{Shcherbakov} et~al.}{2012}]{Shcherbakov_2012}
{Shcherbakov} R.~V.,  {Penna} R.~F.,   {McKinney} J.~C.,  2012, \mn@doi [\apj]
  {10.1088/0004-637X/755/2/133}, \href
  {https://ui.adsabs.harvard.edu/abs/2012ApJ...755..133S} {755, 133}

\bibitem[\protect\citeauthoryear{{Sorathia}, {Krolik}  \& {Hawley}}{{Sorathia}
  et~al.}{2013}]{sorathia2013_MHD_LT}
{Sorathia} K.~A.,  {Krolik} J.~H.,   {Hawley} J.~F.,  2013, \mn@doi [\apj]
  {10.1088/0004-637X/777/1/21}, \href
  {https://ui.adsabs.harvard.edu/abs/2013ApJ...777...21S} {777, 21}

\bibitem[\protect\citeauthoryear{{Stella} \& {Vietri}}{{Stella} \&
  {Vietri}}{1998}]{stella98}
{Stella} L.,  {Vietri} M.,  1998, \mn@doi [\apjl] {10.1086/311075}, \href
  {http://adsabs.harvard.edu/abs/1998ApJ...492L..59S} {492, L59}

\bibitem[\protect\citeauthoryear{{Tchekhovskoy} \& {McKinney}}{{Tchekhovskoy}
  \& {McKinney}}{2012}]{tchekhovskoy12}
{Tchekhovskoy} A.,  {McKinney} J.~C.,  2012, \mn@doi [\mnras]
  {10.1111/j.1745-3933.2012.01256.x}, \href
  {http://adsabs.harvard.edu/abs/2012MNRAS.423L..55T} {423, L55}

\bibitem[\protect\citeauthoryear{{Tchekhovskoy}, {Narayan}  \&
  {McKinney}}{{Tchekhovskoy} et~al.}{2010}]{tchekhovskoy10}
{Tchekhovskoy} A.,  {Narayan} R.,   {McKinney} J.~C.,  2010, \mn@doi [\apj]
  {10.1088/0004-637X/711/1/50}, \href
  {http://adsabs.harvard.edu/abs/2010ApJ...711...50T} {711, 50}

\bibitem[\protect\citeauthoryear{{Tchekhovskoy}, {Narayan}  \&
  {McKinney}}{{Tchekhovskoy} et~al.}{2011}]{tch11}
{Tchekhovskoy} A.,  {Narayan} R.,   {McKinney} J.~C.,  2011, \mn@doi [\mnras]
  {10.1111/j.1745-3933.2011.01147.x}, \href
  {http://adsabs.harvard.edu/abs/2011MNRAS.418L..79T} {418, L79}

\bibitem[\protect\citeauthoryear{{Volonteri}, {Madau}, {Quataert}  \&
  {Rees}}{{Volonteri} et~al.}{2005}]{volonteri05}
{Volonteri} M.,  {Madau} P.,  {Quataert} E.,   {Rees} M.~J.,  2005, \mn@doi
  [\apj] {10.1086/426858}, \href
  {http://adsabs.harvard.edu/abs/2005ApJ...620...69V} {620, 69}

\bibitem[\protect\citeauthoryear{{Walker}, {Hardee}, {Davies}, {Ly}  \&
  {Junor}}{{Walker} et~al.}{2018}]{walker_2018_M87}
{Walker} R.~C.,  {Hardee} P.~E.,  {Davies} F.~B.,  {Ly} C.,   {Junor} W.,
  2018, \mn@doi [\apj] {10.3847/1538-4357/aaafcc}, \href
  {https://ui.adsabs.harvard.edu/abs/2018ApJ...855..128W} {855, 128}

\bibitem[\protect\citeauthoryear{White, Quataert  \& Blaes}{White
  et~al.}{2019}]{White_2019}
White C.~J.,  Quataert E.,   Blaes O.,  2019, \mn@doi [\apj]
  {10.3847/1538-4357/ab089e}, 878, 51

\bibitem[\protect\citeauthoryear{{White}, {Dexter}, {Blaes}  \&
  {Quataert}}{{White} et~al.}{2020}]{White_2020_tiltedimages}
{White} C.~J.,  {Dexter} J.,  {Blaes} O.,   {Quataert} E.,  2020, \mn@doi
  [\apj] {10.3847/1538-4357/ab8463}, \href
  {https://ui.adsabs.harvard.edu/abs/2020ApJ...894...14W} {894, 14}

\bibitem[\protect\citeauthoryear{{Younsi}, {Wu}  \& {Fuerst}}{{Younsi}
  et~al.}{2012}]{younsi_2012}
{Younsi} Z.,  {Wu} K.,   {Fuerst} S.~V.,  2012, \mn@doi [A\&A]
  {10.1051/0004-6361/201219599}, 545, A13

\bibitem[\protect\citeauthoryear{{Younsi}, {Zhidenko}, {Rezzolla}, {Konoplya}
  \& {Mizuno}}{{Younsi} et~al.}{2016}]{younsi_2016}
{Younsi} Z.,  {Zhidenko} A.,  {Rezzolla} L.,  {Konoplya} R.,   {Mizuno} Y.,
  2016, \mn@doi [\prd] {10.1103/PhysRevD.94.084025}, \href
  {https://ui.adsabs.harvard.edu/abs/2016PhRvD..94h4025Y} {94, 084025}

\bibitem[\protect\citeauthoryear{{Younsi}, {Porth}, {Mizuno}, {Fromm}  \&
  {Olivares}}{{Younsi} et~al.}{2019}]{younsi_2019_polarizedbhoss}
{Younsi} Z.,  {Porth} O.,  {Mizuno} Y.,  {Fromm} C.~M.,   {Olivares} H.,  2019,
  arXiv e-prints, \href {https://ui.adsabs.harvard.edu/abs/2019arXiv190709196Y}
  {p. arXiv:1907.09196}

\bibitem[\protect\citeauthoryear{{van den Eijnden}, {Ingram}, {Uttley},
  {Motta}, {Belloni}  \& {Gardenier}}{{van den Eijnden}
  et~al.}{2017}]{vandeneijnden17}
{van den Eijnden} J.,  {Ingram} A.,  {Uttley} P.,  {Motta} S.~E.,  {Belloni}
  T.~M.,   {Gardenier} D.~W.,  2017, \mn@doi [\mnras] {10.1093/mnras/stw2634},
  \href {http://adsabs.harvard.edu/abs/2017MNRAS.464.2643V} {464, 2643}

\makeatother
\end{thebibliography}



\appendix
\section{Considering disk- and jet-dominated emission}
\label{sec:other_Rhigh_images}

Here we discuss the GRRT images for $R_{\rm h}$ values of 1 and 100, for the two observer positions mentioned in the text, namely, edge-on to the outer disk in Sec.~\ref{sec:images_edgeon} and $17^{\circ}$ offset to the bottom large-scale jet in Sec.~\ref{sec:images_M87}.

\subsection{Edge-on case}
\label{sec:other_Rhigh_images_edgeon}

\begin{figure} 
    \includegraphics[width=\columnwidth]{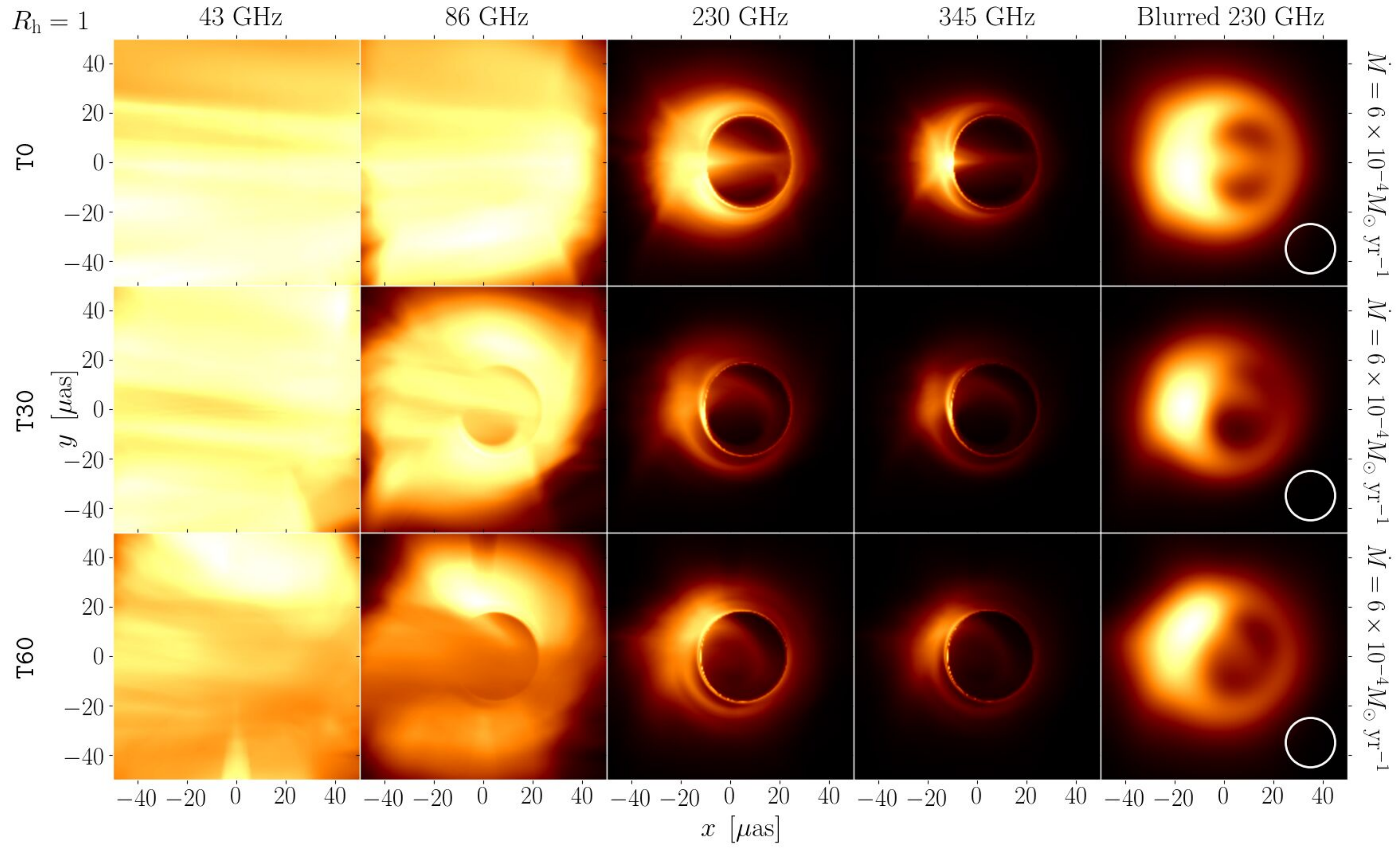}
    \caption{Time-averaged edge-on GRRT images of misaligned BH disk/jet systems using $R_{\rm h}$=1 calculated over [99960, 100960] $t_{\rm g}$ and zoomed-in to the inner $100\times 100~\upmu {\rm as}^2$ region.
    The camera is positioned at $(\theta_0, \varphi_0)$=$(90^{\circ},0^{\circ})$. From left to right: images at frequencies of 43 GHz, 86 GHz, 230 GHz and 345 GHz, along with the convolved 230 GHz image.
    From top to bottom: underlying models \tilta{}, \tiltb{} and \tiltc{}.
    With $R_{\rm h}=1$, the electron temperature is taken to be the same as that of the ion temperature everywhere, hence the disk shines brighter than the jet, which usually has a lower ion temperature.} 
    \label{fig:images_avg_rh1}
\end{figure}
\begin{figure} 
    \includegraphics[width=\columnwidth]{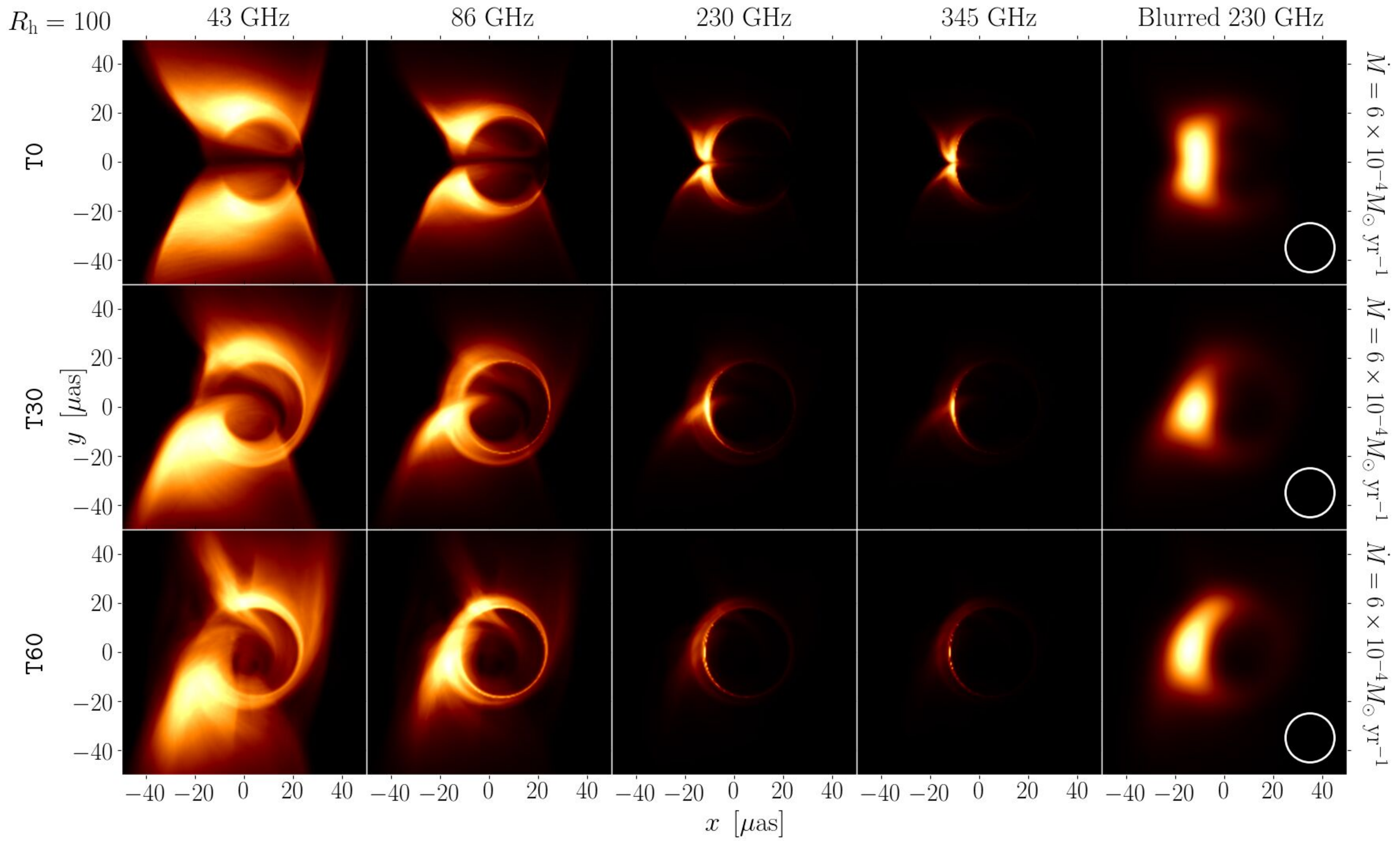}
    \caption{Same as Fig.~\ref{fig:images_avg_rh1} (edge-on case), but with $R_{\rm h}$=100.
    Here the jet contribution is more significant as the disk electron temperature is much smaller.
    The streak in front of the shadow is barely visible at 230 GHz and 345 GHz, in contrast to Fig.~\ref{fig:images_avg_rh1}.}
    \label{fig:images_avg_rh100}
\end{figure}

For $R_{\rm h}=1$, we have disk-dominated emission as the electron temperature is set to be the same as the ion temperature [see eqn.~\eqref{eqn:temp_ratio}], which is higher in the disk as compared to the jet.
The disk warp is visible as the outer accretion disk becomes increasingly optically thin, with the jet-dominated emission for $R_{\rm h}=100$. 
Higher $R_{\rm h}$ images exhibit a curved streak similar to the one discussed above for $R_{\rm h}=1$, but now offset from the shadow as the feature originates in the plunging streams and the jet edge. 

\subsection{M87 case}
\label{sec:other_Rhigh_images_M87}

\begin{figure} 
    \includegraphics[width=\columnwidth]{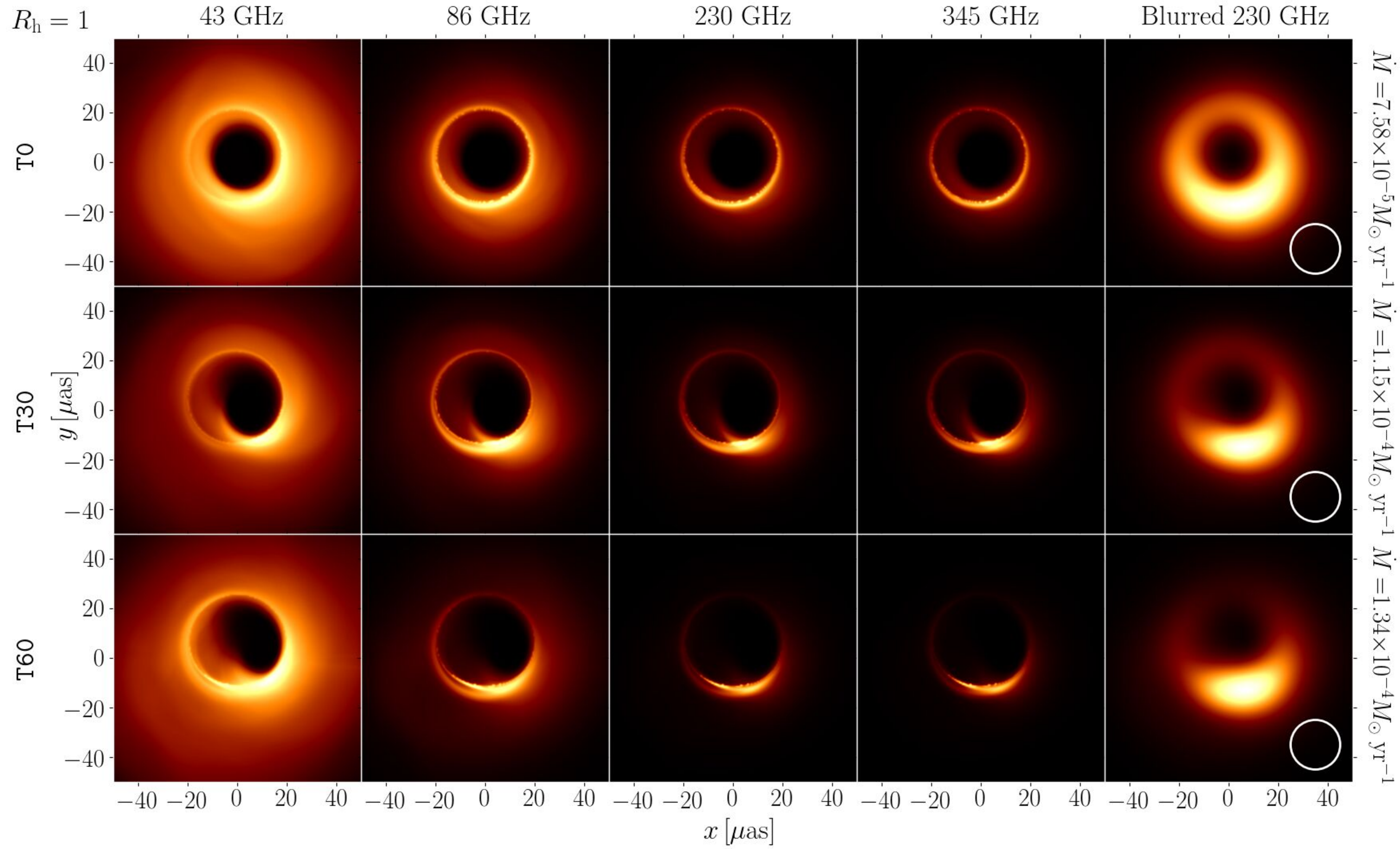}
    \caption{Time-averaged GRRT images of misaligned BH disk/jets targeted towards M87 (nearly face-on).
    Same as Fig.~\ref{fig:images_avg_m87_rh10}, but with $R_{\rm h}$=1.
    For $R_{\rm h}$=1, the disk emission dominates the image in the form of the photon ring, which is clearly distinguishable at all frequencies shown here.
    The blurred 230 GHz images look comparable to their $R_{\rm h}$=10 versions, while at 43 GHz, there is almost no sign of the jet.
    }
    \label{fig:images_avg_m87_rh1}
\end{figure}
\begin{figure} 
    \includegraphics[width=\columnwidth]{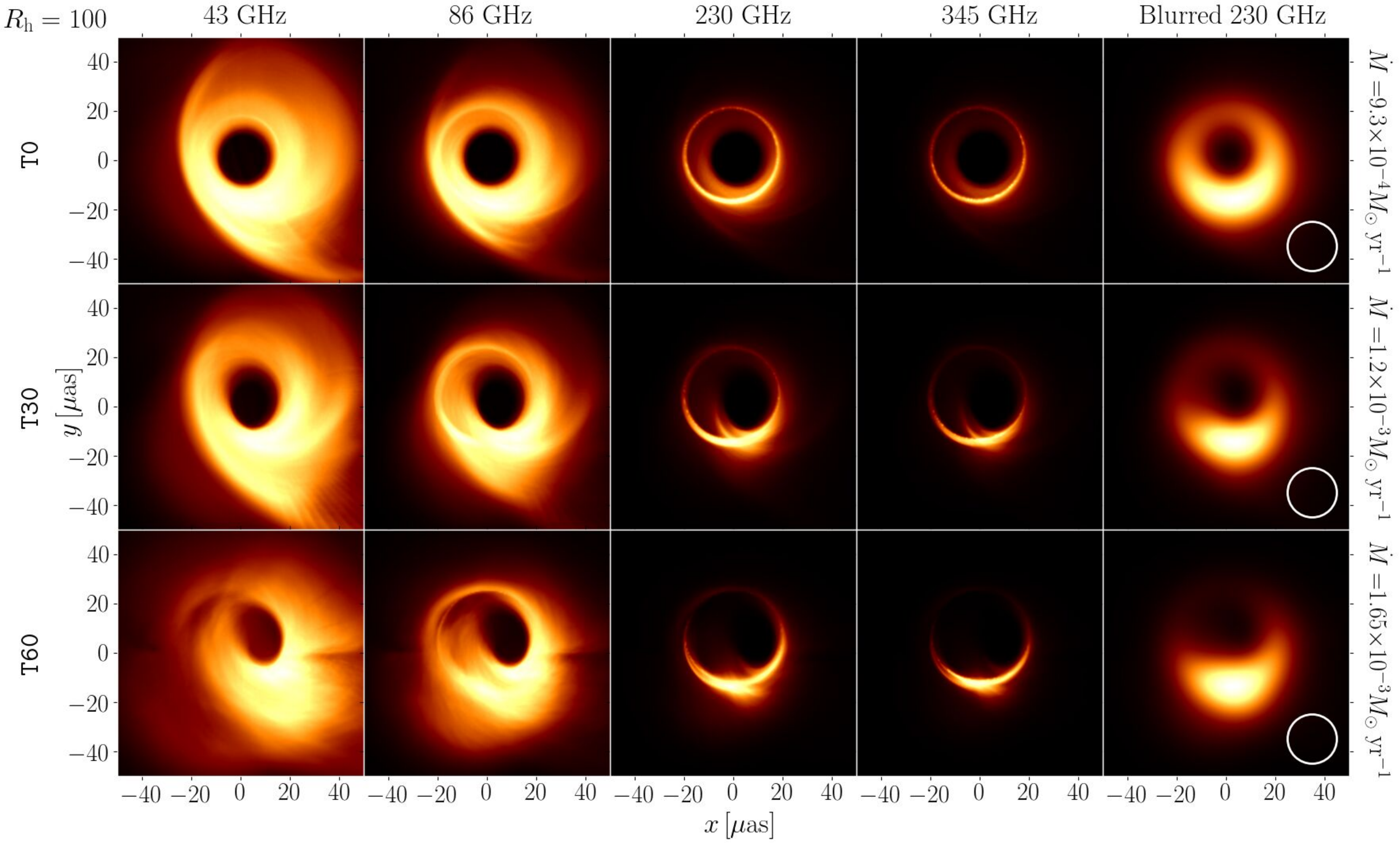}
    \caption{Same as Fig.~\ref{fig:images_avg_m87_rh1} but with $R_{\rm h}$=100. The southern crescent is slightly extended for the tilted cases due to the boosted emission of the streak cutting across the face of the shadow. The dark feature seen in the righthand side of the 43 GHz image of \tiltc{} is an artefact due to the jet sheath crossing the polar axis of the GRMHD grid.
    }
    \label{fig:images_avg_m87_rh100}
\end{figure}

The most prominent difference between $R_{\rm h}$=1 and $R_{\rm h}$=100 images lies in their brightest features at 43 GHz: the photon ring in the case of $R_{\rm h}$=1 and the extended streak in the case of $R_{\rm h}$=100.
The 43 GHz image features an optically thick accretion flow in front of the shadow region, which is absent in the corresponding $R_{\rm h}$=1 image, indicating that the streak originates in the sheath region.  

\section{Extended timing analysis: power spectra and structure functions}
\label{sec:timing_extended}

\begin{figure} 
    \includegraphics[width=\columnwidth]{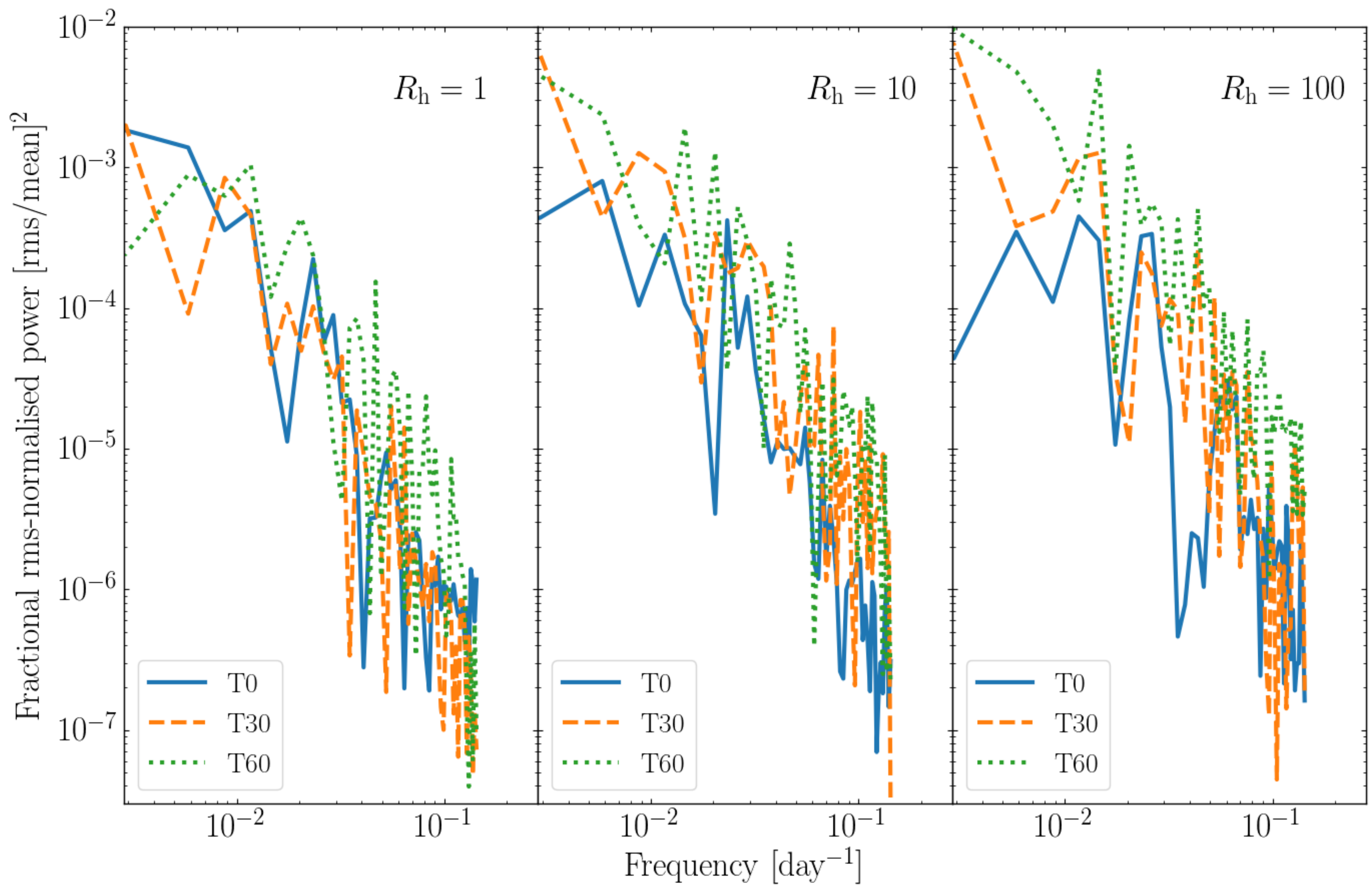}
    \caption{We show the fractional root-mean-square (rms) normalised power spectra for the 230 GHz M87 lightcurve for models \tilta{}, \tiltb{} and \tiltc{} using $R_{\rm h}=[1,~10,~100]$. We see that the short timescale variability is described well by red noise (power~$\propto$~frequency$^{-2}$). However, the long timescale variability, characterised by white noise (power independent of frequency), is not captured due to the short time duration of our lightcurves.
    }
    \label{fig:frms}
\end{figure}

\begin{figure} 
    \includegraphics[width=\columnwidth]{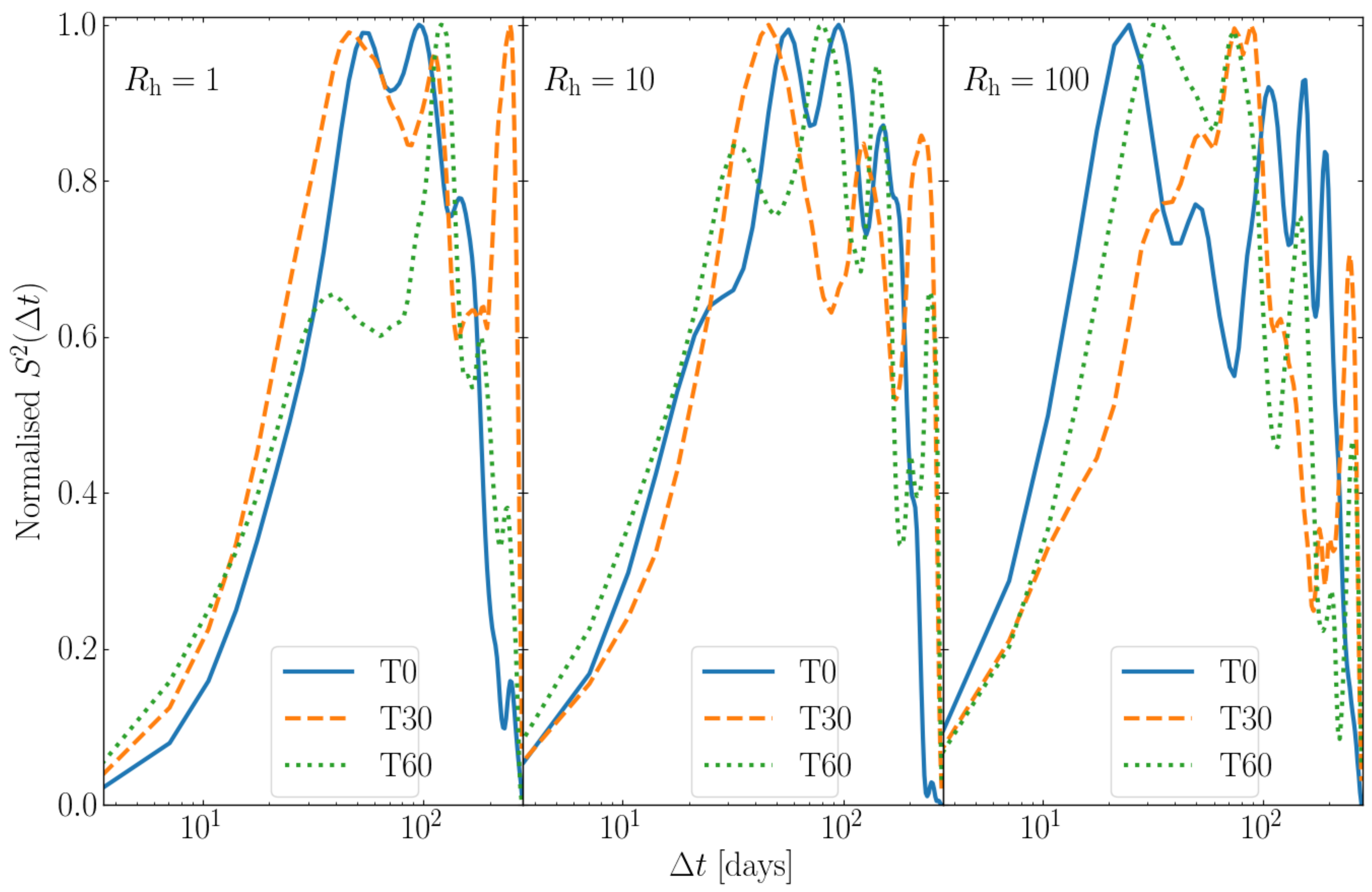}
    \caption{We show the structure function $S^{2}(\Delta t)$ (see eqn.~\eqref{eqn:structure_function}) for each tilt and $R_{\rm h}$ model. The timescale $\Delta t$ spans from $10\,t_{\rm g}\simeq3.54$ days to $1000\,t_{\rm g}\simeq354$ days. The structure functions look very similar except for model \tilta{} with $R_{\rm h}=100$, where the saturation to white noise seems to be achieved at shorter timescales as compared to the other models. Additionally, structure functions suffer from a turnover at timescales comparable to the time duration of the lightcurve, which is what we see for $\Delta t \gtrsim 150$ days.
    }
    \label{fig:sf}
\end{figure}

\begin{table}
\begin{center}
\renewcommand{\arraystretch}{1.6}
\begin{tabularx}{0.8\columnwidth}{l | c c c}
\hline\hline
\vspace*{0mm}
& & $ {\rm rms\%,} \, \tau_{\rm rms} \,{\rm[days]}$ & \\
Model & $R_{\rm h}$=1 & $R_{\rm h}$=10 & $R_{\rm h}$=100\\
\hline
\tilta{} & 7.4, 202.6  & 6.5, 157.3 & 6.0, 43.6 \\
\tiltb{} & 7.4, 130.7 & 12.4, 343.0 & 12.4, 343.0 \\
\tiltc{} & 10.9, 86.0 & 14.2, 231.7 & 16.8, 228.1 \\
\hline\hline
\end{tabularx}
\end{center}
\caption{The fractional rms amplitude and best-fit characteristic timescale ($\tau_{\rm rms}$ in days) from the power spectra (Fig.~\ref{fig:frms}) constructed from the 230 GHz M87 lightcurve for each disk/jet model and $R_{\rm h}$ model (Fig.~\ref{fig:tilt_lightcurves_M87}). The rms\% shows that higher tilt angles and more jet contribution in the emission results in larger variability, with the characteristic timescale of a few 100 days. Each power spectrum is fitted with a white noise component and a red noise component, joined smoothly together by a modified sigmoid function (see eqn.~\eqref{eqn:sigmoid}). The best fit for $\tau_{\rm rms}$ is found using a least-squares method.}
\label{tab:rms_fits}
\end{table}

In this section, we perform a Fourier transform of the 230 GHz lightcurves from Sec~\ref{sec:timing} for our M87 models.
Figure~\ref{fig:frms} shows the fractional root-mean-square (rms) normalised power spectra for each of our models, using $R_{\rm h}$=1, 10 and 100.
The cadence $(\Delta t)_{\rm min}$ of our images is $10~t_{\rm g}\equiv3.54$ days, when scaled to M87, which enables us to capture the short timescale variability in the emission.
Each power spectrum can be well-fitted by a white noise component, where the power is independent of the variability frequency, and a red noise component, where the power scales as frequency$^{-2}$, smoothly joined together by a modified sigmoid function.
The modified sigmoid function $S_{\rm var}$ is defined as a function of the variability frequency ($\nu_{\rm var} := 1/\Delta t$):
\begin{equation}
    S_{\rm var} (\nu_{\rm var} , n_{\rm var} ):=\left\{1+\exp{\left[-\left(\tau_{\rm rms} \nu_{\rm var} \right)^{n_{\rm var}}\right]}\right\}^{-1}\,,
\label{eqn:sigmoid}
\end{equation}
\noindent where $\tau_{\rm rms}$ is defined as the characteristic timescale where the variability transitions from red noise, which dominates at shorter timescales, to white noise, which is more characteristic of long timescales.
We use $n_{\rm var} =6$ to reduce the transition zone.
Table~\ref{tab:rms_fits} shows the fractional rms variability amplitude calculated from the power spectra and the corresponding characteristic transition timescale $\tau_{\rm rms}$, determined using a least-squares fitting method.
The overall trend is that with higher initial disk tilt, the rms-percentage (rms\%) increases, while higher $R_{\rm h}$ values result in larger rms\% as well (except for \tilta{}, where the rms\% is relatively unchanged).
This suggests that introducing disk misalignment and/or a higher jet contribution to the total emission results in more variability on longer timescales. Of course, the caveat is that the slope of -2 for the power law is the steepest slope one can measure reliably with a boxcar window Fourier method (which is what we used), where steeper slopes ($\leq -2$) get converted to slope=-2 due to low-frequency leakage.

\begin{table}
\begin{center}
\renewcommand{\arraystretch}{1.6}
\begin{tabularx}{0.55\columnwidth}{l | c c c}
\hline\hline
\vspace*{0mm}
& & $\tau_{\rm sf} \, {\rm [days]}$ & \\
Model & $R_{\rm h}$=1 & $R_{\rm h}$=10 & $R_{\rm h}$=100\\
\hline
\tilta{} & 26.3 & 22.8 & 10.7 \\
\tiltb{} & 14.0 & 33.1 & 34.9 \\
\tiltc{} & 38.3 & 19.9 & 12.7 \\
\hline\hline
\end{tabularx}
\end{center}
\caption{The best-fit characteristic timescale ($\tau_{\rm sf}$ in days) from the structure functions (Fig.~\ref{fig:sf}) constructed from the 230 GHz M87 lightcurve for each disk/jet model and $R_{\rm h}$ model (Fig.~\ref{fig:tilt_lightcurves_M87}), fitted with eqn.~\eqref{eqn:DRW} using a least-squares method.
The bounding box for the fitting is [3.54, 150] days, with the lower end indicating the cadence of the data sampling. The upper end is chosen such that we avoid the turnover of the structure function at timescales comparable to the time duration of the lightcurve ($\sim$1 year).}
\label{tab:sf_fits}
\end{table}

\citet{Bower_2015_submm_M87} analysed the variability of M87's 1.3mm ($\equiv$230 GHz) lightcurve over a period of 10 years using the Submillimeter Array (SMA) and found the characteristic timescale to be approximately $45^{+61}_{-24}$ days, modelling the lightcurve $s(t)$ as a damped random walk (DRW) process.
They used a structure function $S^{2}(\Delta t)$ of the form:

\begin{equation}
    S^{2}(\Delta t)=\frac{1}{N}\sum_{}^{}\left[s(t)-s(t+\Delta t)^2\right]\,,
    \label{eqn:structure_function}
\end{equation}

\noindent where $N$ is the number of data points in the lightcurve and the summation is taken over all lightcurve data points.
Figure.~$2$ in \citet[][]{Bower_2015_submm_M87} showed that the structure function for M87 increases steadily on short timescales due to red noise, saturating at long timescales due to white noise. This structure function was modelled as:

\begin{equation}
    S^{2}_{\rm DRW}(\Delta t)=S^{2}_{\infty}\left[1-\exp{\left(-\Delta t/\tau_{\rm sf}\right)}\right]\,,
    \label{eqn:DRW}
\end{equation}

\noindent with $S^{2}_{\infty}$ is the power in the lightcurve on long timescales and $\tau_{\rm sf}$ is the characteristic timescale for the transition from red to white noise.
\citet{Bower_2015_submm_M87} used a Bayesian approach to model the lightcurve due to the noisy nature of observed data and irregular spacing of the data points in time \citep[following ][]{Dexter_2014}. 

Here we choose to adopt the simpler approach of \citet[][]{Chael_2019}, since our sampling cadence is constant, and directly fit the structure functions (shown in Fig.~\ref{fig:sf}) for each tilt and $R_{\rm h}$ model with eqn.~\eqref{eqn:DRW} using a least-squares method.
Using this technique, we see that $\tau_{\rm sf}$ is of the order of a few tens of days (Table~\ref{tab:sf_fits}), in stark contrast to the $\tau_{\rm rms}$ values (Table~\ref{tab:rms_fits}).
This result exposes the primary caveat of our modelling approach for both the power spectra and the structure functions: the short time duration of our analysed lightcurves.
Figure 9 of \citet[][]{Chael_2019} shows that white noise is achieved beyond timescales of 300 days, while the lightcurves cover 4.8 years, significantly longer than the duration of our lightcurves. 

\citet[][]{Emmanoulopoulos_2010} showed that one cannot reliably estimate timescales with the structure function, which is illustrated by the incompatibility 
between $\tau_{\rm sf}$ and $\tau_{\rm rms}$. Our power spectra from the year-long lightcurve cannot explain $\tau_{\rm sf}$ values as we clearly have a sufficiently long-enough duration to see the noise transition from red to white on timescales of 10-20 days, suggesting that the timescales are longer than that given by the structure functions. However, we must note that we have very few power estimates at the lowest frequencies, and their statistical error (arising from the stochastic nature of the light curve) is equal to the powers themselves, which makes it difficult to find a good fit for 
the timescale. Further, the use of least-squares minimisation is unsuitable for finding the turnover frequency as the error distribution is exponential rather than 
Gaussian in behaviour. However, the least-squares technique is no doubt the most straightforward fitting method and provides good order-of-magnitude estimates.

\section{A glossary of images from our models}
\label{sec:flyby_T0_T60}

\begin{figure*}
    \includegraphics[width=\textwidth]{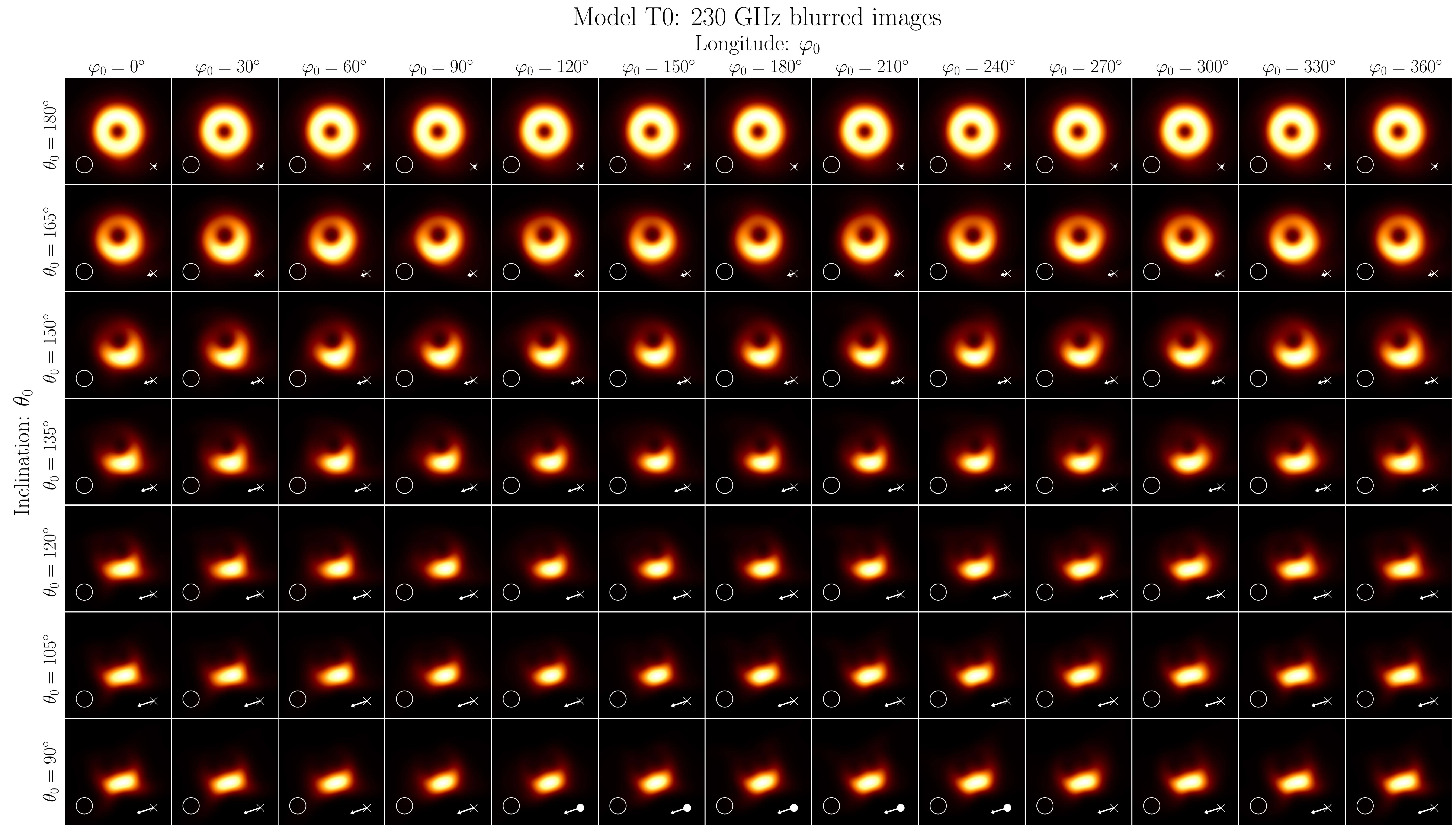}\vspace*{5mm}
    \includegraphics[width=\textwidth]{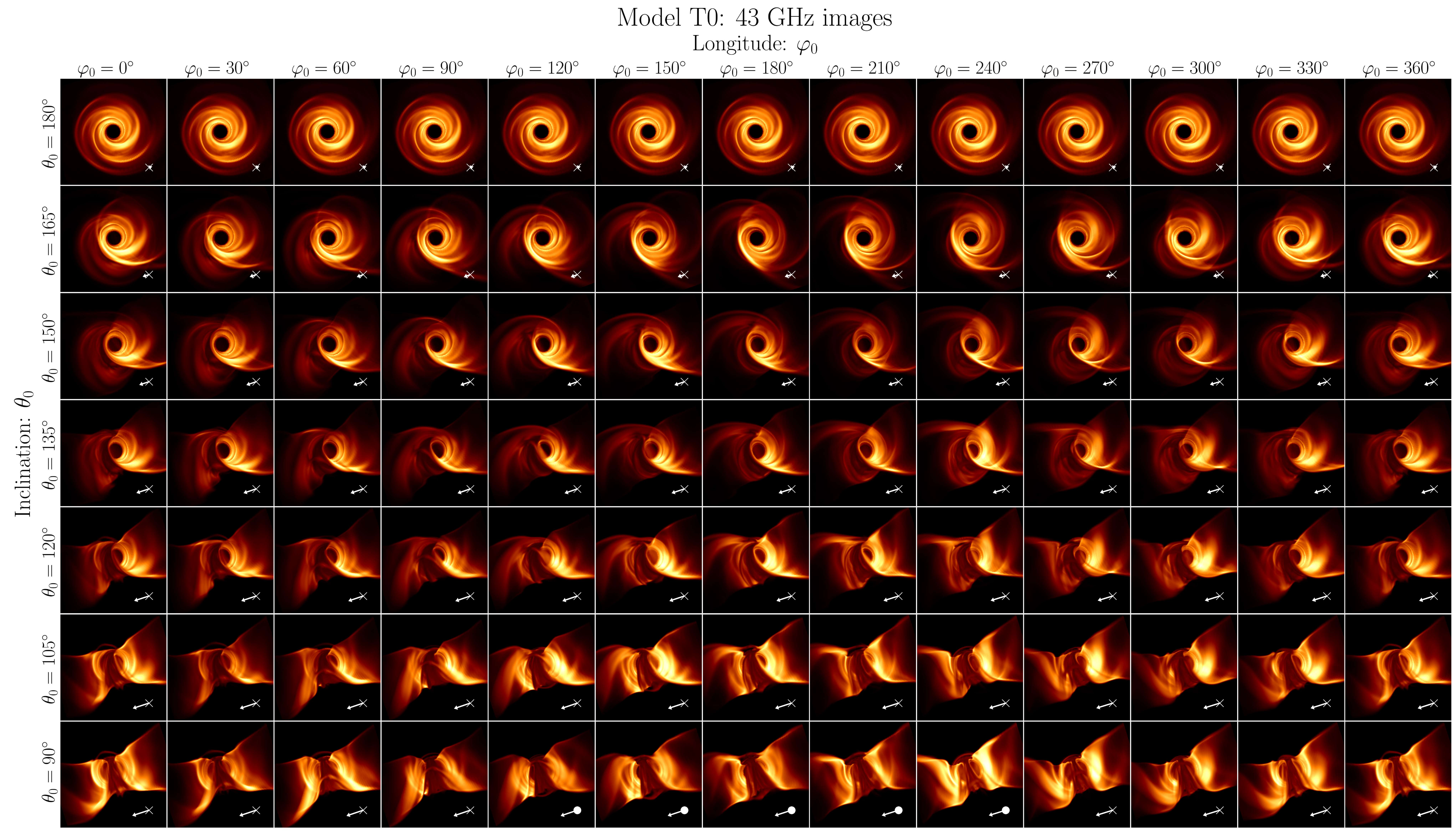}
    \caption{Single snapshot images at 230 GHz and 43 GHz scaled to M87 and rotated to fit the M87 jet PA for different values of the observer inclination $\theta_0$ and longitude $\varphi_0$, for aligned model \tilta{} with $R_{\rm h}=10$. The box sizes are $110\times 110~ \upmu {\rm as}^2$ and $150\times 150 ~\upmu {\rm as}^2$ for the 230 and the 43 GHz images respectively.
    The white arrow indicates the BH spin vector direction with the arrow length illustrating the spin projection onto the image plane.
    The X (O) indicates whether the BH spin vector is pointing into (out of) the image plane. These images are similar to the ones considered in \citet[][]{EHTPaperV}, where GRMHD models were used to interpret the M87 image.
    Refer to Sec.~\ref{sec:flyby} for further details.
    }
    \label{fig:T0_flyby}
\end{figure*}

\begin{figure*}
    \includegraphics[width=\textwidth]{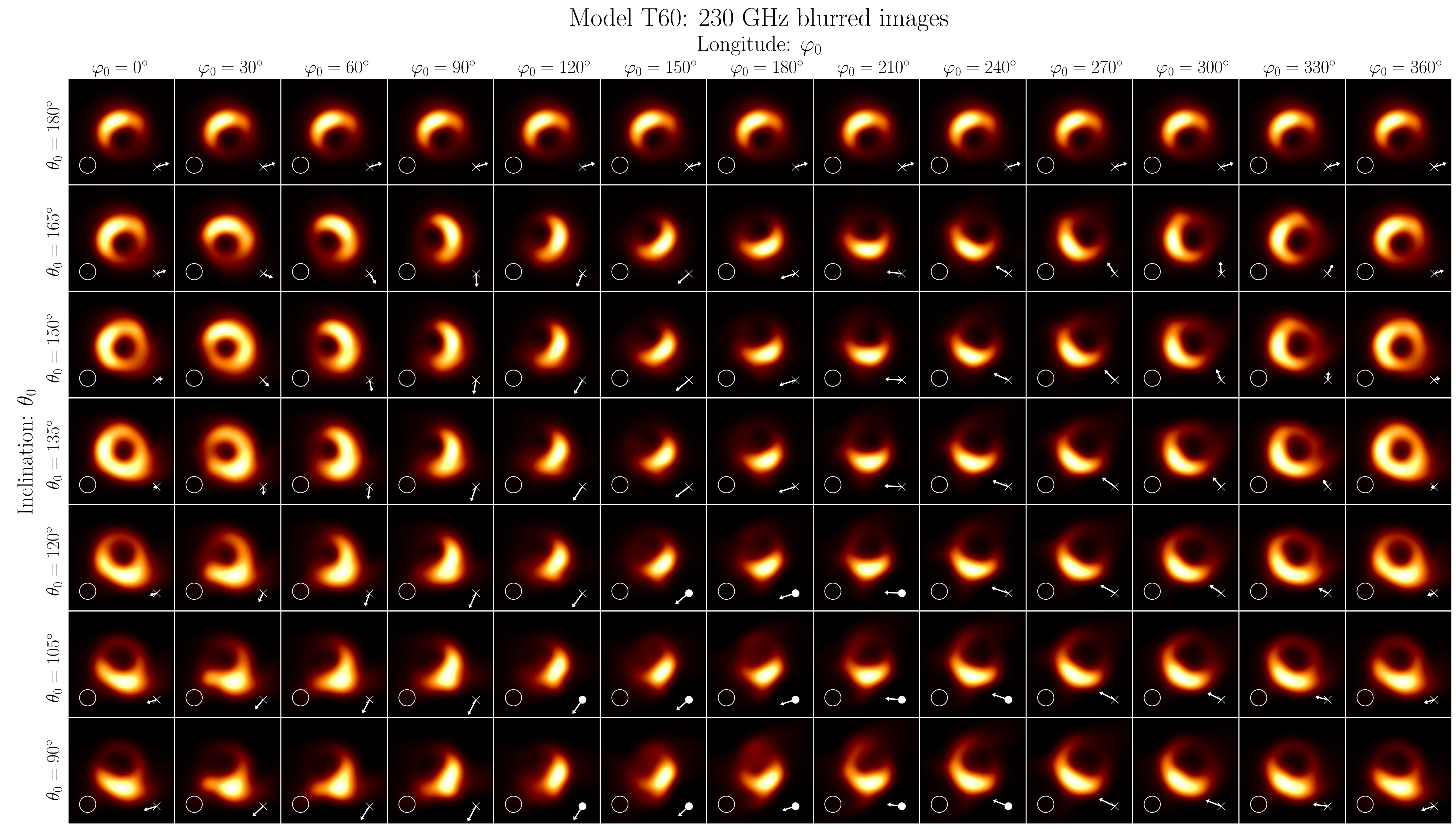}\vspace*{5mm}
    \includegraphics[width=\textwidth]{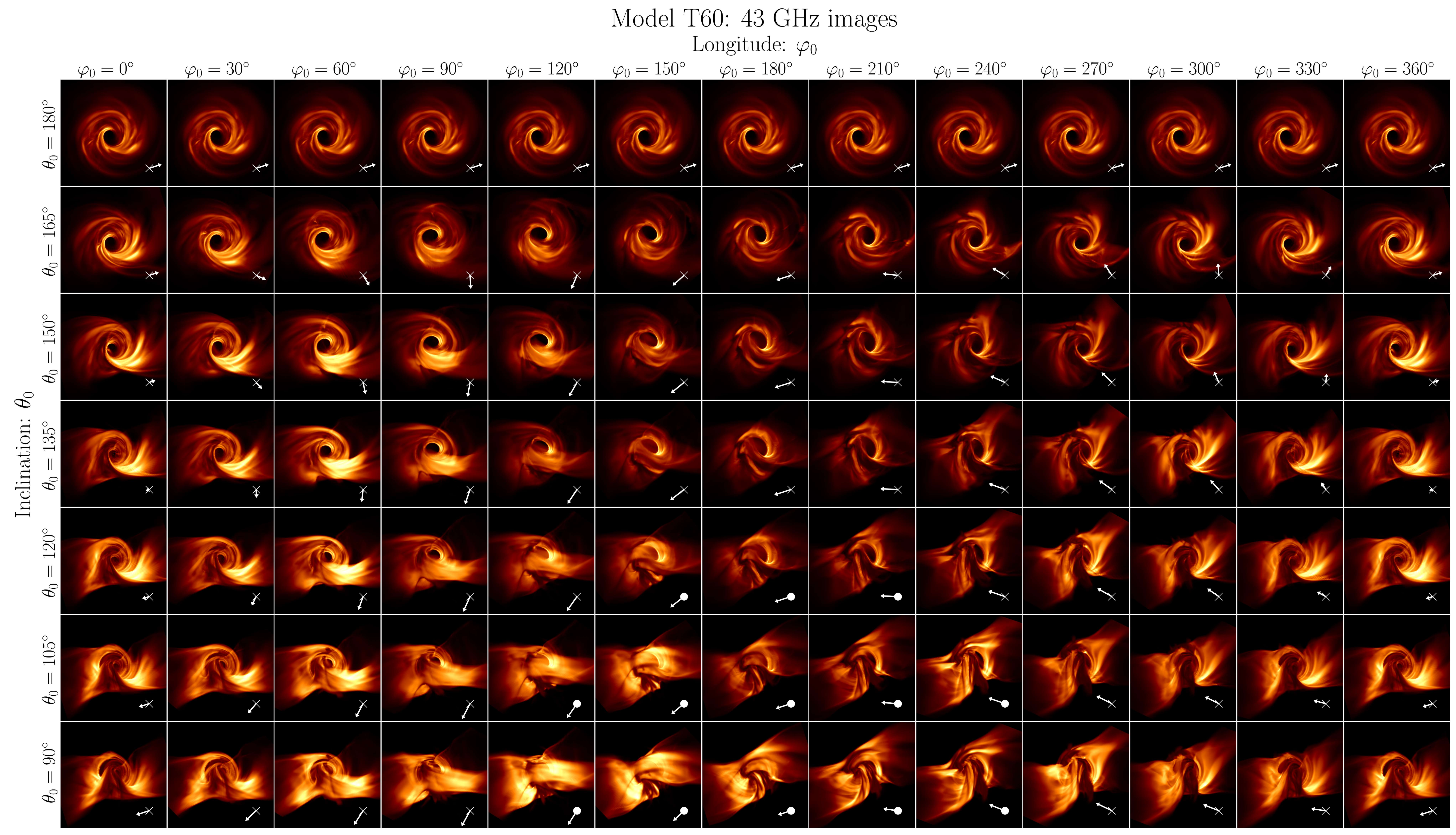}
    \caption{Single snapshot images at 230 GHz and 43 GHz scaled to M87 and rotated to fit the M87 jet PA for different values of the observer inclination $\theta_0$ and longitude $\varphi_0$, for tilted model \tiltc{} with $R_{\rm h}=10$.
    The box sizes are $110\times 110 ~\upmu {\rm as}^2$ and $150\times 150~\upmu {\rm as}^2$ for the 230 and the 43 GHz images respectively.
    The arrow direction and length indicates the BH spin direction and projection onto the image plane, with X (O) meaning into (out of) the plane.
    The warping of the disk and jet significantly alter the 43 GHz images.
    }
    \label{fig:T60_flyby}
\end{figure*}
 
Here we provide a collection of images at 230 GHz and 43 GHz for the aligned model \tilta{} (Fig.~\ref{fig:T0_flyby}) and misaligned model \tiltc{} (Fig.~\ref{fig:T60_flyby}).
Fig.~\ref{fig:T0_flyby} clearly shows that images for an aligned BH disk/jet system is independent of the camera longitude $\varphi_0$, while Fig.~\ref{fig:T60_flyby} further illustrates the powerful effect of misalignment of BH images, with more pronounced disk/jet warping as compared to Fig.~\ref{fig:T30_flyby}.

We also note that the images for the camera inclination $\theta_0=0^{\circ}$ do not change with $\varphi_0$ due to the direction of the camera ``up'' vector.
In order to illustrate the orientation of the camera, we can consider an analogue of the situation: imagine a person walking on the surface of the Earth with their eyes always directed towards the North Pole.
In this analogy, their feet are pointing downwards, towards the centre of the Earth: this vector can be thought of as the camera ``viewing'' vector that is normal to the image plane, pointing into the plane.
Their eyes are always looking straight ahead, towards North: this is the same as the camera ``up'' vector which is the positive y-axis in each of our images.
In each M87 image, the camera ``up'' vector is rotated such that the large scale GRMHD jet projected onto the image plane fits the M87 jet PA.
At $\theta_0=0^{\circ}$ (i.e., at the North Pole), the definition of a longitude fails, and hence, changing $\varphi_0$ does not change the orientation of the image.

\bsp    
\label{lastpage}
\end{document}